\documentclass[12pt]{article}
\usepackage{amsfonts}
\usepackage{amsmath}
\usepackage{amssymb}
\usepackage{array}
\usepackage{bigints}
\usepackage{booktabs}
\usepackage[nosort]{cite}
\usepackage{color}
\usepackage{dsfont}
\usepackage{float}
\usepackage{framed}
\usepackage{graphicx}
\usepackage{indentfirst}
\usepackage{mathrsfs}
\usepackage{multirow}
\usepackage{pdflscape}
\usepackage{setspace}
\usepackage{subdepth}
\usepackage{subfig}
\usepackage{titlesec}
\usepackage[dotinlabels]{titletoc}
\usepackage{wrapfig}
\usepackage[all]{xy}
\usepackage{young}
\usepackage[vcentermath]{youngtab}
\usepackage{relsize}
\usepackage{stackengine}

\usepackage{hyperref}
\hypersetup{colorlinks=true}
\hypersetup{linkcolor=black}
\hypersetup{citecolor=black}
\hypersetup{urlcolor=black}

\numberwithin{equation}{section}

\usepackage[left=2.5cm,right=2.5cm,top=2.5cm,bottom=3cm]{geometry}
\titleformat{\section}{\normalfont\bfseries}{\thesection.}{4pt}{}
\titlespacing{\section}{0pt}{20pt}{6pt}

\titleformat{\subsection}{\normalfont\itshape}{\thesubsection.}{4pt}{}
\titlespacing{\subsection}{0pt}{15pt}{6pt}

\titleformat{\subsubsection}{\normalfont}{\thesubsubsection.}{4pt}{}
\titlespacing{\subsubsection}{0pt}{15pt}{6pt}

\def\ie{\begin{equation}\begin{aligned}}
\def\fe{\end{aligned}\end{equation}}

\def\tilde{\widetilde}
\def\t{\tilde}
\def\hat{\widehat}

\def\bar{\overline}
\def\b{\bar}

\def\half{{1 \over 2}}
\def\d{\partial}

\def\ep{\varepsilon}
\def\1{{\mathds 1}}

\def\Im{\mathop{\rm Im}}

\DeclareMathOperator{\Tr}{\mathrm{Tr}}

\def\alphadot{{\dot\alpha}}

\newcommand{\Z}{{\mathbb Z}}

\newcommand{\R}{{\mathbb R}}

\def\SL{{\mathscr L}}

\def\CA{{\mathcal A}}
\def\CB{{\mathcal B}}
\def\CC{{\mathcal C}}
\def\CD{{\mathcal D}}
\def\CE{{\mathcal E}}

\def\CJ{{\mathcal J}}

\def\CM{{\mathcal M}}
\def\CN{{\mathcal N}}
\def\CO{{\mathcal O}}

\def\CR{{\mathcal R}}
\def\CS{{\mathcal S}}
\def\CT{{\mathcal T}}

\def\CV{{\mathcal V}}
\def\CW{{\mathcal W}}

\DeclareFontShape{OT1}{cmr}{mx}{n}%
    {<->cmr10}{}
\newcommand{\mytitlefont}{\fontseries{mx}\selectfont}
\DeclareMathAlphabet{\titlemath}{OT1}{cmr}{mx}{n}

\begin{document}


\begin{titlepage}

\begin{center}

~\\[2cm]

{\fontsize{26pt}{0pt} \mytitlefont Deformations of Superconformal Theories}

~\\[0.5cm]

Clay C\'{o}rdova,$^{1}$ Thomas T.~Dumitrescu,$^2$ and Kenneth Intriligator\,$^3$

~\\[0.1cm]

$^1$~{\it School of Natural Sciences, Institute for Advanced Study, Princeton, NJ 08540, USA}

$^2$~{\it Department of Physics, Harvard University, Cambridge, MA 02138, USA}

$^3$ {\it Department of Physics, University of California, San Diego, La Jolla, CA 92093, USA}

~\\[0.8cm]

\end{center}

\noindent We classify possible supersymmetry-preserving relevant, marginal, and irrelevant deformations of unitary superconformal theories in~$d \geq 3$ dimensions. Our method only relies on symmetries and unitarity. Hence, the results are model independent and do not require a Lagrangian description. Two unifying themes emerge: first, many theories admit deformations that reside in multiplets together with conserved currents. Such deformations can lead to modifications of the supersymmetry algebra by central and non-central charges.  Second, many theories with a sufficient amount of supersymmetry do not admit relevant or marginal deformations, and some admit neither.  The classification is complicated by the fact that short superconformal multiplets display a rich variety of sporadic phenomena, including supersymmetric deformations that reside in the middle of a multiplet. We illustrate our results with examples in diverse dimensions. In particular, we explain how the classification of irrelevant supersymmetric deformations can be used to derive known and new constraints on moduli-space effective actions.

\vfill

\begin{flushleft}
February 2016
\end{flushleft}

\end{titlepage}


\tableofcontents

\section{Introduction}

In this paper we consider unitary superconformal field theories (SCFTs) in~$3 \leq d \leq 6$ spacetime dimensions.\footnote{~SCFTs also exist in~$d=1,2$. They are particularly well-studied in~$d=2$, where the superconformal algebra is typically enhanced to a super-Virasoro algebra.} Our main result is a classification of their possible relevant, irrelevant, and marginal operator deformations that preserve the non-conformal Poincar\'e supersymmetries and Lorentz invariance, but not necessarily conformal symmetry. These deformations are tabulated in section~\ref{sec:tables}, which is self-contained. The classification utilizes the fact that the deforming operators  reside in unitary representations of the superconformal symmetry, which are much more constrained that representations of Poincar\'e supersymmetry.\footnote{~As we will discuss below, one consequence of this fact is that deformations of SCFTs constitute a proper subset of the deformations that can arise in more general supersymmetric theories.} Since we only rely on general properties of these representations, our results are model independent and do not require a Lagrangian.

\subsection{Deformations of Conformal Field Theories}

Quantum field theories can be thought of as renormalization group (RG) flows from short distances in the UV to long distances in the IR. The endpoints of such flows are RG fixed points, and hence scale invariant. In relativistic theories, it is common to further assume that the fixed-point theory is a  conformal field theory (CFT), whose spacetime symmetry is enhanced to 
the conformal algebra~$\frak{so}(d,2)$.\footnote{~In~$d=2$ spacetime dimensions, this enhancement follows from unitarity and Poincar\'e invariance~\cite{Zamolodchikov:1986gt,Polchinski:1987dy}. See~\cite{Nakayama:2013is} for a review of what is known in other spacetime dimensions, and~\cite{Luty:2012ww, Fortin:2012hn, Dymarsky:2013pqa,Bzowski:2014qja,Dymarsky:2014zja,Dymarsky:2015jia} for developments in~$d=4$.} In addition to free CFTs, which exist in every dimension, there is compelling evidence for a vast landscape of interacting CFTs in diverse dimensions.\footnote{~There are no known interacting CFTs in~$d > 6$.} Many of these theories are non-Lagrangian, i.e.~they do not possess a known presentation in terms of fields and a Lagrangian, and in some cases the existence of such a presentation is believed to be unlikely.  

Given a CFT, we would like to analyze the nearby quantum field theories that can be obtained by deforming it, i.e.~we would like to analyze the possible RG flows in the vicinity of the corresponding fixed point. Broadly speaking, such deformations fall into three classes: 

\begin{itemize} 
\item[\it 1.)] {\it  Adding local operators to the Lagrangian:}~This is the most common way to modify the dynamics of a theory, where the Lagrangian~$\SL$ is deformed as follows,
\begin{equation}
\delta \SL = g \CO~.\label{deform}
\end{equation}
Here~$g$ is a (typically running) coupling constant and~$\CO$ is a local operator in the original, undeformed CFT at~$g = 0$.  Note that the deformation~$\delta \SL$ in~\eqref{deform} can always be defined using conformal perturbation theory,\footnote{~Conformal perturbation theory should be valid in a sufficiently small neighborhood of the CFT. In general, we expect it to break down eventually, and it is not clear to what extent it provides a non-perturbative definition of the deformed theory. See however the discussion in~\cite{Kastor:1988ef}.} even if the original CFT is non-Lagrangian and $\SL$ is not known, or perhaps does not exist.

\item[\it 2.)]  {\it Gauging a global symmetry:} In the most familiar case, the symmetry is a continuous flavor symmetry with a conserved one-form current~$j_\mu$, but it could also be discrete or a higher-form global symmetry (see for instance~\cite{Gaiotto:2014kfa} and references therein). The gauging procedure involves projecting out some degrees of freedom from the original theory (those that are not gauge invariant) and adding new ones, which arise from the gauge fields, and it typically involves a choice of continuous or discrete coupling constants. Gauging a global symmetry may be obstructed by anomalies or lead to a theory with a Landau pole that is not UV complete. Note that gauging cannot be understood as an operator deformation~\eqref{deform}. A similar comment applies to Chern-Simons terms, which are not gauge-invariant local operators. 

\item[\it 3.)] {\it Moving onto a moduli space of vacua:} In $d>2$ non-compact spacetime dimensions, a CFT may possess a non-trivial moduli space of vacua, which is continuously connected to the conformal vacuum at the origin. This is the case for many superconformal theories. Deforming away from the conformal vacuum involves tuning the boundary conditions at spatial infinity and leads to vacuum expectation values for some fields, which generate a scale and break conformal symmetry spontaneously. Unlike the deformations in~$1.)$ and~$2.)$, which modify the dynamics of the theory at short distances, moving along a moduli space of vacua represents a modification in the deep IR, via boundary conditions. Nevertheless, one can consider an RG flow that interpolates between the UV physics of the CFT at the origin and the IR physics on the moduli space of vacua.
\end{itemize}
In this paper, we will almost exclusively focus on deformations of type~$1.)$, i.e.~adding local operators to the Lagrangian as in~\eqref{deform}.

The local operators must reside in representations of the conformal algebra $\frak{so}(d,2)$. They can be labeled by their weights under the maximal compact subalgebra~$\frak{so}(d) \times \frak{so}(2)$. Here the~$\frak{so}(d)$ weight specifies the (Wick-rotated) Lorentz representation, and the~$\frak{so}(2)$ eigenvalue is related to the scaling dimension~$\Delta$, see e.g.~\cite{Minwalla:1997ka,Dolan:2002zh,Rychkov:2016iqz} for more detail.  All unitary irreducible representations of~$\frak{so}(d,2)$ possess an operator~$\CO$, known as the conformal primary, of lowest scaling dimension~$\Delta_\CO$. It transforms in an irreducible representation~$L_\CO$ of the Lorentz group~$\frak{so}(d)$. The conformal primary~$\CO$ is thus annihilated (as an operator at the origin~$x^\mu =0$, or as a state in radial quantization) by the special conformal generators~$K_\mu$, which have scaling dimenion~$-1$. All other operators in the same~$\frak{so}(d,2)$ multiplet are descendants of~$\CO$, obtained by acting on it with an arbitrary number of spacetime derivatives~$P_\mu \sim \d_\mu$ with scaling dimension~$+1$. By contrast, the conformal primary~$\CO$ cannot be written as a total derivative of a well-defined, local operator.

There is a natural inner product on all CFT operators, which is defined by their two-point functions in flat space, or equivalently by the the inner product on the Hilbert space of states in radial quantization. In unitary theories, all primary and descendant operators must nave non-negative norms with respect to this inner product. This leads to unitarity bounds for the scaling dimension~$\Delta_\CO$ in terms of the Lorentz representation~$L_\CO$, see e.g.~\cite{Mack:1975je,Minwalla:1997ka,Rychkov:2016iqz}  
\begin{equation}\label{cftub}
\Delta_\CO \geq f\left(L_\CO\right)~.
\end{equation}
When the bound is saturated, the representation has null states, i.e.~zero-norm descendants that can be consistently removed from the representation.

The possible operator deformations~\eqref{deform} can be understood using the structure of unitary~$\frak{so}(d,2)$ representations. First note that the deformation~$\CO$ should be a conformal primary. Descendants are total derivatives of well-defined operators, and adding them to the Lagrangian leads to boundary terms that do not modify the bulk dynamics. Similarly, we do not consider deformations by the identity operator~$\1$, since these only modify the vacuum energy, but not the dynamics. If we further restrict~$\CO$ to be an~$\frak{so}(d)$ scalar, as we will do throughout most of this paper, then the deformation preserves Lorentz symmetry. In this case the strongest unitarity bound~\eqref{cftub} comes from demanding that the norm of the descendant~$\square \CO$ be non-negative,
\begin{equation}\label{scalarub}
 \Delta_\CO \geq {d-2\over 2}~.
\end{equation}
The bound is saturated if~$\CO$ is a free scalar field satisfying~$\square \CO = 0$. 

The qualitative properties of the deformation depend on the value of~$\Delta_{\CO}$ relative to the spacetime dimension $d$. This is the standard distinction between relevant, irrelevant, and marginal operators:
\begin{itemize}
\item {\it Relevant deformations ($\Delta_{\CO}<d$):} Here the CFT at~$g=0$ is the UV fixed point of an RG flow that is initiated by turning on the deformation. The relevant coupling~$g$ grows in the IR, and conformal perturbation theory in $g$ is expected to break down eventually. 

\item {\it Irrelevant deformations ($\Delta_{\CO}>d$):} In this case the CFT is the IR fixed point of an RG flow along which the irrelevant coupling~$g$ flows to zero. The deformed theory can be interpreted as an effective field theory that typically requires a UV completion. 

\item {\it Marginal deformations ($\Delta_{\CO}=d$):} These preserve conformal invariance at leading order in the coupling~$g$ and can therefore lead to a nearby fixed point. They may be further subdivided into marginally relevant, irrelevant, or exactly marginal, to indicate the direction of the RG flow once higher-order corrections are taken into account.
\end{itemize}
The only restriction on the possible deformations that follows directly from the structure of~$\frak{so}(d,2)$ representations is the unitarity bound~\eqref{scalarub}, which constrains the possible scaling dimensions of relevant deformations.  The more detailed question of which deformations actually exist in a given CFT, and when they lead to well-behaved RG flows, cannot be answered using only representation theory. 

\subsection{Superconformal Theories}

We will use the structure of unitary superconformal multiplets to analyze the possible supersymmetric deformations of unitary superconformal theories in~$3 \leq d \leq 6$ spacetime dimensions, generalizing the analysis of~\cite{Green:2010da,Argyres:2015ffa}. By this we mean deformations of the form~\eqref{deform} that preserve all Poincar\'e~$Q$-supersymmetries, but not necessarily the superconformal~$S$-supersymmetries. From now on, we will always use the term deformations to refer to such supersymmetric operator deformations.  

Together, the~$Q$- and the~$S$-supersymmetries anticommute to the superconformal algebra, whose bosonic subalgebra contains the conformal algebra~$\frak{so}(d,2)$, as well as an~$R$-symmetry algebra. Unlike the Poincar\'e supersymmetry algebra, which exists in all dimensions, superconformal algebras are highly constrained: they do not exist in~$d \geq 7$ dimensions, and in~$3 \leq d \leq 6$ dimensions the only consistent superconformal algebras are given by~\cite{Nahm:1977tg} (see also~\cite{Minwalla:1997ka} for a nice discussion),
\begin{align}\label{scftalgs}
d=3 & \qquad \frak{osp}(\CN| 4) \; \supset \; \frak{so}(3,2)\times \frak{so}(\CN)_R~, \nonumber \\[4pt]
d=4 & \hskip11pt \begin{cases}
\frak{su}(2,2|\CN) \; \supset \; \frak{so}(4,2)\times \frak{su}(\CN)_R\times \frak{u}(1)_R~,\quad \CN \neq 4~, \\ 
\frak{psu}(2,2|4)\supset \frak{so}(4,2)\times \frak{su}(4)_R~,\quad \CN =4~,
\end{cases} 
\nonumber
\\[4pt]
d=5& \qquad \frak{f}(4) \; \supset \; \frak{so}(5,2)\times \frak{su}(2)_R~, \quad \CN = 1~, \nonumber \\[4pt]
d=6 & \qquad \frak{osp}(6,2|\CN ) \; \supset \; \frak{so}(6,2)\times \frak{sp}(2\CN )_R~. 
\end{align}
In every case, we have indicated the maximal bosonic subalgebra, which factorizes into the superconformal algebra~$\frak{so}(d,2)$ and the~$R$-symmetry. As usual, $\CN \in \Z_{\geq 1}$ is a positive integer that indicates the number of supercharges in units of the minimal amount of supersymmetry that is possible in a given dimension. We use~$N_Q$ to denote the total number of independent supercharges. In~$d = 3,4,5,6$ dimensions, minimal~$\CN=1$ supersymmetry corresponds to~$N_Q = 2,4,8,8$ supercharges, respectively. For~$d = 5$, there is a unique superconformal algebra, with~$\CN=1$ supersymmetry; theories with more supersymmetry (e.g.~$\CN=2$ maximally supersymmetric Yang-Mills theory) exist, but cannot be superconformal. By contrast, the superconformal algebras in~$d = 3,4,6$ come in infinite families, labeled by a positive integer~$\CN$. However, it can be shown~\cite{multiplets} that interacting superconformal field theories only exist for~$\CN \leq 8, 4, 2$ in~$d = 3,4,6$ dimensions, respectively, and hence we will only discuss these values of~$\CN$. Note that SCFTs in six dimensions are often referred to as~$(\CN,0)$ theories. There is compelling evidence for the existence of many interacting SCFTs in these allowed ranges of~$d$ and~$\CN$. 

We will make extensive use of known facts about unitary representations of the superconformal algebras in~\eqref{scftalgs}, especially results from~\cite{Dobrev:1985qv,Minwalla:1997ka,Ferrara:2000xg,Dobrev:2002dt,Dolan:2002zh,Bhattacharya:2008zy,multiplets},\footnote{~In addition to references dedicated to the representation theory of the superconformal algebras, there are also many supergravity papers that considered these representations from the perspective of what was later understood to be the holographic AdS duals. See for instance~\cite{Gunaydin:1984fk,Gunaydin:1985tc,Gunaydin:1984wc} for a discussion of~$\half$-BPS multiplets in maximally supersymmetric SCFTs in~$d = 4,3,6$ dimensions, respectively. We will not attempt the challenging task of assembling a complete set of references.} which we briefly review. Each unitary irreducible representation of a superconformal algebra decomposes into a finite number of irreducible representations of the bosonic subalgebra, which consists of the conformal algebra~$\frak{so}(d,2)$ and the~$R$-symmetry. In other words, the superconformal multiplet decomposes into a supermultiplet of conformal representations.  Since we are interested in deformations of the form~\eqref{deform}, we will only consider conformal primaries~$\CO$, which are labeled by their Lorentz- and~$R$-symmetry representations~$L_\CO$ and~$R_\CO$, as well as their scaling dimension~$\Delta_\CO$,
\begin{equation}\label{primarylabel}
\CO \in \left[L_\CO\right]^{\left(R_\CO\right)}_{\Delta_\CO}~.
\end{equation}
Throughout the paper, we will use integer-valued Dynkin labels to specify the representations~$L_\CO, R_\CO$ (see section~\ref{sec:tables} for more details).  

Every unitary superconformal multiplet contains a unique conformal primary~$\CV$ of lowest scaling dimension~$\Delta_\CV$, which transforms irreducibly under the~$R$-symmetry. The operator~$\CV$ is known as the superconformal primary, or simply as the bottom component, of the multiplet. It is annihilated by the superconformal generators of negative scaling dimension (i.e. the~$S$-supersymmetries, with dimension~$-\half$, and the special conformal generators~$K_\mu$, with dimension~$-1$). The other operators in the superconformal multiplet (the other conformal primaries, and all conformal descendants) are superconformal descendants of~$\CV$, i.e.~they are obtained by acting on~$\CV$ with any number of~$Q$-supersymmetries, whose scaling dimension is~$+\half$. Demanding that all of these operators have non-negative norm leads to unitarity bounds for the superconformal primary. These bounds are are stronger than the bosonic unitarity bounds~\eqref{cftub} because there are more~$Q$-descendants than~$P_\mu$-descendants, all of whose norms must be non-negative. Schematically, 
\begin{equation}\label{scftub}
\Delta_\CV \geq f(L_\CV, R_\CV)~.
\end{equation}
Whenever such a bound is saturated, the representation has null states. These must themselves form a superconformal representation (though not necessarily a unitary one) that can be consistently removed from the multiplet. We will refer to all representations with null states as short, and those without null states as long.  A superconformal representation is completely determined by the quantum numbers of its superconformal primary~$\CV$. As a result, the multiplet is typically described by specifying the quantum numbers~\eqref{primarylabel} for~$\CV$. 

We are interested in supersymmetry-preserving deformations~\eqref{deform}, i.e.~conformal primaries~$\CO$ that are annihilated by the action of all~$Q$-supersymmetries, up to a total derivative. Schematically,
\begin{equation}
Q \mathcal{O}=\partial_{\mu}\left(\cdots\right)~, \label{topdef}
\end{equation}
where~$(\dots)$ denotes a well-defined operator.  The conformal primary~$\CO$ must therefore transform into a conformal descendant under all~$Q$-supersymmetries. This immediately shows that~$\CO$ cannot be the bottom component (i.e.~the superconformal primary) of its multiplet.\footnote{~The only exception is the identity operator~$\1$, which has already been excluded as a deformation.} We refer to a conformal primary~$\CO$ satisfying~\eqref{topdef} as a top component of its superconformal multiplet. (This definition does not restrict the Lorentz quantum numbers of a top component, but below we will largely focus on top components that are Lorentz scalars.) A classification of all supersymmetric deformations amounts to an enumeration of all possible top components. In order to carry out this task, it is not sufficient to know the list of allowed superconformal primaries that lead to unitary representations. Instead, the following, more detailed information is required:
\begin{itemize}
\item[1.)] The decomposition of all unitary superconformal representations into conformal primaries. This is analogous to the expansion of a superfield into components.

\item[2.)] An understanding of how the conformal primaries map into each other under the action of the~$Q$-supersymmetries, and when they map into a descendent, as in~\eqref{topdef}.
\end{itemize}

\noindent We will heavily draw on the results of~\cite{multiplets}, where a solution to~$1.)$ is presented for all unitary superconformal multiplets in~$3 \leq d \leq 6$ dimensions, generalizing the results of~\cite{Dolan:2002zh} for~$d = 4$. However, the methods of~\cite{multiplets}, do not immediately solve~$2.)$ as well. The problem is in principle straightforward:~it can be solved by explicitly expressing all conformal primaries as~$Q$-descendants of the superconformal primary and, if the multiplet is short, imposing the vanishing of all null states. This head-on approach was used in~\cite{Green:2010da,Argyres:2015ffa} to analyze the deformations of four-dimensional~$\CN=1,2$ SCFTs, but it is prohibitively tedious in many other cases. Here we will carry out a classification of supersymmetric deformations while largely sidestepping this problem. As a result, the completeness of our classification depends on some assumptions that are spelled out in section~\ref{sec:multdef}.

\subsection{Supersymmetric Deformations: Generic and Sporadic Phenomena}

There are several familiar classes of top components, and hence supersymmetric deformations, that can be described in a uniform manner. Given a superconformal primary~$\CV$, its descendants are obtained by acting with the~$Q$-supersymmetries. Operators of the form~$Q^\ell \CV$, which are obtained by acting with~$\ell$ supercharges on~$\CV$ are said to reside at level~$\ell$. The expression~$Q^\ell \CV$ should always be understood as~$\ell$ nested (anti-) commutators. Since we are only interested in conformal primaries, we can drop all spacetime derivatives, $\partial _\mu \sim P_\mu \sim 0$, so that the~$Q$-supercharges effectively anticommute,
\begin{equation}
\{Q_i, Q_j\} \sim 0~, \qquad i, j = 1, \ldots, N_Q~.\label{QQanti}
\end{equation}
Here~$N_Q$ is the total number of supercharges. By Fermi statistics, conformal primaries only occur at levels~$0 \leq \ell \leq \ell_\text{max}$, where~$\ell_\text{max}$ must satisfy the bound~$\ell_\text{max} \leq N_Q$. This bound is saturated for long multiplets, without null states, for which~$Q^{N_Q} \CV$ is the unique top component. Its quantum numbers are the same as those of~$\CV$, because~$Q^{N_Q}$ is a Lorentz and~$R$-symmetry singlet. This leads to the generalized supersymmetric~$D$-term deformation, 
\begin{equation}
\SL _D= Q^{N _Q}\CV~,\label{Llong}
\end{equation}
which is a Lorentz singlet if the superconformal primary~$\CV$ is a Lorentz singlet; its dimension is $\Delta (\SL _D)=\half N_Q+\Delta _\CV$. When a superspace formulation is available, the generalized~$D$-term~\eqref{Llong} can be written as an integral over all of superspace. A typical example is the K\"ahler potential in four-dimensional theories with~$\CN=1$ supersymmetry and~$N_Q = 4$ supercharges. 

Another common type of deformation is a generalized~$F$-term. It is constructed using a short, $\half$-BPS multiplet, whose superconformal primary~$\CV_\text{BPS}$ is annihilated by half of the supercharges.  Then the~$F$-term deformation is given by the action of the other~$\half N_Q$ supercharges on~$\CV_\text{BPS}$, 
\begin{equation}
\SL _F=Q^{\half N_Q}\CV _{\rm BPS}~.\label{Lshort}
\end{equation}
When a superspace formulation is available, a generalized~$F$-term can be written as an integral over half of superspace. A typical example is the chiral superpotential~$W$ in four-dimensional~$\CN=1$ theories, which satisfies $\bar Q_{\alphadot}W=0$ and leads to the~$F$-term deformation~$\SL _F=Q^2W$.\footnote{~Equivalently, we can impose the constraint~$\b D_\alphadot W = 0$ in superspace and write the~$F$-term deformation as~$\SL_F = \int d^2 \theta \, W$.} (The Hermitian conjugate deformation~$\b Q^2 \b W$ is also an~$F$-term.) The detailed structure of~$\half$-BPS multiplets changes for different~$d$ and~$\CN$, but they often lead to generalized~$F$-term deformations. Also, we will see below that some theories admit different types of~$F$-term deformations that reside in distinct~$\half$-BPS multiplets. 

The~$D$- and~$F$-term deformations are generic: they are constructed using multiplets that exist for all (or most) values of~$d$ and~$\CN$, and for a variety of quantum numbers. By contrast, there are deformations that reside in special, typically very short multiplets and only occur sporadically, i.e.~only for certain values of~$d$ and~$\CN$, and only when the quantum numbers of the superconformal primary take certain small values. 

As an example of such sporadic behavior, consider the stress-tensor multiplet of~$\CN=4$ SCFTs in three dimensions. (It is an~$A_2[0]_1^{(0;0)}$ multiplet, see table~\ref{tab:3DN4}.) According to~\eqref{scftalgs}, the~$R$-symmetry is~$\frak{so}(4)_R = \frak{su}(2)_R \times \frak{su}(2)'_R$, and the Lorentz algebra is also~$\frak{su}(2)$, so that we can label operators as~$[j]^{(R\,; R')}_\Delta$. Here~$R, R', j \in \Z_{\geq 0}$ are integer-valued Dynkin labels for the~$ \frak{su}(2)_R \times \frak{su}(2)'_R$ symmetry and the~$\frak{su}(2)$ Lorentz algebra.\footnote{~The standard half-integral~$\frak{su}(2)$ spins are given by~${R \over 2}, {R' \over 2}, {j \over 2}$.} The supercharges~$Q_\alpha^{i, i'}$ transform as a trifundamental~$[1]^{(1;1)}_{1/2}$. The decomposition of the stress-tensor multiplet into conformal primaries takes the following form,
 \begin{equation}
\xymatrix  @R=1pc {
 *++[F]{{[0]}^{(0;0)}_1} \ar[r]^-Q&  *++[F]{{[1]}^{(1;1)}_{3/2}} \ar[r]^-Q&  *++[F]{{[2]}^{(2;0)}_2 \oplus {[2]}^{(0;2)}_2\oplus {[0]}^{(0;0)}_2} \ar[r]^-Q&  *++[F]{{[3]}^{(1;1)}_{5/2}}\ar[r]^-Q&  *++[F]{{[4]}^{(0;0)}_3} }
 \label{3dn4em}
 \end{equation}
Here the operators~$[2]_2^{(2; 0) \oplus (0;2)}$ at level~$\ell=2$ are the~$\frak{su}(2)_R \times \frak{su}(2)'_R$ currents, the operator~$[3]^{(1;1)}_{5/2}$ at level~$\ell=3$ is the supersymmetry current, and the operator~$[4]^{(0;0)}_3$ at level~$\ell=4$ is the stress tensor. This multiplet has two top components:
\begin{itemize}
\item The stress-tensor~$[4]^{(0;0)}_3$ at level~$\ell=4$ is clearly a (Lorentz non-invariant) top component, because there are no conformal primaries at~$\ell=5$.  
\item The Lorentz scalar~$[0]^{(0;0)}_2$ at~$\ell=2$ is also a top component, even though it occurs in the middle of the multiplet. Acting on it with the~$Q$-supercharges leads to an operator with quantum numbers~$[1]^{(1;1)}_{3/2}$, but there is no such conformal primary at~$\ell=3$. 
\end{itemize}
The scalar top component at~$\ell=2$ gives rise to a relevant deformation of the theory with scaling dimension~$\Delta = 2$, just like a fermion mass term. Since it occurs in the stress-tensor multiplet, this relevant deformation exists for all three-dimensional~$\CN=4$ SCFTs, and we will refer to it as a universal mass. Its existence invalidates the standard lore that supersymmetric deformations necessarily reside at the highest level of a multiplet. (As we will explain in section~\ref{sec:multdef}, this lore is correct for suitably generic multiplets.) Similar universal mass deformations, which reside in the middle of stress-tensor multiplets, exist in three-dimensional theories with~$\CN \geq 5$ supersymmetry. Such deformations are further discussed in section~\ref{sec:defst}. As we review there, they lead to exotic deformations of the non-conformal supersymmetry algebra that includes the~$R$-symmetry generators, even though they do not commute with the supercharges. 

\renewcommand{\arraystretch}{1.8}
\renewcommand\tabcolsep{7.5pt}
\begin{table}
  \centering
  \begin{tabular}{|c |c|c|c|c| }
\hline
{\bf $d$} & {$\CN$} &  \multicolumn{1}{c}{\bf Relevant} &  \multicolumn{1}{|c|}{\bf Marginal} & {\bf Irrelevant $\Delta_\text{min}$ } \\
\hline
\hline
\multirow{6}{*}{$d = 3$} 
 & $\CN =1$ & $D$-term & $D$-term &$\Delta_{\mathrm{min}}>3$ \\
 \cline{2-5}
  & $\CN =2$ & Flavor Current, $F$-term & $F$-term &$\Delta_{\mathrm{min}}>3$ \\
\cline{2-5} 
  & $\CN =3$ & Flavor Current & --- & 4  \\
 \cline{2-5}
  & $\CN =4$ & Stress Tensor, Flavor Current & --- & 4    \\
  \cline{2-5}
  & $5\leq \CN \leq 6$ & Stress Tensor & --- &  5  \\
\cline{2-5}
& $\CN = 8$ & Stress Tensor & --- & 6  \\
\hline
\hline
\multirow{4}{*}{$d = 4$}  & $\CN =1$ & $F$-term & $F$-term &$\Delta_{\mathrm{min}}>4$ \\
\cline{2-5}
& $\CN =2$ & Flavor Current, $F$-term & $F$-term &$\Delta_{\mathrm{min}}>4$  \\
\cline{2-5}
 & $\CN =3$ & --- & --- &$\Delta_{\mathrm{min}}>4$  \\
\cline{2-5}
&  $\CN =4$ & --- & Stress Tensor & 8    \\
\hline
\hline
$d = 5$ & $\CN =1$ & Flavor Current & --- &8    \\
\hline
\hline
\multirow{2}{*}{$d = 6$} & $\CN =(1,0)$ & --- & --- &10 \\
\cline{2-5}
& $\CN =(2,0)$ & --- & --- &12   \\
\hline
\end{tabular}
  \caption{Supersymmetric deformations of interacting SCFTs in~$3 \leq d \leq 6$ dimensions. Empty entries indicate that the corresponding type of deformation does not exist. For relevant and marginal deformations, we give a crude indication of what kind of multiplet the deformation resides in. For irrelevant deformations, we list the smallest possible scaling dimension~$\Delta_\text{min}$. Additional details can be found in section~\ref{sec:tables}. As we review there, $\CN=7$ theories in three dimensions do not exist, which is why there is no corresponding entry.}
  \label{tab:intro}
\end{table} 

The main result of this paper is a classification of all Lorentz-invariant, supersymmetric deformations that can arise for SCFTs in~$3 \leq d \leq 6$ dimensions. The full classification is tabulated in section~\ref{sec:tables}, and a brief summary appears in~table~\ref{tab:intro}. Even at this level of detail, two unifying themes emerge:
\begin{itemize}
\item[\it 1.)] {\it Many theories possess special deformations that reside in multiplets together with conserved currents.} We have already mentioned the universal mass deformations for~$\CN\geq 4$ theories in three dimensions, which reside in stress-tensor multiplets. Similarly, in theories with~$N_Q = 8$ supercharges in~$d = 3,4,5$ dimensions, as well as~$\CN=2,3$ theories in~$d = 3$, the multiplet containing a conserved flavor current also contains a relevant supersymmetric deformation of dimension~$d-1$, which we will refer to as a flavor mass. These deformations modify the supersymmetry algebra by conventional central charges.  Finally, four-dimensional~$\CN=4$ theories necessarily possess an exactly marginal supersymmetric deformation, which resides in the stress-tensor multiplet.

\item[\it 2.)] {\it Many theories with a sufficient amount of supersymmetry do not admit relevant or marginal deformations.} For instance, $\CN\geq 3$ theories in three dimensions and SCFTs in five dimensions do not possess marginal deformations, and their only relevant deformations are the universal and flavor mass deformations mentioned in~$1.)$ above. Similarly, genuine~$\CN=3$ theories in four dimensions and~$\CN=(1,0)$ or~$\CN=(2,0)$ theories in six dimensions admit neither relevant nor marginal deformations, and hence they are isolated. Note that all of these theories possess supersymmetric irrelevant deformations, e.g.~$D$-terms residing in long multiplets. In theories with enough supersymmetry, irrelevant deformations can also reside in short multiplets. 
\end{itemize}

Some of our results are well known, or overlap with results that have recently been obtained by other authors. The possible deformations of four-dimensional~$\CN=1$ and~$\CN=2$ theories were classified in~\cite{Green:2010da,Argyres:2015ffa}. The fact that~$\CN=4$ theories in three dimensions and $\CN=(1,0)$ theories in six dimensions do not possess marginal deformations was independently found in~\cite{Louis:2014gxa,Louis:2015mka}, while the absence of relevant or marginal deformations in genuine~$\CN=3$ theories in four dimensions was observed in~\cite{Aharony:2015oyb}.

\subsection{Outline}

In section~\ref{sec:multdef}, we explore aspects of long and short superconformal representations, and their decomposition into supermultiplets of conformal primaries. We use various examples to illustrate possible sporadic phenomena. This leads to the notion of manifest versus accidental top components, and allows us to state the assumptions that underly our classification of supersymmetric deformations. 

Section~\ref{sec:tables} contains our main results in table form. For all values of~$d$ and~$\CN$ that can lead to interacting SCFTs, we summarize the possible shortening conditions for unitary superconformal multiplets and the possible supersymmetric deformations that preserve Lorentz invariance. The subsections describing different~$d$ and~$\CN$ are essentially self-contained and may be read independently. 

In section~\ref{sec:currdefs}, we discuss Lorentz-invariant deformations that reside in superconformal multiplets together with conserved currents, focusing on flavor currents and the stress tensor. Such deformations can lead to a modified supersymmetry algebra, which may contain central or non-central charges. The latter are particularly interesting, since they naively contradict the supersymmetric extension~\cite{Haag:1974qh} of the Coleman-Mandula theorem~\cite{Coleman:1967ad}. We also use flavor mass deformations to illustrate the fact that deformations which preserve supersymmetry at leading order need not do so at higher order. 

Section~\ref{sec:examples} contains various applications and examples. In particular, we use the classification of irrelevant supersymmetric deformations to constrain the low-energy effective Lagrangians that describe different moduli spaces of supersymmetric vacua. We also comment on the status of Fayet-Iliopoulos (FI) terms in different dimensions. In~$d \geq 4$, such terms cannot arise as deformations of SCFTs, even though they are common in supersymmetric theories with abelian gauge fields. Finally, we briefly discuss supersymmetric deformations that break Lorentz invariance.

\section{Superconformal Multiplets and Supersymmetric Deformations}

\label{sec:multdef}

As discussed in the introduction, the problem of classifying supersymmetric deformations amounts to identifying top components of superconformal multiplets, i.e.~conformal primaries that are annihilated by all~$Q$-supersymmetries up to a total derivative, as in~\eqref{topdef}. Since total spacetime derivatives play no role in this discussion they can be dropped without repercussion so that the~$Q$-supersymmetries anticommute, as in~\eqref{QQanti}, 
\begin{equation}
\{Q_i, Q_j\} \sim 0~, \qquad i, j = 1, \ldots, N_Q~.\label{QQantirep}
\end{equation}
As discussed around~\eqref{Llong}, it follows from~\eqref{QQantirep} that a superconformal multiplet can only contain a finite number of conformal primaries, which must occur at levels~$0 \leq \ell \leq \ell_\text{max}$, where~$\ell_\text{max} \leq N_Q$ by Fermi statistics. Thus, every multiplet contains at least one top component, which resides at level~$\ell_\text{max}$. In this section we will explore supermultiplets with a unique top component, as well as others that possess multiple top components. This will enable us to precisely formulate our classification scheme for supersymmetric deformations. 

\subsection{Long Multiplets and the Racah-Speiser Algorithm}

\label{sec:longrs}

Long multiplets do not possess any null states, i.e.~the supercharges~$Q_i$ do not satisfy any relations other than Fermi statistics~\eqref{QQantirep} when acting on the superconformal primary~$\CV$.  The primary~$\CV$ transforms irreducibly under the Lorentz- and~$R$-symmetry, and the 
independent conformal primaries at level~$\ell$ of a long multiplet transform in the reducible representation
\begin{equation} 
\left(\wedge^{\ell} \CR_{Q}\right) \otimes \mathcal{V}~. \label{wedge}
\end{equation}
Here~$\CR_Q$ is the Lorentz- and~$R$-symmetry representation of the supercharges and~$\wedge^\ell \CR_Q$ denotes its~$\ell$-fold totally antisymmetric wedge power.\footnote{~Note that~$\CR_Q$ may be a reducible representation, as in four dimensions and in three-dimensional~$\CN=2$ theories (see section~\ref{sec:tables}).} It follows from the antisymmetry of the wedge power that this multiplet has a unique top component at level
\begin{equation}
\ell_\text{max} = \dim \CR_Q = N_Q~.
\end{equation}
Since the maximal wedge power of~$\CR_Q$ transforms as a Lorentz and~$R$-symmetry singlet, the top component~$Q^{N_Q} \CV$ has the same Lorentz and~$R$-symmetry quantum numbers as the superconformal primary~$\CV$, but its dimension is~$\Delta_\CV + \half N_Q$. 

The structure of long multiplets is conceptually straightforward. The short multiplets are more complicated.  It is useful to use the Racah-Speiser (RS) algorithm for decomposing tensor products of Lie-algebra representations, which was applied to superconformal multiplets in~\cite{Dolan:2002zh,Kinney:2005ej,Bhattacharya:2008zy} and plays a crucial role in~\cite{multiplets}. Here we will only briefly sketch the algorithm and use it to illustrate various general features of superconformal multiplets. In broad strokes, the RS construction of a long multiplet proceeds as follows:
\begin{itemize}
\item Select the highest-weight state~$\CV_\text{h.w.} \in \CV$ of the superconformal primary with respect to both the Lorentz and the~$R$-symmetry. 

\item At each level~$\ell$, consider all sequences of~$\ell$ supercharges acting on the highest weight state~$\CV_\text{h.w.}$, which are distinct up to rearrangements using~\eqref{QQantirep}, 
\begin{equation}
Q_{i_{1}}Q_{i_{2}}\cdots Q_{i_{\ell}}\CV_\text{h.w.}~. \label{rs1}
\end{equation}
Adding the Lorentz and~$R$-symmetry weights of the supercharges in~\eqref{rs1} to those of~$\CV_\text{h.w.}$ for all such sequences leads to a set of RS trial weights~$\CW_\text{RS}^{(\ell)}$ at level~$\ell$. 

As long as the representation~$\CV$ is sufficiently large, the RS algorithm states that the highest weights of all irreducible representations that occur in~\eqref{wedge} are in one-to-one correspondence with the RS trial weights~$\CW_\text{RS}^{(\ell)}$. 

\item When the representation~$\CV$ is too small, the bijection between irreducible subrepresentations of~\eqref{wedge} and RS trial weights in~$\CW_\text{RS}^{(\ell)}$ can fail. This happens when one or several trial weights cannot be highest weights of an irreducible representation, because some of their Dynkin labels are negative. In this case the RS algorithm states that these states should be removed, possibly at the expense of also removing other weights from~$\CW_\text{RS}^{(\ell)}$, or adding new ones, according to a precise set of group-theoretic rules. 

\end{itemize}

As a simple example, consider long multiplets in three-dimensional~$\CN=1$ SCFTs. According to~\eqref{scftalgs}, the~$R$-symmetry is trivial and the Lorentz symmetry is~$\frak{su}(2)$. The supercharges~$Q_\alpha \, (\alpha = \pm)$ transform in a Lorentz doublet, which we denote as~$\CR_Q = [1]$. (As in~\eqref{3dn4em}, we use integer-valued~$\frak{su}(2)$ Dynkin labels.) If the Lorentz representation of the superconformal primary~$\CV$ is~$[n]$, it can be represented by an~$n$-index symmetric spinor~$\CV_{(\alpha_1 \, \cdots \, \alpha_n)}$ with~$\alpha_{i=1, \ldots, n} = \pm$. The RS trial states~\eqref{rs1} and their weights~$\CW_\text{RS}^{(\ell)}$ are then  
\begin{align}\label{tdnisoneexample}
\ell = 0: & \qquad \CV_\text{h.w.} = \CV_{++\cdots +}~, \hskip41pt  \CW^{(0)}_\text{RS} = \big\{[n] \big\}~,\cr
\ell = 1: & \qquad Q_{+}\CV_\text{h.w.}~,~Q_{-}\CV_\text{h.w.}~,  \qquad  \CW^{(1)}_\text{RS} = \big\{[n+1], [n-1]\big\}~, \cr
\ell = 2: & \qquad Q_{+}Q_{-}\CV_\text{h.w.}~, \hskip60pt    \CW^{(2)}_\text{RS} = \big\{[n]\big\}~.
\end{align}
For~$n\geq 1$, the Lorentz representations of conformal primaries occurring at level~$\ell$ precisely agree with~$\CW_\text{RS}^{(\ell)}$. However, when~$n = 0$ only the~$[1]$ representation occurs at~$\ell = 1$, while the~$[-1]$ representation is removed by the RS algorithm. It is important to note that the RS trial states~\eqref{rs1} generally do not coincide with the true highest-weight states of the corresponding representations. For instance, the true highest-weight state of the~$[n-1]$ representation at~$\ell = 1$ in~\eqref{tdnisoneexample} is
\begin{equation}\label{truehw}
Q_- \CV_{++\cdots +} - Q_+ \CV_{-+\cdots +}~,
\end{equation}
rather than just~$Q_- \CV_{++\cdots +}$. 

A powerful simplification afforded by the RS algorithm is that it only involves the simple trial states~\eqref{rs1}, rather than the (generally very complicated) true highest weight states that arise in the decomposition of a product representation into irreducible subrepresentations, i.e.~it bypasses the full Clebsch-Gordan problem. However, this aspect also obscures how the~$Q$-supersymmetries map different conformal primaries into each other. For instance, the structure of the trial states in~\eqref{tdnisoneexample} incorrectly suggests that~$Q_-$ maps the superconformal primary~$[n]$ into~$[n-1]$, but not~$[n+1]$. In fact, the state~$Q_-  \CV_\text{h.w.}$ can be written as a sum of~\eqref{truehw}, which is the highest-weight state of the~$[n-1]$ representation, and~$Q_- \CV_{++\cdots +} + Q_+ \CV_{-+\cdots +}$, which belongs to the~$[n+1]$ representation (but is not its highest-weight state). 

We are not aware of a simple, group-theoretic principle that predicts the possible transitions between different conformal primaries that can be achieved by acting with the~$Q$-supersymmetries. It is clear that acting with a supercharge~$Q$ on a conformal primary~$\CO$ at level~$\ell$ can only give rise to operators~$\CO'$ at level~$\ell+1$ whose Lorentz and~$R$-symmetry representation occurs in the tensor product of the supercharge representation~$\CR_Q$ with~$\CO$,
\begin{equation}\label{tp}
\CO' \subset \CR_Q \otimes \CO~.
\end{equation}
However, a given irreducible subrepresentation~$\CO'$ in this tensor product may fail to occur in the image of~$\CO$ under the action of~$Q$, for one of two reasons:
\begin{itemize}
\item[1.)] The representation~$\CO'$ does not occur at level~$\ell+1$ in the multiplet. This could be due to Fermi statistics or, if the multiplet is short, due to null states. 
\item[2.)] The representation~$\CO'$ occurs at level~$\ell+1$, but the transition~$Q: \CO \rightarrow \CO'$ does not occur, even though it is allowed by group theory. 
\end{itemize}
The occurrence of~$2.)$ depends on the detailed structure of~$\CO$ and~$\CO'$, written as~$Q$-descendants of the superconformal primary, i.e.~on their full Clebsch-Gordan decomposition. 

In order to illustrate this phenomenon, we consider a long multiplet in three-dimensional $\CN=4$ theories. As in the discussion around~\eqref{3dn4em}, the supercharges~$Q_\alpha^{i,i'}$ transform in the trifundamental~$[1]^{(1;1)}_{1/2}$ of the~$\frak{su}(2)_R \times \frak{su}(2)_R'$ symmetry and the~$\frak{su}(2)$ Lorentz symmetry. Here~$i , i', \alpha = \pm$ are doublet indices for the respective~$\frak{su}(2)$'s. For simplicity, we take the superconformal primary~$\CV \in  [0]^{(0;0)}$ to be a singlet. We examine the true highest weight states~$\CS, \CO, \CO'$ of three conformal primaries that occur at levels~$\ell = 2,3,4$ in the multiplet,
\begin{align}\label{nostrongmapex}
\ell = 2: & \qquad \CS = \left(Q_+^{+,+} Q_-^{-,-} + Q_+^{-,-} Q_-^{+,+} - Q_+^{+,-} Q_-^{-,+} - Q_+^{-,+} Q_-^{+,-}\right) \CV \;  \in \; [0]^{(0;0)} ~,\cr
\ell = 3: & \qquad \CO = Q_+^{+,+} \CS \; \in \; [1]^{(1;1)}~,\cr
\ell = 4: & \qquad \CO' = Q_+^{+,+} Q_+^{+,-}Q_+^{-,+} Q_-^{+,+} \CV \; \in\; [2]^{(2;2)}~.   
\end{align}
Note that the transition~$Q: \CO \rightarrow \CO'$ does not occur, because~$Q_+^{+,+} \CO = \left(Q_+^{+,+} \right)^2 \CS = 0$ by Fermi statistics, despite the fact that the representation~$[2]^{(2;2)}$ of~$\CO'$ occurs in the tensor product $\CR_Q \otimes \CO = [1]^{(1;1)} \otimes [1]^{(1;1)}$. (In this example~$\CO'$ is the only operator at~$\ell = 4$ that transforms in the~$[2]^{(2;2)}$ representation.)

The fact that some transitions do not occur even through they are consistent with all quantum numbers raises the possibility of accidental top components, which cannot be inferred from the decomposition of a superconformal multiplet into conformal primaries. As we argued around~\eqref{wedge} above, this does not occur in long multiplets, which have unique top components. We will now examine the short multiplets. 

\subsection{Short Multiplets and Manifest Top Components}

\label{sec:sm}

Short multiplets possess null states, which must be removed from the representation. In some cases, this can be done by simply dropping some of the supercharges and constructing a long multiplet using the remaining ones. (See for instance~\cite{Dolan:2002zh} for a discussion in four dimensions.) Since the resulting multiplets are essentially long multiplets constructed using a reduced set of~$Q$-supersymmetries, they have unique top components. More generally, the null states lead to (potentially very complicated) relations, which must be solved explicitly.\footnote{~See for instance appendix~C of~\cite{Beem:2014kka} for some intricate examples in six-dimensional~$(2,0)$ theories.} Such multiplets may possess additional top components, which can be categorized as follows:
\begin{itemize}
\item[\it 1.)] {\it Manifest Top Components:} These are conformal primaries~$\CO$ that are necessarily mapped into descendants by the~$Q$-supersymmetries because of quantum numbers. If~$\CO$ resides at level~$\ell$, then none of the conformal primaries at level~$\ell+1$ occur in the tensor product~$\CR_Q \otimes \CO$. All primaries that reside at the highest level~$\ell_\text{max}$ of a multiplet (so that there are no  conformal primaries at level~$\ell_\text{max} +1$) are examples of manifest top components. The universal mass deformation in three dimensions, which was discussed around~\eqref{3dn4em}, is also a manifest top component, even though it resides in the middle of its multiplet, i.e.~not at level~$\ell_\text{max}$.

\item[\it 2.)] {\it Accidental Top Components:} As discussed after~\eqref{nostrongmapex}, these are hypothetical conformal primaries at level~$\ell$ that are mapped into descendants, even though there are conformal primaries at level~$\ell+1$ whose quantum numbers occur in the tensor product~$\CR_Q \otimes \CO$. We do not know any examples of such accidental top components, and we suspect they do not exist, but we have not ruled them out systematically.\footnote{~In any given example, it is straightforward to check for accidental top components by explicitly solving the Clebsch-Gordan problem. We have implemented this numerically for a variety of superconformal representations and confirmed the absence of accidental top components in those cases.} 
\end{itemize}
\noindent Note that representations with multiple top components do not respect the reflection symmetry of the Clifford algebra~\eqref{QQantirep} between occupied and unoccupied levels (i.e.~particles and holes), which exchanges the~$Q$- and~$S$-supersymmetries.  

In this paper we will only discuss manifest top components, which can be analyzed using the decomposition of a superconformal multiplet into conformal primaries. 
In the remainder of this section, we illustrate various properties of manifest top components in simple examples.  The vast majority of manifest top components reside at the highest level~$\ell_\text{max}$ of a multiplet. To our knowledge, the only Lorentz-invariant deformations that reside in the middle of a multiplet are the universal mass deformations in three-dimensional theories with~$\CN\geq 4$ supersymmetry (see the discussion around~\eqref{3dn4em} and in section~\ref{sec:defst}). 

Suitably generic multiplets have a single (generally not Lorentz-invariant) operator at level~$\ell_\text{max}$. For long multiplets this was discussed around~\eqref{wedge} above. Here we consider an example of a generic short multiplet in three-dimensional~$\CN=2$ theories, where the supercharges~$Q_\alpha$ and~$\b Q_\alpha$ carry~$\frak{u}(1)_R$ charges~$-1$ and~$+1$,  i.e.~they transform reducibly as~$[1]_{1/2}^{(-1)} \oplus [1]_{1/2}^{(1)}$. Here~$[j]^{(r)}_\Delta$ denotes an operator of Lorentz spin~$\half j \in \half \Z_{\geq 0}$, $R$-charge~$r$, and scaling dimension~$\Delta$. Consider an~$A_1\b A_1[j]^{(0)}_{\half j +1}$ multiplet, which obeys a shortening condition of type~$A_1$ with respect to~$Q_\alpha$ (see table~\ref{tab:3DN2C}) and a shortening condition of type~$\b A_1$ with respect to~$\b Q_\alpha$ (see table~\ref{tab:3DN2AC}), with generic Lorentz spin and vanishing~$R$-charge, $r=0$.  The superconformal multiplet decomposes into the following conformal primaries:
\begin{equation}
\xymatrix @C=7pc @R=7pc @!0 @dr {
*++[F]{[j]_{\half j+1}^{(0)}}  \ar[r]|--{{~\b Q~}} \ar[d]|--{~Q~} 
& *++[F]{[j+1]_{\half (j+3)}^{(+1)}}  
\ar[d]|--{~Q~}
\\
*++[F]{[j+1]_{\half (j+3)}^{(-1)}} 
\ar[r]|--{~\b Q~}
& *++[F]{[j+2]_{\half j+2}^{(0)}}
}
\label{3dn2gg}
\end{equation}
\noindent Note that there is a unique operator at~$\ell_\text{max} = 2$. This multiplet exists for any~$j \geq 1$. It only contains conserved currents (generally with high spin). Taking into account the conservation laws leads to~$4+4$ independent operators, independent of~$j$. The case~$j = 2$ is the superconformal stress-tensor multiplet. 

As the Lorentz and~$R$-symmetry quantum numbers of a short multiplet are specialized to small values, we encounter a host of sporadic phenomena that can result in additional top components. We have analyzed these phenomena on a case-by-case basis, by relying on the explicit construction of unitary superconformal multiplets in~\cite{multiplets}. As an example, consider an~$A_2 \b A_2[0]_1^{(0)}$ flavor current multiplet in three-dimensional~$\CN =2$ theories, which obeys a~$Q$-shortening condition of type~$A_2$ (see table~\ref{tab:3DN2C}) and a~$\b Q$-shortening condition of type~$\b A_2$ (see table~\ref{tab:3DN2AC}). It can be viewed as the specialization of the~$A_1\b A_1[j]^{(0)}_{\half j+1}$ multiplets discussed above to~$j = 0$. As indicated by the subscripts on~$A$ and~$\b A$, the primary null states jump from~$\ell=1$ to~$\ell=2$. The component decomposition of the~$A_2 \b A_2[0]_1^{(0)}$ multiplet is well known,
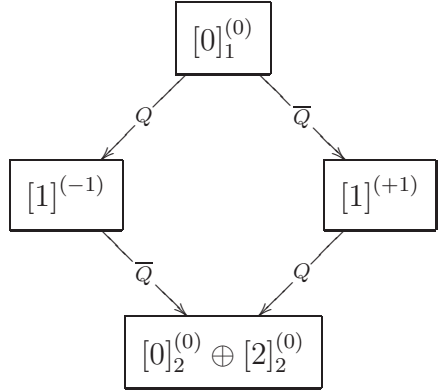
\begin{equation*}
\xymatrix @C=7pc @R=7pc @!0 @dr {
*++[F]{[0]_1^{(0)}}  \ar[r]|--{{~\b Q~}} \ar[d]|--{~Q~} 
& *++[F]{[1]^{(+1)}}  
\ar[d]|--{~Q~}
\\
*++[F]{[1]^{(-1)}} 
\ar[r]|--{~\b Q~}
& *++[F]{[0]_2^{(0)} \oplus [2]_2^{(0)}}
}
\label{3dn2g}
\end{equation*}
\noindent
There are now two manifest top components at~$\ell_\text{max} = 2$. The conserved flavor current~$[2]_2^{(0)}$ is the generic top component, i.e.~the specialization of the top component~$[j+2]^{(0)}_{\half j+2}$ in~\eqref{3dn2gg} to~$j=0$. The additional scalar~$[0]_2^{(0)}$ is special to~$j =0$. It gives rise to the Lorentz-invariant flavor mass deformation in these theories, which is further discussed in section~\ref{sec:flavormass}. 

As in the previous example, many multiplets with multiple top components contain conserved currents, but that is not always the case. Consider an example in four-dimensional $\CN=2$ SCFTs. Operators are labeled as~$[\, j; \b j\,]^{(R \, ; \, r)}_\Delta$, where~$j, \b j$ and~$R$ are Dynkin labels for the~$\frak{so}(3,1) = \frak{su}(2) \times \b{\frak{su}(2)}$ Lorentz symmetry and the~$\frak{su}(2)_R$ symmetry, while~$r$ is the~$\frak{u}(1)_R$ charge and~$\Delta$ is the scaling dimension. The supercharges~$Q_\alpha$ and~$\b Q_\alphadot$ transform as~$[1;0]^{(1;-1)}_{1/2} \oplus [0;1]^{(1;+1)}_{1/2}$. Consider an~$A_2\b A_2[0;0]^{(R; 0)}$ multiplet, which obeys a shortening condition of type~$A_2$ with respect to~$Q_\alpha$ (see table~\ref{tab:3DN2C}) and a shortening condition of type~$\b A_2$ with respect to~$\b Q_\alphadot$ (see table~\ref{tab:3DN2AC}). The superconformal primary is given by~$\CV = [0;0]^{(R; 0)}_{R+2}$, and any~$R \in \Z_{\geq 0}$ is allowed. For sufficiently large~$R$, there is a single top component at level~$\ell_\text{max} = 6$, which transforms as~$[1;1]^{(R-2\,; \,0)}_{R+5}$. (For generic~$R$, the multiplet is tabulated in equation~(4.37) of~\cite{Dolan:2002zh}.) However, when~$R = 0,1$ this top component disappears, and the multiplet undergoes further shortening, i.e.~$\ell_\text{max}$ decreases. The case~$R = 0$ is the stress-tensor multiplet, which also contains the conserved~$R$-symmetry and supersymmetry currents; the stress tensor is the unique top component at~$\ell_\text{max} = 4$. However, for~$R = 1$ the multiplet has two manifest top components, at~$\ell_\text{max} = 5$, as seen in its explicit decomposition into conformal primaries:
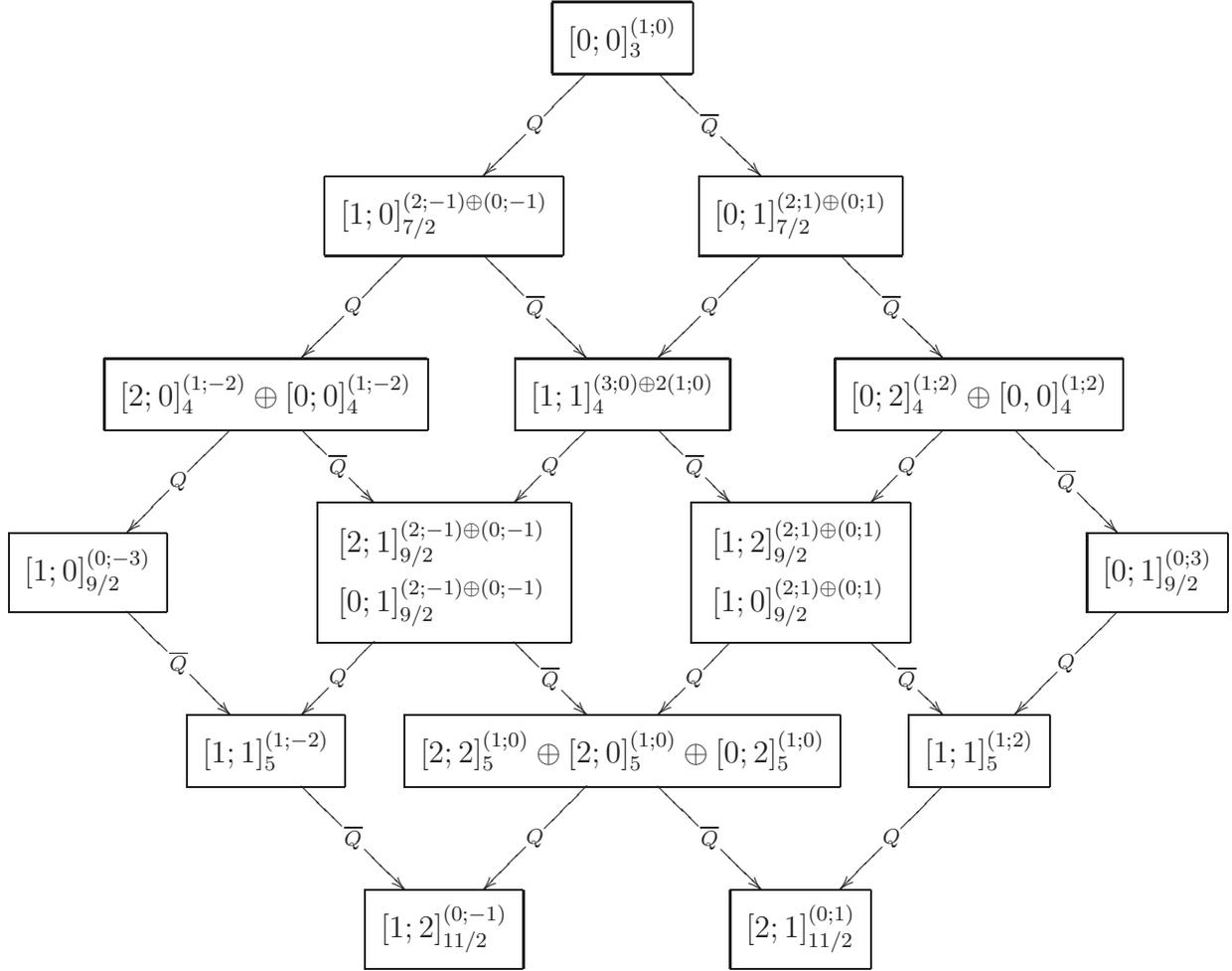
\begin{align*}
\xymatrix @C=8pc @R=8pc @!0 @dr {
*++[F]{[0;0]^{(1;0)}_3} 
\ar@{}[r]^(.16){}="a"^(.82){}="b" \ar|--{{~\b Q~}} "a";"b"
\ar@{}[d]^(.16){}="a"^(.82){}="b" \ar|--{{~Q~}} "a";"b"
& *++[F]{ [0;1]^{(2; 1) \oplus (0;1)}_{7/2}} 
\ar@{}[r]^(.18){}="a"^(.84){}="b" \ar|--{{~\b Q~}} "a";"b"
\ar@{}[d]^(.18){}="a"^(.84){}="b" \ar|--{{~Q~}} "a";"b"
& *++[F]{ [0; 2]^{(1;2)}_4 \oplus [0,0]^{(1;2)}_4} 
\ar@{}[r]^(.16){}="a"^(.82){}="b" \ar|--{{~\b Q~}} "a";"b"
\ar@{}[d]^(.16){}="a"^(.65){}="b" \ar|--{{~Q~}} "a";"b"
& *++[F]{[0;1]^{(0;3)}_{9/2}} 
\ar@{}[d]^(.18){}="a"^(.84){}="b" \ar|--{{~Q~}} "a";"b"
\\
*++[F]{[1;0]^{(2; -1) \oplus (0;-1)}_{7/2}} 
\ar@{}[r]^(.18){}="a"^(.84){}="b" \ar|--{{~\b Q~}} "a";"b"
\ar@{}[d]^(.18){}="a"^(.84){}="b" \ar|--{{~Q~}} "a";"b"
& *++[F]{[1;1]^{(3;0) \oplus 2(1;0)}_4} 
\ar@{}[r]^(.16){}="a"^(.65){}="b" \ar|--{{~\b Q~}} "a";"b"
\ar@{}[d]^(.16){}="a"^(.65){}="b" \ar|--{{~Q~}} "a";"b"
& *++[F]{ \begin{aligned} &  [1;2]^{(2;1) \oplus(0;1)}_{9/2}~\\
&  [1;0]^{(2;1) \oplus (0;1)}_{9/2}
\end{aligned}}  
\ar@{}[r]^(.35){}="a"^(.84){}="b" \ar|--{{~\b Q~}} "a";"b"
\ar@{}[d]^(.35){}="a"^(.84){}="b" \ar|--{{~Q~}} "a";"b"
& *++[F]{[1;1]^{(1;2)}_{5}} 
\ar@{}[d]^(.16){}="a"^(.82){}="b" \ar|--{{~Q~}} "a";"b"
\\
*++[F]{ [2; 0]^{(1;-2)}_4 \oplus [0;0]^{(1;-2)}_4}
\ar@{}[r]^(.16){}="a"^(.65){}="b" \ar|--{{~\b Q~}} "a";"b"
\ar@{}[d]^(.16){}="a"^(.82){}="b" \ar|--{{~Q~}} "a";"b"
& *++[F]{\begin{aligned} &  [2;1]^{(2;-1) \oplus(0;-1)}_{9/2}~\\
&  [0;1]^{(2;-1) \oplus (0;-1)}_{9/2}
\end{aligned}} \ar[d]|--{~Q~}\ar[r]|--{{~\b Q~}} 
& *++[F]{ [2;2]^{(1;0)}_5 \oplus [2;0]^{(1;0)}_5 \oplus [0;2]^{(1;0) }_5} \ar[d]|--{~Q~} \ar[r]|--{{~\b Q~}} 
& *++[F]{[2;1]_{11/2}^{(0;1)}} \\
*++[F]{[1;0]^{(0;-3)}_{9/2}} \ar[r]|--{{~\b Q~}} 
& *++[F]{[1;1]^{(1;-2)}_{5}} \ar[r]|--{{~\b Q~}}
& *++[F]{[1;2]_{11/2}^{(0;-1)}}
}
\label{noncurrent}
\end{align*}
Using bosonic conformal unitarity bounds (see for instance section~2.5 of~\cite{Minwalla:1997ka}), it can be checked that this multiplet does not contain any conserved currents.

\section{Tables of Supersymmetric Deformations}

\label{sec:tables}

In this section we tabulate all Lorentz-invariant supersymmetric deformations of interacting SCFTs in~$3 \leq d \leq 6$ dimensions.  The subsections describing the results for different values of~$d$ and~$\CN$ are largely self-contained and can be read independently. In each case we briefly summarize our conventions and review the Lorentz and~$R$-symmetry transformation properties of the supercharges. As was already stated in the introduction, we always use integer-valued Dynkin labels to denote Lie-algebra representations.\footnote{~For the rank-$r$ odd and even orthogonal algebras~$\frak{so}(2r+1)$ and~$\frak{so}(2r)$, the relation between  Dynkin labels~$R_i \in \Z$ and orthogonal labels~$h_i \in \half \Z$ (which are, for instance, used in~\cite{Bhattacharya:2008zy,Minwalla:1997ka}) is given by
\begin{align*}
& \frak{so}(2r+1)~:~~h_i = R_i + R_{i+1} + \cdots + R_{r-1} + \half R_r \quad (i = 1, \ldots, r-1)~, \qquad h_r = \half R~.\\
& \frak{so}(2r)~:~~h_i = R_i + R_{i+1} + \cdots + R_{r-2} + {R_{r-1} + R_r \over 2} \quad (i = 1, \ldots, r-2)~, \\
& \qquad \qquad~\, h_{r-1} = {R_{r-1} + R_r \over 2}~, \qquad h_r = {R_{r-1} - R_r \over 2}~.
\end{align*}
 } 
 
For each~$d$ and~$\CN$, we summarize the possible unitarity superconformal multiplets, relying on the results of~\cite{Dobrev:1985qv,Minwalla:1997ka,Ferrara:2000xg,Dobrev:2002dt,Dolan:2002zh,Bhattacharya:2008zy}. We use a streamlined labeling scheme for superconformal representations that uniformly covers all values of~$d$ and~$\CN$.  (See~\cite{multiplets} for a detailed discussion.) Multiplets are denoted by capital letters that indicate whether they satisfy any shortening conditions. Long multiplets are always denoted by~$L$, while the letters~$A, B, C, D$ indicate short multiplets. $A$-type multiplets exist for all values of~$d$ and~$\CN$. They reside at the threshold to the continuum of long multiplets, and their Lorentz or~$R$-symmetry quantum numbers are not restricted. By contrast, the letters~$B, C, D$ denote families of short multiplets that are isolated from the continuum and whose Lorentz or~$R$-symmetry quantum numbers are restricted. The notation is chosen such that~$A, B, C, D$-type multiplets with the same Lorentz and~$R$-symmetry quantum numbers are ordered according to their scaling dimension: $\Delta_A > \Delta_B > \Delta_C > \Delta_D$. 

Short multiplets have null states, which descend from a primary null state whose quantum numbers are uniquely fixed by those of the superconformal primary. We will use a subscript~$\ell$ to denote the level of the primary null state, e.g.~$A_\ell$ denotes an~$A$-type shortening condition whose primary null state resides at level~$\ell$. In~$d = 4$ and in three-dimensional~$\CN=2$ theories there are independent~$Q$ and~$\b Q$ supercharges, both of which give rise to shortening conditions. In these theories, we denote multiplets by a pair of capital letters (one unbarred and one barred) to indicate the~$Q, \b Q$ null states, e.g.~$L \b B_1$ or~$A_1 \b A_1$. 

For every value of~$d$ and~$\CN$, we list the superconformal shortening conditions allowed by unitarity, the possible Lorentz and~$R$-symmetry quantum numbers of the superconformal primary, the restrictions on its scaling dimension imposed by unitarity, and the quantum numbers of the primary null state. In theories with~$Q$ and~$\b Q$ supercharges, we independently list the corresponding shortening conditions, which must be combined in a consistent fashion to obtain a sensible superconformal multiplet. 

In each case, we then tabulate (and briefly comment on) all Lorentz-invariant supersymmetric deformations. In these tables, we indicate both the superconformal primary of the multiplet containing the deformation, as well as the deformation itself. Here we would like to make some general comments, which apply for all values of~$d$ and~$\CN$.
\begin{itemize}
\item In this section, we only discuss Lorentz-invariant deformations.\footnote{~See section~\ref{sec:nolorentz} for some examples of supersymmetric deformations that break Lorentz invariance.} As can be seen from the tables below, the superconformal primaries of the multiplets that harbor such deformations are also always Lorentz scalars. In order to streamline the presentation, we will therefore omit the (trivial) Lorentz quantum numbers from the deformation tables. 

\item We shift the quantum numbers of the superconformal primaries by constant offsets, to make the quantum numbers of the deformations as uniform as possible. This facilitates the comparison of deformations that reside in different multiplets. 

\item The deformations are ordered according to the level at which they reside in their respective multiplets. Every table starts with deformations that reside in the shortest possible multiplets and ends with generalized~$D$-term deformations, which reside in long multiplets.

\item For some values of~$d$ and~$\CN$, we find deformations residing in multiplets that also contain additional supersymmetry currents. We will not include such deformations in our tables, since they can be thought of as deformations of a theory with enhanced supersymmetry. Similarly, we will not tabulate deformations residing in multiplets that also contain higher-spin currents, since such theories are expected to be free~\cite{Maldacena:2011jn}. See~\cite{multiplets} for a systematic discussion of superconformal multiplets with conserved currents. 

\item Some deformations are related by Hermitian conjugation. We indicate conjugate pairs by including a common symbol, e.g.~$(*)$ or~$(\star)$, in the `comments' column of the deformation tables. We similarly indicate deformations that are related by mirror symmetry or~$\frak{so}(8)_R$ triality in three-dimensional~$\CN=4$ or~$\CN=8$ theories.

\end{itemize}

\subsection{Three Dimensions}

In this subsection we list all Lorentz-invariant deformations of three-dimensional~ SCFTs with~$1 \leq \CN \leq 6$ and~$\CN=8$ supersymmetry. Unitarity SCFTs with~$\CN \geq 9$ exist, but are necessarily free, because the stress-tensor multiplet also contains higher-spin currents~\cite{multiplets}. Genuine theories with~$\CN=7$ supersymmetry do not exist: they always enhance to~$\CN=8$, because the~$\CN=7$ stress-tensor multiplet contains eight, rather than seven, supersymmetry currents~\cite{Bashkirov:2011fr,multiplets}. The pertinent superconformal algebras and their unitary representations are briefly summarized below. (See for instance~\cite{Minwalla:1997ka,Bhattacharya:2008zy,multiplets} and references therein for additional details.) Throughout, representations of the~$\frak{so}(3) = \frak{su}(2)$ Lorentz algebra are denoted by
\begin{equation}\label{Lor3d }
[j]~, \qquad j \in \Z_{\geq 0}~.
\end{equation} 
Here~$j$ is an integer-valued~$\frak{su}(2)$ Dynkin label, so that the~$[j]$-representation is~$(j+1)$-dimensional. (The conventional half-integral~$\frak{su}(2)$ spin is~$j \over 2$.) We write~$[j]_\Delta$ whenever we wish to indicate the scaling dimension~$\Delta$.

\subsubsection{$d = 3$,~$\CN=1$}

The~$\CN=1$ superconformal algebra is~$\frak{osp}(1|4)$, which does not contain an~$R$-symmetry. The~$Q$-supersymmetries transform as \begin{equation}
Q \in [1]_{1/2}~, \qquad N_Q = 2~.
\end{equation}
The superconformal unitarity bounds and shortening conditions are summarized in table~\ref{tab:3DN1}. 

\smallskip

\renewcommand{\arraystretch}{1.6}
\renewcommand\tabcolsep{6pt}
\begin{table}[H]
  \centering
  \begin{tabular}{ |c|lr| l|l| }
\hline
{\bf Name} &  \multicolumn{2}{c}{\bf Primary} &  \multicolumn{1}{|c|}{\bf Unitarity Bound} & \multicolumn{1}{c|}{\bf Null State } \\
\hline
\hline
$L$ & $[j]_{\Delta}~,$& $j\geq1$ &$\Delta>\frac{1}{2} \,j+1$ & \multicolumn{1}{c|}{$-$} \\
\hline
$L'$ & $[0]_{\Delta}$& $$ &$\Delta>\frac{1}{2}$ & \multicolumn{1}{c|}{$-$} \\
\hline 
\hline
$A_{1}$ & $[j]_{\Delta}~,$& $j\geq1$ &$\Delta=\frac{1}{2} \, j+1$ & $[j-1]_{\Delta+1/2}$ \\
\hline 
$A_{2}'$ & $[0]_{\Delta}$& $$ &$\Delta=\frac{1}{2}$ & $[0]_{\Delta+1}$ \\
\hline
\hline
$B_{1}$ & $[0]_{\Delta}$& $$ &$\Delta=0$ & $[1]_{\Delta+1/2}$ \\
\hline
\end{tabular}
  \caption{Shortening conditions in three-dimensional~$\CN=1$ SCFTs.}
  \label{tab:3DN1}
\end{table} 

\noindent As is summarized in table~\ref{tab:3DN1D}, the only Lorentz-invariant supersymmetric deformations of three-dimensional~$\CN=1$ SCFTs are~$D$-terms, which reside in long~$L'$ multiplets. They can be relevant, irrelevant, or marginal. Since they reside in long multiplets, we generally do not expect marginal deformations to remain exactly marginal beyond leading order. 
\medskip\renewcommand{\arraystretch}{1.7}
\renewcommand\tabcolsep{8pt}
\begin{table}[H]
  \centering
  \begin{tabular}{|c|c|c| }
\hline
\multicolumn{1}{|c}{\bf Primary $\mathcal{O}$} &  \multicolumn{1}{|c}{\bf Deformation $\delta \SL$} & \multicolumn{1}{|c|}{\bf Comments} \\
\hline
\hline
 $L' \left\{\Delta_\CO > \frac{1}{2} \right\}$  &$Q^{2}\mathcal{O} \in \left\{\Delta >\frac{3}{2}\right\}$ & \multicolumn{1}{c|}{$D$-Term} \\
\hline
\end{tabular}
  \caption{Deformations of three-dimensional~$\mathcal{N}=1$ SCFTs.}
  \label{tab:3DN1D}
\end{table} 

\subsubsection{$d = 3$,~$\CN=2$}

\label{sec:3dn2def}

The~$\CN=2$ superconformal algebra is~$\frak{osp}(2|4)$, hence the~$R$-symmetry is~$\frak{so}(2)_R \simeq \frak{u}(1)_R$. Operators of~$R$-charge~$r \in \R$ are denoted by~$(r)$. There are independent~$Q$ and~$\b Q$ supersymmetries, which transform as 
\begin{equation}
Q \in [1]_{1/2}^{(-1)}~, \qquad \b Q \in [1]_{1/2}^{(1)}~, \qquad N_Q = 4~.
\end{equation}
Superconformal multiplets obey unitarity bounds and shortening conditions with respect to both~$Q$ and~$\b Q$, which are summarized in tables~\ref{tab:3DN2C} and~\ref{tab:3DN2AC}, respectively. As a result, they are labeled by a pair of capital letters. For instance, a generic chiral multiplet (annihilated by all~$\b Q$ supercharges) is denoted by~$L\b B_1[0]^{(r)}_r$. Consistency of the~$L$ and~$\b B_1$ shortening conditions in tables~\ref{tab:3DN2C} and~\ref{tab:3DN2AC} requires that~$r > \half$. By contrast, a free scalar field satisfies~$\Delta = r = \half$ and resides in an~$A_2 \b B_1[0]_{1/2}^{(1/2)}$ multiplet, which is annihilated by~$Q^2$ as well as all~$\b Q$ supercharges. Conserved flavor currents reside in an~$A_2 \b A_2[0]_1^{(0)}$ multiplet, while the stress-tensor multiplet is given by~$A_1 \b A_1[2]_2^{(0)}$.
\medskip
\renewcommand{\arraystretch}{1.5}
\renewcommand\tabcolsep{8pt}
\begin{table}[H]
  \centering
  \begin{tabular}{ |c|lr| l|l| }
\hline
{\bf Name} &  \multicolumn{2}{c}{\bf Primary} &  \multicolumn{1}{|c|}{\bf Unitarity Bound} & \multicolumn{1}{c|}{\bf $Q$ Null State } \\
\hline
\hline
$L$ & $[j]_{\Delta}^{(r)}$&  &$\Delta>\frac{1}{2} \,j-r+1$ & \multicolumn{1}{c|}{$-$} \\
\hline
\hline
$A_{1}$ & $[j]_{\Delta}^{(r)}~,$& $j\geq1$ &$\Delta=\frac{1}{2} \, j-r+1$ & $[j-1]_{\Delta+1/2}^{(r-1)}$ \\
\hline 
$A_{2}$ & $[0]_{\Delta}^{(r)}$& $$ &$\Delta=1-r$ & $[0]_{\Delta+1}^{(r-2)}$ \\
\hline
\hline
$B_{1}$ & $[0]_{\Delta}^{(r)}$& $$ &$\Delta=-r$ & $[1]_{\Delta+1/2}^{(r-1)}$ \\
\hline
\end{tabular}
  \caption{$Q$ shortening conditions in three-dimensional~$\CN=2$ SCFTs.}
  \label{tab:3DN2C}
\end{table} 
\renewcommand{\arraystretch}{1.5}
\renewcommand\tabcolsep{8pt}
\begin{table}[H]
  \centering
  \begin{tabular}{ |c|lr| l|l| }
\hline
{\bf Name} &  \multicolumn{2}{c}{\bf Primary} &  \multicolumn{1}{|c|}{\bf Unitarity Bound} & \multicolumn{1}{c|}{\bf $\b Q$ Null State } \\
\hline
\hline
$\overline{L}$ & $[j]_{\Delta}^{(r)}$&  &$\Delta>\frac{1}{2} \,j+r+1$ & \multicolumn{1}{c|}{$-$} \\
\hline
\hline
$\overline{A}_{1}$ & $[j]_{\Delta}^{(r)}~,$& $j\geq1$ &$\Delta=\frac{1}{2} \, j+r+1$ & $[j-1]_{\Delta+1/2}^{(r+1)}$ \\
\hline 
$\overline{A}_{2}$ & $[0]_{\Delta}^{(r)}$& $$ &$\Delta=1+r$ & $[0]_{\Delta+1}^{(r+2)}$ \\
\hline
\hline
$\overline{B}_{1}$ & $[0]_{\Delta}^{(r)}$& $$ &$\Delta=r$ & $[1]_{\Delta+1/2}^{(r+1)}$ \\
\hline
\end{tabular}
  \caption{$\b Q$ shortening conditions in three-dimensional~$\CN=2$ SCFTs.}
  \label{tab:3DN2AC}
\end{table} 
The Lorentz-invariant supersymmetric deformations of three-dimensional~$\CN=2$ SCFTs are summarized in table~\ref{tab:3DN2D}. The~$F$-term deformations reside in chiral~$L\b B_1$ and anti-chiral~$B_1 \b L$ multiplets, which are related by complex conjugation. (This is indicated by the symbol~$(*)$ in table~\ref{tab:3DN2D}.) Depending on their~$R$-charge, they may be relevant, irrelevant, or marginal. As in four-dimensional~$\CN=1$ theories (see section~\ref{sec:4dn1defs} below), marginal deformations are exactly marginal if and only if they do not break any flavor symmetries~\cite{Green:2010da}, because the chiral~$L \b B_1[0]^{(2)}_2$ multiplet containing the marginal deformation (and its complex conjugate) can pair up with an~$A_2 \b A_2[0]^{(0)}_1$ flavor current multiplet to form a~$L\b L[0]^{(0)}$ long multiplet. 
\renewcommand{\arraystretch}{1.5}
\renewcommand\tabcolsep{10pt}
\begin{table}[H]
  \centering
  \begin{tabular}{ |c|c|c| }
\hline
  \multicolumn{1}{|c|}{\bf Primary $\mathcal{O}$} &  \multicolumn{1}{c}{\bf Deformation $\delta \SL$} &\multicolumn{1}{|c|}{\bf Comments}\\
\hline
\hline
 \multirow{ 2}{*}{$A_{2}\overline{A}_{2} \left\{ \stackanchor{$(0)$}{$\Delta_\CO =1$}\right \}$}&  \multirow{ 2}{*}{$Q\overline{Q}\mathcal{O} \in \left\{ \stackanchor{$(0)$}{$\Delta =2$}\right \}$} & \multirow{ 2}{*}{Flavor Current} \\
&&\\
\hline
 \multirow{ 2}{*}{$L\overline{B}_{1} \left\{ \stackanchor{$(r+2)~,~r>-{3 \over 2}$}{$\Delta_\CO =2+r$}\right \}$}&  \multirow{ 2}{*}{$Q^{2}\mathcal{O} \in \left\{ \stackanchor{$(r)~,~r>-{3\over 2}$}{$\Delta =3+r > {3 \over 2}$}\right \}$} & \multirow{ 2}{*}{$F$-Term~$(*)$} \\
&&\\
\hline
 \multirow{ 2}{*}{$B_{1}\overline{L}\left\{ \stackanchor{$(r-2)~,~r < {3 \over 2}$}{$\Delta_\CO =2-r$}\right \}$}&  \multirow{ 2}{*}{$\overline{Q}^{2}\mathcal{O} \in \left\{ \stackanchor{$(r)~,~r<{3 \over 2}$}{$\Delta =3-r>{3 \over 2}$}\right \}$} & \multirow{ 2}{*}{$F$-Term~$(*)$} \\
&&\\
\hline
\multirow{ 2}{*}{$L\overline{L} \left\{ \stackanchor{$(r)$}{$\Delta_\CO >1+|r|$}\right \}$}&  \multirow{ 2}{*}{$Q^{2}\overline{Q}^{2}\mathcal{O} \in \left\{ \stackanchor{$(r)$}{$\Delta >3+|r|$}\right \}$} & \multirow{ 2}{*}{$D$-Term} \\
&&\\
\hline
\end{tabular}
  \caption{Deformations of three-dimensional~$\mathcal{N}=2$ SCFTs. Here~$r \in \R$ denotes the~$\frak{u}(1)_R$ charge of the deformation.}
  \label{tab:3DN2D}
\end{table} 

\subsubsection{$d=3$,~$\CN=3$}

The~$\CN=3$ superconformal algebra is~$\frak{osp}(3|4)$, so that there is a~$\frak{so}(3)_R \simeq \frak{su}(2)_R$ symmetry. The~$R$-charges are denoted by~$(R)$, where~$R \in \Z_{\geq 0}$ is an~$\frak{su}(2)_R$ Dynkin label. The~$Q$-supersymmetries transform in the vector representation~$\bf 3$ of~$\frak{so}(3)_R$,
 \begin{equation}
Q \in [1]_{1/2}^{(2)}~, \qquad N_Q = 6~.
\end{equation}
The superconformal unitarity bounds and shortening conditions are summarized in table~\ref{tab:3DN3}. For instance, $B_1[0]^{(1)}_{1/2}$ is a free hypermultiplet, and~$B_1[0]^{(2)}_1$ contains a conserved flavor current. 

\medskip

\renewcommand{\arraystretch}{1.7}
\renewcommand\tabcolsep{8pt}
\begin{table}[H]
  \centering
  \begin{tabular}{ |c|lr| l|l| }
\hline
{\bf Name} &  \multicolumn{2}{c}{\bf Primary} &  \multicolumn{1}{|c|}{\bf Unitarity Bound} & \multicolumn{1}{c|}{\bf Null State } \\
\hline
\hline
$L$ & $[j]_{\Delta}^{\left(R\right)}$&  &$\Delta>\frac{1}{2} \,j+\half R+1$ & \multicolumn{1}{c|}{$-$} \\
\hline
\hline
$A_{1}$ & $[j]_{\Delta}^{\left(R\right)}~,$& $j\geq1$ &$\Delta=\frac{1}{2} \, j+\half R+1$ & $[j-1]_{\Delta+1/2}^{(R+2)}$ \\
\hline 
$A_{2}$ & $[0]_{\Delta}^{\left(R\right)}$& $$ &$\Delta=\half R+1$ & $[0]_{\Delta+1}^{(R+4)}$ \\
\hline
\hline
$B_{1}$ & $[0]_{\Delta}^{\left(R\right)}$& $$ &$\Delta=\half R$ & $[1]_{\Delta+1/2}^{(R+2)}$ \\
\hline
\end{tabular}
  \caption{Shortening conditions in three-dimensional~$\CN=3$ SCFTs.}
  \label{tab:3DN3}
\end{table} 

The Lorentz-invariant supersymmetric deformations of three-dimensional~$\CN=3$ SCFTs are summarized in table~\ref{tab:3DN3D}. The only exception is a relevant deformation residing in an~$A_2[0]^{(0)}_1$ multiplet, which contains an extra supersymmetry current  that enhances~$\CN=3$ to~$\CN=4$. The relevant deformation is the universal mass deformation residing in the~$\CN=4$ stress-tensor multiplet (see sections~\ref{sec:3dN4defs} and~\ref{sec:defst}). Note that~$\CN=3$ theories never have marginal deformations, and in genuine~$\CN=3$ theories the only relevant deformations are flavor masses residing in flavor current multiplets (see section~\ref{sec:flavormass}). 
\smallskip
\renewcommand{\arraystretch}{1.5}
\renewcommand\tabcolsep{8pt}
\begin{table}[H]
  \centering
  \begin{tabular}{ |c|c|c| }
\hline
 \multicolumn{1}{|c|}{\bf Primary $\mathcal{O}$} &  \multicolumn{1}{c}{\bf Deformation $\delta \SL$} &\multicolumn{1}{|c|}{\bf Comments}\\
\hline
\hline
 \multirow{ 2}{*}{$B_{1} \left\{ \stackanchor{$(R =2)$}{$\Delta_\CO =1$}\right \}$}&  \multirow{ 2}{*}{$Q^{2}\mathcal{O} \in \left\{ \stackanchor{$(R=2)$}{$\Delta =2$}\right \}$} & \multirow{ 2}{*}{Flavor Current } \\
&&\\
\hline
 \multirow{ 2}{*}{$B_{1} \left\{ \stackanchor{$(R+4)$}{$\Delta_\CO =2+\frac{1}{2}R$}\right \}$}&  \multirow{ 2}{*}{$Q^{4}\mathcal{O} \in \left\{ \stackanchor{$(R)$}{$\Delta =4+\frac{1}{2}R$}\right \}$} & \multirow{ 2}{*}{$-$} \\
&&\\
\hline
 \multirow{ 2}{*}{$L \left\{ \stackanchor{$(R)$}{$\Delta_\CO >1+\frac{1}{2}R$}\right \}$}&  \multirow{ 2}{*}{$Q^{6}\mathcal{O} \in \left\{ \stackanchor{$(R)$}{$\Delta >4+\frac{1}{2}R$}\right \}$} & \multirow{ 2}{*}{$D$-Term} \\
&&\\
\hline
\end{tabular}
  \caption{Deformations of three-dimensional~$\CN=3$ SCFTs. The~$\frak{su}(2)_R$ Dynkin label~$R \in \Z_{\geq 0}$ denotes the~$R$-charge of the deformation.}
  \label{tab:3DN3D}
\end{table} 

\subsubsection{$d=3$,~$\CN=4$}

\label{sec:3dN4defs}

The~$\CN=4$ superconformal algebra is~$\frak{osp}(4|4)$, hence the~$R$-symmetry is~$\frak{so}(4)_R \simeq \frak{su}(2)_R \times \frak{su}(2)'_R$. Its representations are denoted by~$(R\,; R')$, where~$R, R' \in \Z_{\geq 0}$ are Dynkin labels for~$\frak{su}(2)_R$ and~$\frak{su}(2)'_R$, respectively. For example, $(1;0)$ and~$(0;1)$ are the left- and right-handed spinors~$\bf 2$ and~$\bf 2'$ of~$\frak{so}(4)_R$, while~$(1;1)$ is its vector representation~$\bf 4$. Note that the~$\frak{su}(2)_R$ and~$\frak{su}(2)'_R$ factors of the~$R$-symmetry algebra are inert under complex conjugation. However, they are exchanged by the action of mirror symmetry~$M$, which is an outer automorphism of the~$\CN=4$ superconformal algebra. (It need not be a symmetry of the field theory, although it can be.) The~$Q$-supersymmetries transform as
 \begin{equation}
Q \in [1]_{1/2}^{(1;1)}~~, \qquad N_Q = 8~.
\end{equation}
The superconformal unitarity bounds and shortening conditions are summarized in table~\ref{tab:3DN4}. For instance, $B_1[0]^{(1;0)}_{1/2}$ is a free hypermultiplet, and~$B_1[0]^{(0;1)}_{1/2}$ is a free twisted hypermultiplet. The two multiplets are exchanged by the mirror automorphism~$M$.  By contrast, the stress-tensor multiplet~$A_2[0]^{(0;0)}_1$ is invariant under~$M$.
\smallskip
\renewcommand{\arraystretch}{1.5}
\renewcommand\tabcolsep{8pt}
\begin{table}[H]
  \centering
  \begin{tabular}{ |c|lr| l|l| }
\hline
{\bf Name} &  \multicolumn{2}{c|}{\bf Primary} &  \multicolumn{1}{c}{\bf Unitarity Bound} & \multicolumn{1}{|c|}{\bf Null State } \\
\hline
\hline
$L$ & $[j]_{\Delta}^{(R\,;\,R')}$&  &$\Delta>\frac{1}{2} \,j+\half \left(R+R'\right)+1$ & \multicolumn{1}{c|}{$-$} \\
\hline
\hline
$A_{1}$ & $[j]_{\Delta}^{(R\,;\,R')}~,$& $j\geq1$ &$\Delta=\frac{1}{2} \, j+\half \left(R+R'\right)+1$ & $[j-1]_{\Delta+1/2}^{(R+1\,;\, R'+1)}$ \\
\hline 
$A_{2}$ & $[0]_{\Delta}^{(R\,;\,R')}$& $$ &$\Delta=\half \left(R+R'\right)+1$ & $[0]_{\Delta+1}^{(R+2\,;\, R'+2)}$ \\
\hline
\hline
$B_{1}$ & $[0]_{\Delta}^{(R\,;\,R')}$& $$ &$\Delta=\half \left(R+R'\right)$ & $[1]_{\Delta+1/2}^{(R+1\,;\, R'+1)}$ \\
\hline
\end{tabular}
  \caption{Shortening conditions in three-dimensional~$\CN=4$ SCFTs.}
  \label{tab:3DN4}
\end{table} 

The Lorentz-invariant supersymmetric deformations of three-dimensional~$\CN=4$ SCFTs are summarized in table~\ref{tab:3DN4D}. There are no marginal deformations, as was observed holographically in~\cite{Louis:2014gxa}. The only relevant deformations are flavor masses residing in flavor current multiplets, or universal masses residing in the stress-tensor multiplet (see section~\ref{sec:currdefs}). Note that the two flavor mass deformations are exchanged by mirror symmetry (this is indicated by the symbol~$(M)$ in table~\ref{tab:3DN4D}), and likewise for the two~$F$-term deformations (as indicated by~$(\t M)$ in table~\ref{tab:3DN4D}). The only deformation that does not appear in table~\ref{tab:3DN4D} resides in a~$B_1[0]^{(1;1)}_1$ multiplet, which contains an additional supersymmetry current that enhances~$\CN=4$ to~$\CN=5$. It is an additional universal mass deformation that resides in the~$\CN=5$ stress-tensor multiplet (see sections~\ref{sec:3dN5defs} and~\ref{sec:defst}).

\medskip 
\renewcommand{\arraystretch}{1.7}
\renewcommand\tabcolsep{10pt}
\begin{table}[H]
  \centering
  \begin{tabular}{ |c|c|c| }
\hline
  \multicolumn{1}{|c|}{\bf Primary $\mathcal{O}$} &  \multicolumn{1}{c}{\bf Deformation $\delta \SL$} &\multicolumn{1}{|c|}{\bf Comments}\\
\hline
\hline
 \multirow{ 2}{*}{$B_{1} \left\{ \stackanchor{$(2\,;0)$}{$\Delta_\CO =1$}\right \}$}&  \multirow{ 2}{*}{$Q^{2}\mathcal{O} \in \left\{ \stackanchor{$(0\,; 2)$}{$\Delta =2$}\right \}$} & \multirow{ 2}{*}{Flavor Current~$(M)$} \\
&&\\
\hline
 \multirow{ 2}{*}{$B_{1} \left\{ \stackanchor{$(0\,;2)$}{$\Delta_\CO =1$}\right \}$}&  \multirow{ 2}{*}{$Q^{2}\mathcal{O} \in \left\{ \stackanchor{$(2\,;0)$}{$\Delta =2$}\right \}$} & \multirow{ 2}{*}{Flavor Current~$(M)$} \\
&&\\
\hline
 \multirow{ 2}{*}{$A_{2} \left\{ \stackanchor{$(0\,;0)$}{$\Delta_\CO =1$}\right \}$}&  \multirow{ 2}{*}{$Q^{2}\mathcal{O} \in \left\{ \stackanchor{$(0\,;0)$}{$\Delta =2$}\right \}$} & \multirow{ 2}{*}{Stress Tensor} \\
&&\\
\hline
\multirow{ 2}{*}{$B_{1} \left\{ \stackanchor{$(R+4\,;0)$}{$\Delta_\CO =2+\frac{1}{2}R$}\right \}$}&  \multirow{ 2}{*}{$Q^{4}\mathcal{O} \in \left\{ \stackanchor{$(R\,; 0)$}{$\Delta =4+\frac{1}{2}R$}\right \}$} & \multirow{ 2}{*}{$F$-Term~$(\t M)$} \\
&&\\
\hline
 \multirow{ 2}{*}{$B_{1} \left\{ \stackanchor{$(0\,;R'+4)$}{$\Delta_\CO =2+\frac{1}{2}R'$}\right \}$}&  \multirow{ 2}{*}{$Q^{4}\mathcal{O} \in \left\{ \stackanchor{$(0\,;R')$}{$\Delta =4+\frac{1}{2}R'$}\right \}$} & \multirow{ 2}{*}{$F$-Term~$(\t M)$} \\
&&\\
\hline
 \multirow{ 2}{*}{$B_{1}\left\{ \stackanchor{$(R+2\,;R'+2)$}{$\Delta_\CO =2+\frac{1}{2}(R+R')$}\right \}$}&  \multirow{ 2}{*}{$Q^{6}\mathcal{O} \in \left\{ \stackanchor{$(R\,;R')$}{$\Delta =5+\frac{1}{2}(R+R')$}\right \}$} & \multirow{ 2}{*}{$-$} \\
&&\\
\hline
 \multirow{ 2}{*}{$L \left\{ \stackanchor{$(R\,;R')$}{$\Delta_\CO >1+\frac{1}{2}\left(R+R'\right)$}\right \}$}&  \multirow{ 2}{*}{$Q^{8}\mathcal{O} \in \left\{ \stackanchor{$\left(R\,;R'\right)$}{$\Delta >5+\frac{1}{2}\left(R+R'\right)$}\right \}$} & \multirow{ 2}{*}{$D$-Term} \\
&&\\
\hline
\end{tabular}
  \caption{Deformations of three-dimensional~$\CN=4$ SCFTs. The~$\frak{su}(2)_R \times \frak{su}(2)'_R$ Dynkin labels~$R,R' \in \Z_{\geq 0}$ denote the~$R$-charges of the deformation.}
  \label{tab:3DN4D}
\end{table} 

\subsubsection{$d=3$,~$\CN=5$}

\label{sec:3dN5defs}

The~$\CN=5$ superconformal algebra is~$\frak{osp}(5|4)$ and therefore the~$R$-symmetry is~$\frak{so}(5)_R$. Its representations are denoted by~$(R_1,R_2)$, where~$R_1, R_2 \in \Z_{\geq 0}$ are~$\frak{so}(5)_R$ Dynkin labels. For example, $(1,0)$ is the vector representation~$\bf 5$, while~$(0,1)$ is the spinor representation~$\bf 4$.\footnote{~Note that the corresponding~$\frak{sp}(4) \simeq \frak{so}(5)$ Dynkin labels are reversed, e.g.~$(1,0)$ is the~$\bf 4$ of~$\frak{sp}(4)$.}  The~$Q$-supersymmetries transform as
 \begin{equation}
Q \in [1]_{1/2}^{(1, 0)}~, \qquad N_Q = 10~.
\end{equation}
The superconformal unitarity bounds and shortening conditions are summarized in table~\ref{tab:3DN5}. For example, $B_1[0]^{(0,1)}_{1/2}$ is a free hypermultiplet and~$B_1[0]^{(1,0)}_1$ is the stress-tensor multiplet. 
\medskip
\renewcommand{\arraystretch}{1.7}
\renewcommand\tabcolsep{8pt}
\begin{table}[H]
  \centering
  \begin{tabular}{ |c|lr| l|l| }
\hline
{\bf Name} &  \multicolumn{2}{c}{\bf Primary} &  \multicolumn{1}{|c|}{\bf Unitarity Bound} & \multicolumn{1}{c|}{\bf Null State } \\
\hline
\hline
$L$ & $[j]_{\Delta}^{(R_1,R_2)}$&  &$\Delta>\frac{1}{2} \,j+R_1+\half R_2+1$ & \multicolumn{1}{c|}{$-$} \\
\hline
\hline
$A_{1}$ & $[j]_{\Delta}^{(R_1,R_2)}~,$& $j\geq1$ &$\Delta=\frac{1}{2} \, j+R_1+\half R_2+1$ & $[j-1]_{\Delta+1/2}^{(R_1+1,R_2)}$ \\
\hline 
$A_{2}$ & $[0]_{\Delta}^{(R_1,R_2)}$& $$ &$\Delta=R_1+\half R_2+1$ & $[0]_{\Delta+1}^{(R_1+2,R_2)}$ \\
\hline
\hline
$B_{1}$ & $[0]_{\Delta}^{(R_1,R_2)}$& $$ &$\Delta=R_1+\half R_2$ & $[1]_{\Delta+1/2}^{(R_1+1,R_2)}$ \\
\hline
\end{tabular}
  \caption{Shortening conditions in three-dimensional~$\CN=5$ SCFTs.}
  \label{tab:3DN5}
\end{table}

The Lorentz-invariant supersymmetric deformations of three-dimensional~$\CN=5$ SCFTs are summarized in table~\ref{tab:3DN5D}. There are no marginal deformations, and the only relevant deformations are universal masses residing in the stress-tensor multiplet (see section~\ref{sec:defst}). Two relevant deformations have been omitted from table~\ref{tab:3DN5D}. One resides in a~$B_1[0]^{(0,2)}_1$ multiplet, which contains an extra supersymmetry current that enhances~$\CN=5$ to~$\CN=6$ (see section~\ref{sec:3dn6defs}), and the other one belongs to an~$A_2[0]^{(0,0)}_1$ multiplet, which contains higher spin currents. 
\medskip
\renewcommand{\arraystretch}{1.7}
\renewcommand\tabcolsep{8pt}
\begin{table}[H]
  \centering
  \begin{tabular}{ |c|c|c| }
\hline
  \multicolumn{1}{|c|}{\bf Primary $\mathcal{O}$} &  \multicolumn{1}{c}{\bf Deformation $\delta \SL$} &\multicolumn{1}{|c|}{\bf Comments}\\
\hline
\hline
 \multirow{ 2}{*}{$B_{1} \left\{ \stackanchor{$(1,0)$}{$\Delta_\CO =1$}\right \}$}&  \multirow{ 2}{*}{$Q^{2}\mathcal{O} \in \left\{ \stackanchor{$(1,0)$}{$\Delta =2$}\right \}$} & \multirow{ 2}{*}{Stress Tensor} \\
&&\\
\hline
\multirow{ 2}{*}{$B_{1} \left\{ \stackanchor{$(0, R_{2}+4)$}{$\Delta_\CO =2+\frac{1}{2}R_{2}$}\right \}$}&  \multirow{ 2}{*}{$Q^{6}\mathcal{O} \in \left\{ \stackanchor{$(0,R_{2})$}{$\Delta =5+\frac{1}{2}R_{2}$}\right \}$} & \multirow{ 2}{*}{$-$} \\
&&\\
\hline
 \multirow{ 2}{*}{$B_{1} \left\{ \stackanchor{$(R_{1}+2,R_{2})$}{$\Delta_\CO =2+R_{1}+\frac{1}{2}R_{2}$}\right \}$}&  \multirow{ 2}{*}{$Q^{8}\mathcal{O} \in \left\{ \stackanchor{$(R_{1},R_{2})$}{$\Delta =6+R_{1}+\frac{1}{2}R_{2}$}\right \}$} & \multirow{ 2}{*}{$-$} \\
&&\\
\hline
 \multirow{ 2}{*}{$L \left\{ \stackanchor{$(R_{1},R_{2})$}{$\Delta_\CO >1+R_{1}+\frac{1}{2}R_{2}$}\right \}$}&  \multirow{ 2}{*}{$Q^{10}\mathcal{O} \in \left\{ \stackanchor{$(R_{1},R_{2})$}{$\Delta >6+R_{1}+\frac{1}{2}R_{2}$}\right \}$} & \multirow{ 2}{*}{$D$-Term} \\
&&\\
\hline
\end{tabular}
 \caption{Deformations of three-dimensional~$\CN=5$ SCFTs. The~$\frak{so}(5)_R$ Dynkin labels $R_1,R_2 \in \Z_{\geq 0}$ denote the~$R$-charges of the deformation.}
  \label{tab:3DN5D}
\end{table}

\subsubsection{$d=3$,~$\CN=6$}

\label{sec:3dn6defs}

The~$\CN=6$ superconformal algebra is~$\frak{osp}(6|4)$ and thus the~$R$-symmetry is~$\frak{so}(6)_R$. The~$R$-symmetry representations are denoted by~$(R_1,R_2,R_3)$, where~$R_1, R_2,R_3 \in \Z_{\geq 0}$ are $\frak{so}(6)_R$ Dynkin labels. Therefore~$(1,0,0)$ is the vector representation~$\bf 6$, while~$(0,1,0)$ and~$(0,0,1)$ are the two chiral spinor representations~$\bf 4$ and~$\bf \b 4$, which are related by complex conjugation.\footnote{~Note that the Dynkin labels of the isomorphic~$\frak{so}(6)$ and~$\frak{su}(4)$ algebras are related by a permutation. For instance, the~$(0,1,0)$ and~$(0,0,1)$ chiral spinor representations of~$\frak{so}(6)$ correspond to the fundamental~$(1,0,0)$ and anti-fundamental~$(0,0,1)$ of~$\frak{su}(4)$, while the vector~$(1,0,0)$ of~$\frak{so}(6)$ is the~$(0,1,0)$ of~$\frak{su}(4)$.} The~$Q$-supersymmetries transform as
 \begin{equation}
Q \in [1]_{1/2}^{(1,0,0)}~, \qquad N_Q = 12~.
\end{equation}
The superconformal unitarity bounds and shortening conditions are summarized in table~\ref{tab:3DN6}. For example, $B_1[0]^{(0,1,0)}_{1/2}$, $B_1[0]^{(0,0,1)}_{1/2}$ are complex conjugate free hypermultiplets, and~$B_1[0]^{(0,1,1)}_1$ is the stress-tensor multiplet. 
\medskip

\renewcommand{\arraystretch}{1.7}
\renewcommand\tabcolsep{8pt}
\begin{table}[H]
  \centering
  \begin{tabular}{ |c|lr| l|l| }
\hline
{\bf Name} &  \multicolumn{2}{c}{\bf Primary} &  \multicolumn{1}{|c|}{\bf Unitarity Bound} & \multicolumn{1}{c|}{\bf Null State } \\
\hline
\hline
$L$ & $[j]_{\Delta}^{(R_1,R_2,R_3)}$&  &$\Delta>\frac{1}{2} \,j+R_1+\half \left(R_2+R_3\right)+1$ & \multicolumn{1}{c|}{$-$} \\
\hline
\hline
$A_{1}$ & $[j]_{\Delta}^{(R_1,R_2,R_3)}~,$& $j\geq1$ &$\Delta=\frac{1}{2} \, j+R_1+\half \left(R_2+R_3\right)+1$ & $[j-1]_{\Delta+1/2}^{(R_1+1,R_2,R_3)}$ \\
\hline 
$A_{2}$ & $[0]_{\Delta}^{(R_1,R_2,R_3)}$& $$ &$\Delta=R_1+\half \left(R_2+R_3\right)+1$ & $[0]_{\Delta+1}^{(R_1+2,R_2,R_3)}$ \\
\hline
\hline
$B_{1}$ & $[0]_{\Delta}^{(R_1,R_2,R_3)}$& $$ &$\Delta=R_1+\half\left(R_2+R_3\right)$ & $[1]_{\Delta+1/2}^{(R_1+1,R_2,R_3)}$ \\
\hline
\end{tabular}
  \caption{Shortening conditions in three-dimensional~$\CN=6$ SCFTs.}
  \label{tab:3DN6}
\end{table}

The Lorentz-invariant supersymmetric deformations of three-dimensional~$\CN=6$ SCFTs are summarized in table~\ref{tab:3DN6D}. There are no marginal deformations, and the only relevant deformations are universal masses residing in the stress-tensor multiplet (see section~\ref{sec:defst}). Note that the two~$F$-term deformations are related by complex conjugation. (This is indicated by the symbol~$(*)$ in table~\ref{tab:3DN6D}.) The following multiplets contain relevant deformations, but have been omitted from table~\ref{tab:3DN6D}: the~$B_1[0]^{(0,0,2)}_1$ multiplet and its complex conjugate~$B_1[0]^{(0,2,0)}_1$ contain two extra supersymmetry currents that enhance~$\CN=6$ to~$\CN=8$ (see section~\ref{sec:d3n8defs}), consistent with the fact that there are no genuniue~$\CN=7$ SCFTs, while the~$A_2[0]^{(0,0,0)}_1$ and~$B_1[0]^{(1,0,0)}_1$ multiplets contain higher spin currents.

\medskip
\renewcommand{\arraystretch}{1.7}
\renewcommand\tabcolsep{8pt}
\begin{table}[H]
  \centering
  \begin{tabular}{ |c|c|c| }
\hline
  \multicolumn{1}{|c|}{\bf Primary $\mathcal{O}$} &  \multicolumn{1}{c}{\bf Deformation $\delta \SL$} &\multicolumn{1}{|c|}{\bf Comments}\\
\hline
\hline
 \multirow{ 2}{*}{$B_{1} \left\{ \stackanchor{$(0,1,1)$}{$\Delta_\CO =1$}\right \}$}&  \multirow{ 2}{*}{$Q^{2}\mathcal{O} \in \left\{ \stackanchor{$(0,1,1)$}{$\Delta =2$}\right \}$} & \multirow{ 2}{*}{Stress Tensor} \\
&&\\
\hline
\multirow{ 2}{*}{$B_{1} \left\{ \stackanchor{$(0, R_{2}+4,0)$}{$\Delta_\CO =2+\frac{1}{2}R_{2}$}\right \}$}&  \multirow{ 2}{*}{$Q^{6}\mathcal{O} \in \left\{ \stackanchor{$(0, R_{2},0)$}{$\Delta =5+\frac{1}{2}R_{2}$}\right \}$} & \multirow{ 2}{*}{$F$-Term~$(*)$} \\
&&\\
\hline
\multirow{ 2}{*}{$B_{1} \left\{ \stackanchor{$(0,0, R_{3}+4)$}{$\Delta_\CO =2+\frac{1}{2}R_{3}$}\right \}$}&  \multirow{ 2}{*}{$Q^{6}\mathcal{O} \in \left\{ \stackanchor{$(0, 0,R_{3})$}{$\Delta =5+\frac{1}{2}R_{3}$}\right \}$} & \multirow{ 2}{*}{$F$-Term~$(*)$} \\
&&\\
\hline
 \multirow{ 2}{*}{$B_{1} \left\{ \stackanchor{$(0,R_{2}+2,R_{3}+2)$}{$\Delta_\CO =2+\frac{1}{2}(R_{2}+R_{3})$}\right \}$}&  \multirow{ 2}{*}{$Q^{8}\mathcal{O} \in \left\{ \stackanchor{$(0,R_{2},R_{3})$}{$\Delta =6+\frac{1}{2}(R_{2}+R_{3})$}\right \}$} & \multirow{ 2}{*}{$-$} \\
&&\\
\hline
 \multirow{ 2}{*}{$B_{1} \left\{ \stackanchor{$(R_{1}+2,R_{2},R_{3})$}{$\Delta_\CO =2+R_{1}+\frac{1}{2}(R_{2}+R_{3})$}\right \}$}&  \multirow{ 2}{*}{$Q^{10}\mathcal{O} \in \left\{ \stackanchor{$(R_{1},R_{2},R_{3})$}{$\Delta =7+R_{1}+\frac{1}{2}(R_{2}+R_{3})$}\right \}$} & \multirow{ 2}{*}{$-$} \\
&&\\
\hline
 \multirow{ 2}{*}{$L \left\{ \stackanchor{$(R_{1},R_{2},R_{3})$}{$\Delta_\CO >1+R_{1}+\frac{1}{2}(R_{2}+R_{3})$}\right \}$}&  \multirow{ 2}{*}{$Q^{12}\mathcal{O} \in \left\{ \stackanchor{$(R_{1},R_{2},R_{3})$}{$\Delta >7+R_{1}+\frac{1}{2}(R_{2}+R_{3})$}\right \}$} & \multirow{ 2}{*}{$D$-Term} \\
&&\\
\hline
\end{tabular}
 \caption{Deformations of three-dimensional~$\CN=6$ SCFTs. The~$R$-charges of the deformation are denoted by the~$\frak{so}(6)_R$ Dynkin labels~$R_1,R_2,R_3 \in \Z_{\geq 0}$.}

  \label{tab:3DN6D}
\end{table} 

\subsubsection{$d=3$,~$\CN=8$}

\label{sec:d3n8defs}

The~$\CN=8$ superconformal algebra is~$\frak{osp}(8|4)$ and thus the~$R$-symmetry is~$\frak{so}(8)_R$. The~$R$-symmetry representations are denoted by~$(R_1,R_2,R_3,R_4)$, where~$R_1, R_2,R_3,R_4 \in \Z_{\geq 0}$ are $\frak{so}(8)_R$ Dynkin labels. For instance, $(1,0,0,0)$ is the vector representation~$\bf 8_\text{v}$, while~$(0,0,1,0)$ and~$(0,0,0,1)$ are the two chiral spinor representations~$\bf 8_\text{s}$ and~$\bf 8_{c}$. All three representations are real (i.e.~the spinors~$\bf 8_\text{s}$, $\bf 8_\text{c}$ are Majorana-Weyl), and they are permuted by the~$S_3$ triality group, which is an outer automorphism of~$\frak{so}(8)$. We choose a triality frame in which the~$Q$-supersymmetries transform in the vector representation~$\bf 8_\text{v}$,
 \begin{equation}\label{eq:d3n8scs}
Q_\alpha \in [1]_{1/2}^{(1,0,0,0)}~, \qquad N_Q = 16~.
\end{equation}
This choice preserves a~$\Z_2 \subset S_3$ triality subgroup~$T$ that exchanges~$\bf 8_\text{s} \leftrightarrow \bf 8_\text{c}$ and is similar to the mirror automorphism~$M$ of three-dimensional~$\CN=4$ theories discussed in section~\ref{sec:3dN4defs}. The superconformal unitarity bounds and shortening conditions are summarized in table~\ref{tab:3DN8}. For instance, $B_1[0]^{(0,0,1,0)}$ is a free hypermultiplet and~$B_1[0]^{(0,0,0,1)}$ is a free twisted hypermultiplet. The two multiplets are exchanged by~$T$. Similarly, there are two possible stress-tensor multiplets~$B_1[0]^{(0,0,2,0)}_1$ and~$B_1[0]^{(0,0,0,2)}_1$, which are also exchanged by~$T$. An irreducible quantum field theory, without locally decoupled sectors, is expected to possess a unique stress tensor (see for instance~\cite{Maldacena:2011jn}), and hence only one stress-tensor multiplet. Specifying the stress-tensor multiplet therefore completely fixes the triality frame.

\medskip

\renewcommand{\arraystretch}{1.7}
\renewcommand\tabcolsep{6pt}
\begin{table}[H]
  \centering
  \begin{tabular}{ |c|lr| l|l| }
\hline
{\bf Name} &  \multicolumn{2}{c}{\bf Primary} &  \multicolumn{1}{|c|}{\bf Unitarity Bound} & \multicolumn{1}{c|}{\bf Null State } \\
\hline
\hline
$L$ & $[j]_{\Delta}^{(R_1,R_2,R_3,R_4)}$&  &$\Delta>\frac{1}{2} \,j+R_1+R_2+\half \left(R_3+R_4\right)+1$ & \multicolumn{1}{c|}{$-$} \\
\hline
\hline
$A_{1}$ & $[j]_{\Delta}^{(R_1,R_2,R_3,R_4)}~,$& $j\geq1$ &$\Delta=\frac{1}{2} \, j+R_1+R_2+\half \left(R_3+R_4\right)+1$ & $[j-1]_{\Delta+1/2}^{(R_1+1,R_2,R_3,R_4)}$ \\
\hline 
$A_{2}$ & $[0]_{\Delta}^{(R_1,R_2,R_3,R_4)}$& $$ &$\Delta=R_1+R_2+\half \left(R_3+R_4\right)+1$ & $[0]_{\Delta+1}^{(R_1+2,R_2,R_3,R_4)}$ \\
\hline
\hline
$B_{1}$ & $[0]_{\Delta}^{(R_1,R_2,R_3,R_4)}$& $$ &$\Delta=R_1+R_2+\half \left(R_3+R_4\right)$ & $[1]_{\Delta+1/2}^{(R_1+1,R_2,R_3,R_4)}$ \\
\hline
\end{tabular}
  \caption{Shortening conditions in three-dimensional~$\CN=8$ SCFTs.}
  \label{tab:3DN8}
\end{table}

The Lorentz-invariant supersymmetric deformations of three-dimensional~$\CN=8$ SCFTs are summarized in table~\ref{tab:3DN8D}. There are no marginal deformations, and the only relevant deformations are universal masses (see section~\ref{sec:defst}) residing in the two stress-tensor multiplets, which are exchanged by the triality subgroup~$T$. (This is indicated by the symbol~$(T)$ in table~\ref{tab:3DN8D}.) Similarly, the two~$F$-term deformations are also exchanged by~$T$ (indicated by the symbol~$(\t T)$ in table~\ref{tab:3DN8D}). Several multiplets containing relevant deformations have been omitted from table~\ref{tab:3DN8D}:~$A_2[0]^{(0,0,0,0)}_1$, $B_1[0]^{(0,0,1,1)}_2$, $B_1[0]^{(1,0,0,0)}_2$, and~$B_1[0]^{(0,1,0,0)}_2$. Among these, the~$B_1[0]^{(0,0,1,1)}_2$ multiplet is distinguished by the fact that it contains an extra supersymmetry current, which enhances~$\CN=8$ to~$\CN=9$. This is consistent with the fact that~$\CN\geq 9$ theories exist, but are necessarily free~\cite{multiplets}. 

\begin{landscape}

\renewcommand{\arraystretch}{1.7}
\renewcommand\tabcolsep{10pt}
\begin{table}[H]
  \centering
  \begin{tabular}{ |c|c|c| }
\hline
  \multicolumn{1}{|c|}{\bf Primary $\mathcal{O}$} &  \multicolumn{1}{c}{\bf Deformation $\delta \SL$} &\multicolumn{1}{|c|}{\bf Comments}\\
\hline
\hline
 \multirow{ 2}{*}{$B_{1} \left\{ \stackanchor{$(0,0,2,0)$}{$\Delta_\CO =1$}\right \}$}&  \multirow{ 2}{*}{$Q^{2}\mathcal{O} \in \left\{ \stackanchor{$(0,0,0,2)$}{$\Delta =2$}\right \}$} & \multirow{ 2}{*}{Stress Tensor~$(T)$} \\
&&\\
\hline
 \multirow{ 2}{*}{$B_{1} \left\{ \stackanchor{$(0,0,0,2)$}{$\Delta_\CO =1$}\right \}$}&  \multirow{ 2}{*}{$Q^{2}\mathcal{O} \in \left\{ \stackanchor{$(0,0,2,0)$}{$\Delta =2$}\right \}$} & \multirow{ 2}{*}{Stress Tensor~$(T)$} \\
&&\\
\hline
\multirow{ 2}{*}{$B_{1} \left\{ \stackanchor{$(0, 0,R_{3}+4,0)$}{$\Delta_\CO =2+\frac{1}{2}R_{3}$}\right \}$}&  \multirow{ 2}{*}{$Q^{8}\mathcal{O} \in \left\{ \stackanchor{$(0, 0,R_{3},0)$}{$\Delta =6+\frac{1}{2}R_{3}$}\right \}$} & \multirow{ 2}{*}{$F$-Term~$(\t T)$} \\
&&\\
\hline
\multirow{ 2}{*}{$B_{1} \left\{ \stackanchor{$(0, 0,0,R_{4}+4)$}{$\Delta_\CO =2+\frac{1}{2}R_{4}$}\right \}$}&  \multirow{ 2}{*}{$Q^{8}\mathcal{O} \in \left\{ \stackanchor{$(0, 0,0,R_{4})$}{$\Delta =6+\frac{1}{2}R_{4}$}\right \}$} & \multirow{ 2}{*}{$F$-Term~$(\t T)$} \\
&&\\
\hline
\multirow{ 2}{*}{$B_{1} \left\{ \stackanchor{$(0,0, R_{3}+2,R_{4}+2)$}{$\Delta_\CO =2+\frac{1}{2}(R_{3}+R_{4})$}\right \}$}&  \multirow{ 2}{*}{$Q^{10}\mathcal{O} \in \left\{ \stackanchor{$(0, 0,R_{3},R_{4})$}{$\Delta =7+\frac{1}{2}(R_{3}+R_{4})$}\right \}$} & \multirow{ 2}{*}{$-$} \\
&&\\
\hline
 \multirow{ 2}{*}{$B_{1} \left\{ \stackanchor{$(0,R_{2}+2,R_{3},R_{4})$}{$\Delta_\CO =2+R_{2}+\frac{1}{2}(R_{3}+R_{4})$}\right \}$}&  \multirow{ 2}{*}{$Q^{12}\mathcal{O} \in \left\{ \stackanchor{$(0,R_{2},R_{3},R_{4})$}{$\Delta =8+R_{2}+\frac{1}{2}(R_{2}+R_{3})$}\right \}$} & \multirow{ 2}{*}{$-$} \\
&&\\
\hline
 \multirow{ 2}{*}{$B_{1} \left\{ \stackanchor{$(R_{1}+2,R_{2},R_{3},R_{4})$}{$\Delta_\CO =2+R_{1}+R_{2}+\frac{1}{2}(R_{3}+R_{4})$}\right \}$}&  \multirow{ 2}{*}{$Q^{14}\mathcal{O} \in \left\{ \stackanchor{$(R_{1},R_{2},R_{3},R_{4})$}{$\Delta =9+R_{1}+R_{2}+\frac{1}{2}(R_{3}+R_{4})$}\right \}$} & \multirow{ 2}{*}{$-$} \\
&&\\
\hline
 \multirow{ 2}{*}{$L \left\{ \stackanchor{$(R_{1},R_{2},R_{3},R_{4})$}{$\Delta_\CO >1+R_{1}+R_{2}+\frac{1}{2}(R_{3}+R_{4})$}\right \}$}&  \multirow{ 2}{*}{$Q^{16}\mathcal{O} \in \left\{ \stackanchor{$(R_{1},R_{2},R_{3},R_{4})$}{$\Delta >9+R_{1}+R_{2}+\frac{1}{2}(R_{3}+R_{4})$}\right \}$} & \multirow{ 2}{*}{$D$-Term} \\
&&\\
\hline
\end{tabular}
 \caption{Deformations of three-dimensional~$\CN=8$ SCFTs. The~$R$-charges of the deformation are denoted by the~$\frak{so}(8)_R$ Dynkin labels~$R_1,R_2,R_3,R_4 \in \Z_{\geq 0}$.}
  \label{tab:3DN8D}
\end{table} 

\end{landscape}

\subsection{Four Dimensions}

In this subsection we list all Lorentz-invariant supersymmetric deformations of four-dimensional~$1 \leq \CN \leq 4$ SCFTs. (Unitary SCFTs with~$\CN \geq 5$ do not exist, because they do not possess a stress tensor~\cite{multiplets}.) The corresponding superconformal algebras and their unitary representations are briefly summarized below. (See for instance~\cite{Dobrev:1985qv,Minwalla:1997ka,Dolan:2002zh,multiplets} and references therein for more detail.) Throughout, representations of the~$\frak{so}(4) = \frak{su}(2) \times \b{\frak{su}(2)}$ Lorentz algebra are denoted by
\begin{equation}\label{Lor4d}
[\, j; \b j\,]~, \qquad j, \b j \in \Z_{\geq 0}~.
\end{equation} 
Here~$j, \b j$ are integer-valued~$\frak{su}(2)$ Dynkin labels, so that the representation in~\eqref{Lor4d} has dimension~$(j+1)(\b j+1)$. We use~$[\, j; \b j\,]_\Delta$ to indicate the Lorentz quantum numbers of an operator with scaling dimension~$\Delta$.

\subsubsection{$d=4$,~$\CN=1$}

\label{sec:4dn1defs}

The~$\CN=1$ superconformal algebra is~$\frak{su}(2,2|1)$, so that there is a~$\frak{u}(1)_R$ symmetry. Operators of~$R$-charge~$r \in \R$ are denoted by~$(r)$. The~$Q$-supersymmetries transform as \begin{equation}
Q \in [1;0]_{1/2}^{(-1)}~, \qquad \b Q \in [0;1]_{1/2}^{(1)}~, \qquad N_Q = 4~.
\end{equation}
Superconformal multiplets obey unitarity bounds and shortening conditions with respect to both~$Q$ and~$\b Q$, which are summarized in tables~\ref{tab:4DN1C} and~\ref{4DN1AC}, respectively. Consequently, they are labeled by a pair of capital letters, e.g.~$L\b L[\, j ; \b j \,]^{(r)}_\Delta$ is a long multiplet without any null states. A generic chiral multiplet with left spin~$j$  is annihilated by all~$\b Q$ supercharges and denoted by~$L\b B_1[\,j;0]^{(r)}_{3r/2}$. Consistency of the~$L$ and~$\b B_1$ shortening conditions in tables~\ref{tab:4DN1C} and~\ref{4DN1AC} requires that~$r > {2 \over 3} + {1 \over 3} j$. By contrast, a free scalar field with~$j = 0$,  $\Delta = 1$  resides in an~$A_2 \b B_1[0;0]_{1}^{(2/3)}$ multiplet, which is annihilated by~$Q^2$ as well as all~$\b Q$ supercharges. Similarly, $A_1 \b B_1[1;0]^{(1)}_{3/2}$ is a free vector multiplet. Conserved flavor currents reside in an~$A_2 \b A_2[0;0]_2^{(0)}$ multiplet, while the stress-tensor multiplet is given by~$A_1 \b A_1[1;1]_3^{(0)}$.
\renewcommand{\arraystretch}{1.5}
\renewcommand\tabcolsep{8pt}
\begin{table}[H]
  \centering
  \begin{tabular}{ |c|lr| l|l| }
\hline
{\bf Name} &  \multicolumn{2}{c}{\bf Primary} &  \multicolumn{1}{|c|}{\bf Unitarity Bound} & \multicolumn{1}{c|}{\bf $Q$ Null State } \\
\hline
\hline
$L$ & $[\,j;\overline{j}\,]_{\Delta}^{(r)}$&  &$\Delta>2 + j-{3 \over 2} r$ & \multicolumn{1}{c|}{$-$} \\
\hline
\hline
$A_{1}$ & $[\,j;\overline{j}\,]_{\Delta}^{(r)}~,$& $j\geq1$ &$\Delta= 2+j-{3 \over 2} r$ & $[\,j-1;\overline{j}\,]_{\Delta+1/2}^{(r-1)}$ \\
\hline 
$A_{2}$ & $[\,0;\overline{j}\,]_{\Delta}^{(r)}$& $$ &$\Delta=2-{3 \over 2} r$ & $[\,0;\overline{j}\,]_{\Delta+1}^{(r-2)}$ \\
\hline
\hline
$B_{1}$ & $[\,0;\overline{j}\,]_{\Delta}^{(r)}$& $$ &$\Delta=-{3 \over 2} r$ & $[\,1;\overline{j}\,]_{\Delta+1/2}^{(r-1)}$ \\
\hline
\end{tabular}
  \caption{$Q$ shortening conditions in four-dimensional~$\CN=1$ SCFTs.}
  \label{tab:4DN1C}
\end{table} 

\renewcommand{\arraystretch}{1.5}
\renewcommand\tabcolsep{8pt}
\begin{table}[H]
  \centering
  \begin{tabular}{ |c|lr| l|l| }
\hline
{\bf Name} &  \multicolumn{2}{c}{\bf Primary} &  \multicolumn{1}{|c|}{\bf Unitarity Bound} & \multicolumn{1}{c|}{\bf $\b  Q$ Null State } \\
\hline
\hline
$\overline{L}$ & $[\,j;\overline{j}\,]_{\Delta}^{(r)}$&  &$\Delta>2+ \overline{j}+{3 \over 2}r$ & \multicolumn{1}{c|}{$-$} \\
\hline
\hline
$\overline{A}_{1}$ & $[\,j;\overline{j}\,]_{\Delta}^{(r)}~,$& $\overline{j}\geq1$ &$\Delta=2+ \overline{j}+{3 \over 2}r$ & $[\,j;\overline{j}-1\,]_{\Delta+1/2}^{(r+1)}$ \\
\hline 
$\overline{A}_{2}$ & $[\,j;0\,]_{\Delta}^{(r)}$& $$ &$\Delta=2+{3 \over 2}r$ & $[\,j;0\,]_{\Delta+1}^{(r+2)}$ \\
\hline
\hline
$\overline{B}_{1}$ & $[\,j;0\,]_{\Delta}^{(r)}$& $$ &$\Delta={3 \over 2}r$ & $[\,j;1\,]_{\Delta+1/2}^{(r+1)}$ \\
\hline
\end{tabular}
  \caption{$\b Q$ shortening conditions in four-dimensional~$\CN=1$ SCFTs.}
  \label{4DN1AC}
\end{table}

The Lorentz-invariant supersymmetric deformations of four-dimensional~$\CN=1$ SCFTs are summarized in table~\ref{tab:4DN1D}. They were first classified in~\cite{Green:2010da}. The~$F$-term deformations reside in chiral and anti-chiral multiplets (one can think of them as superpotential deformations), which are related by complex conjugation. (This is indicated by the symbol~$(*)$ in table~\ref{tab:4DN1D}.) Depending on their~$R$-charge, they may be relevant, irrelevant, or marginal. As was shown in~\cite{Green:2010da}, marginal deformations are exactly marginal if and only if they do not break any flavor symmetries. Essentially, this is because the chiral~$L \b B_1[0,0]^{(2)}_3$ multiplet containing the marginal deformation (and its complex conjugate anti-chiral multiplet) can pair up with an~$A_2 \b A_2[0,0]^{(0)}_2$ flavor current multiplet to form a~$L\b L[0,0]^{(0)}$ long multiplet.
\medskip
\renewcommand{\arraystretch}{1.5}
\renewcommand\tabcolsep{10pt}
\begin{table}[H]
  \centering
  \begin{tabular}{ |c|c|c| }
\hline
 \multicolumn{1}{|c|}{\bf Primary $\mathcal{O}$} &  \multicolumn{1}{c}{\bf Deformation $\delta \SL$} &\multicolumn{1}{|c|}{\bf Comments}\\
\hline
\hline
 \multirow{ 2}{*}{$L \overline{B}_{1} \left\{ \stackanchor{$(r+2)~,~ r >-{4 \over 3}$}{$\Delta_\CO =3+\frac{3}{2}r$}\right \}$}&  \multirow{ 2}{*}{$Q^{2}\mathcal{O} \in \left\{ \stackanchor{$(r)~,~r > -{4 \over 3}$}{$\Delta =4+\frac{3}{2}r>2$}\right \}$} & \multirow{ 2}{*}{$F$-Term~$(*)$} \\
&&\\
\hline
 \multirow{ 2}{*}{$B_{1} \overline{L}\left\{ \stackanchor{$(r-2)~,~ r <\frac{4}{3}$}{$\Delta_\CO =3-\frac{3}{2}r$}\right \}$}&  \multirow{ 2}{*}{$\overline{Q}^{2}\mathcal{O} \in \left\{ \stackanchor{$(r)~,~r<{4 \over 3}$}{$\Delta =4-\frac{3}{2}r >  2$}\right \}$} & \multirow{ 2}{*}{$F$-Term~$(*)$} \\
&&\\
\hline
 \multirow{ 2}{*}{$L\overline{L} \left\{ \stackanchor{$(r)$}{$\Delta_\CO >2+\frac{3}{2}|r|$}\right \}$}&  \multirow{ 2}{*}{$Q^{2}\overline{Q}^{2}\mathcal{O} \in \left\{ \stackanchor{$(r)$}{$\Delta >4+\frac{3}{2}|r|$}\right \}$} & \multirow{ 2}{*}{$D$-Term} \\
&&\\
\hline
\end{tabular}
  \caption{Deformations of four-dimensional~$\CN=1$ SCFTs. Here~$r \in \R$ denotes the~$R$-charge of the deformation.}
  \label{tab:4DN1D}
\end{table} 

\subsubsection{$d=4$,~$\CN=2$}

\label{sec:d4n2defs}

The~$\CN=2$ superconformal algebra is~$\frak{su}(2,2|2)$, so that there is a~$\frak{su}(2)_R \times \frak{u}(1)_R$ symmetry. The~$R$-charges of an operator are denoted by~$(R\,;\,r)$, where~$R \in \Z_{\geq 0}$ is an~$\frak{su}(2)_R$ Dynkin label, while~$r \in \R$ is the~$\frak{u}(1)_R$ charge. The~$Q$-supersymmetries transform as \begin{equation}
Q \in [1;0]_{1/2}^{(1\,;\,-1)}~, \qquad \b Q \in [0;1]_{1/2}^{(1\,;\,1)}~, \qquad N_Q = 8~.
\end{equation}
Superconformal multiplets obey unitarity bounds and shortening conditions with respect to both~$Q$ and~$\b Q$, which are summarized in tables~\ref{tab:4DN2C} and~\ref{4DN2AC}, respectively. Therefore, they are labeled by a pair of capital letters. For instance, $L\b B_1[0;0]^{(0\,;\, r)}_{r/2}$ is a chiral multiplet of~$\frak{u}(1)_R$ charge~$r$, which is annihilated by all~$\b Q$ supercharges, and~$B_1 \b B_1[0;0]^{(2\,;\, 0)}_2$ is a multiplet that contains a conserved flavor current.\footnote{~In brief, the relation between our labeling scheme (which is somewhat similar to that used in~\cite{Kinney:2005ej}) and the labeling scheme of~\cite{Dolan:2002zh} is as follows (see~\cite{multiplets} for more detail): $ \CA^\Delta_{R, r(j, \b \jmath)} = L\b L[j, \b \jmath]^{(R\,;\,r)}_\Delta$ is a long multiplet, and the short multiplets are given by (note that some of them are referred to as semi-short in~\cite{Dolan:2002zh}),
\begin{align*}
& \CC_{R, r(j, \b \jmath)} = A_{1,2}\b L[j, \b \jmath]^{(R\,;\, r)}~, \qquad \CB_{R, r(0,\b \jmath)} = B_1\b L[0, \b \jmath]^{(R\,;\, r)}~,  \qquad \CE_{r(0,\b \jmath)} = B_1\b L[0, \b \jmath]^{(0\,;\, r)}~,\\[1pt]
& \hat \CC_{R(j, \b \jmath)} = A_{1,2} \b A_{1,2}[j, \b \jmath]^{(R \, ; \, j - \b \jmath)}~, \qquad \CD_{R(0, \b \jmath)} = B_1 \b A_{1,2}[0,\b \jmath]^{(R\,;\, -\b \jmath-2)}~, \qquad \hat \CB_R = B_1\b B_1[0,0]^{(R\,;\, 0)}~.
\end{align*}
Analogous relations for the multiplets~$\b \CC_{R, r(j, \b \jmath)}$, $\b \CB_{R, r(j, 0)}$, $\b \CE_{r(j, 0)}$, and~$\b \CD_{R(j, 0)}$ can be obtained by complex conjugation.
}

\medskip

\renewcommand{\arraystretch}{1.5}
\renewcommand\tabcolsep{8pt}
\begin{table}[H]
  \centering
  \begin{tabular}{ |c|lr| l|l| }
\hline
{\bf Name} &  \multicolumn{2}{c}{\bf Primary} &  \multicolumn{1}{|c|}{\bf Unitarity Bound} & \multicolumn{1}{c|}{\bf $Q$ Null State } \\
\hline
\hline
$L$ & $[\,j;\overline{j}\,]_{\Delta}^{(R\,;\,r)}$&  &$\Delta>2 + j+R-\half r$ & \multicolumn{1}{c|}{$-$} \\
\hline
\hline
$A_{1}$ & $[\,j;\overline{j}\,]_{\Delta}^{(R\,;\,r)}~,$& $j\geq1$ &$\Delta= 2+j+R-\half r$ & $[\,j-1;\overline{j}\,]_{\Delta+1/2}^{(R+1\,;\,r-1)}$ \\
\hline 
$A_{2}$ & $[\,0;\overline{j}\,]_{\Delta}^{(R\,;\,r)}$& $$ &$\Delta=2+R-\half r$ & $[\,0;\overline{j}\,]_{\Delta+1}^{(R+2\,;\, r-2)}$ \\
\hline
\hline
$B_{1}$ & $[\,0;\overline{j}\,]_{\Delta}^{(R\,;\,r)}$& $$ &$\Delta=R-\half r$ & $[\,1;\overline{j}\,]_{\Delta+1/2}^{(R+1\,;\, r-1)}$ \\
\hline
\end{tabular}
  \caption{$Q$ shortening conditions in four-dimensional~$\CN=2$ SCFTs.}

  \label{tab:4DN2C}
\end{table}

\renewcommand{\arraystretch}{1.5}
\renewcommand\tabcolsep{8pt}
\begin{table}[H]
  \centering
  \begin{tabular}{ |c|lr| l|l| }
\hline
{\bf Name} &  \multicolumn{2}{c}{\bf Primary} &  \multicolumn{1}{|c|}{\bf Unitarity Bound} & \multicolumn{1}{c|}{\bf $\b  Q$ Null State } \\
\hline
\hline
$\overline{L}$ & $[\,j;\overline{j}\,]_{\Delta}^{(R\,;\,r)}$&  &$\Delta>2+ \overline{j}+R+\half r$ & \multicolumn{1}{c|}{$-$} \\
\hline
\hline
$\overline{A}_{1}$ & $[\,j;\overline{j}\,]_{\Delta}^{(R\,;\,r)}~,$& $\overline{j}\geq1$ &$\Delta=2+ \overline{j}+R+\half r$ & $[\,j;\overline{j}-1\,]_{\Delta+1/2}^{(R+1\,;\, r+1)}$ \\
\hline 
$\overline{A}_{2}$ & $[\,j;0\,]_{\Delta}^{(R\,;\,r)}$& $$ &$\Delta=2+R+\half r$ & $[\,j;0\,]_{\Delta+1}^{(R+2\,;\, r+2)}$ \\
\hline
\hline
$\overline{B}_{1}$ & $[\,j;0\,]_{\Delta}^{(R\,;\,r)}$& $$ &$\Delta=R+\half r$ & $[\,j;1\,]_{\Delta+1/2}^{(R+1\,;\, r+1)}$ \\
\hline
\end{tabular}
  \caption{$\b Q$ shortening conditions in four-dimensional~$\CN=2$ SCFTs.}
  \label{4DN2AC}
\end{table}

The Lorentz-invariant supersymmetric deformations of four-dimensional~$\CN=2$ SCFTs are summarized in table~\ref{tab:4DN2D}. They were also found in~\cite{Argyres:2015ffa}. In addition to relevant flavor mass deformations (see section~\ref{sec:flavormass}) , there are two kinds of~$F$-term deformations: the former reside in~$B_1 \b B_1$ multiplets and are necessarily irrelevant, while the latter reside in chiral~$L\b B_1$ or anti-chiral~$B_1 \b L$ multiplets. The symbol~$(*)$ in table~\ref{tab:4DN2D} indicates that the chiral and anti-chiral~$F$-terms are related by complex conjugation. Depending on their~$\frak{u}(1)_R$ charge, they may be relevant, irrelevant, or marginal. All marginal~$\CN=2$ preserving deformations must be exactly marginal, because the chiral~$L\b B_1[0;0]^{(0\,;\, 4)}_2$ multiplet containing the marginal deformation is absolutely protected: there is no recombination rule that allows it to pair up with other multiplets into a long multiplet (see for instance~\cite{Dolan:2002zh,Kinney:2005ej}). There are also irrelevant~${1 \over 4}$-BPS deformations that reside in generic~$L \b B_1$ and~$B_1 \b L$ multiplets. (As indicated by the symbol~$(\dagger)$ in table~\ref{tab:4DN2D}, they are related by complex conjugation.)  
\renewcommand{\arraystretch}{1.6}
\renewcommand\tabcolsep{7pt}
\begin{table}[H]
  \centering
  \begin{tabular}{ |c|c|c| }
\hline
 \multicolumn{1}{|c|}{\bf Primary $\mathcal{O}$} &  \multicolumn{1}{c}{\bf Deformation $\delta \SL$} &\multicolumn{1}{|c|}{\bf Comments}\\
\hline
\hline
 \multirow{ 2}{*}{$B_{1} \overline{B}_{1} \left\{\stackanchor{$(2\,;0)$}{$\Delta_\CO =2$}\right \}$}&  \multirow{ 2}{*}{$Q^{2}\mathcal{O}\oplus \overline{Q}^{2}\mathcal{O} \in \left\{ \stackanchor{$(0\,;-2)\oplus (0\,;2)$}{$\Delta =3$}\right \}$} & \multirow{ 2}{*}{Flavor Current} \\
&&\\
\hline
 \multirow{ 2}{*}{$B_{1} \overline{B}_{1} \left\{ \stackanchor{$(R+4\,;\,0)$}{$\Delta_\CO =4+R$}\right \}$}&  \multirow{ 2}{*}{$Q^{2}\overline{Q}^{2}\mathcal{O} \in \left\{ \stackanchor{$(R\,;\,0)$}{$\Delta =6+R$}\right \}$} & \multirow{ 2}{*}{$F$-Term} \\
&&\\
\hline
 \multirow{ 2}{*}{$L \overline{B}_{1}\left\{ \stackanchor{$(0\,;\,r+4)~,~r>-2$}{$\Delta_\CO =2+\frac{1}{2}r$}\right \}$}&  \multirow{ 2}{*}{$Q^{4}\mathcal{O} \in \left\{ \stackanchor{$(0\,;\,r)~,~r>-2$}{$\Delta =4+\frac{1}{2}r>3$}\right \}$} & \multirow{ 2}{*}{$F$-Term~$(*)$} \\
&&\\
\hline
 \multirow{ 2}{*}{$B_{1} \overline{L}\left\{ \stackanchor{$(0\,;\,r-4)~,~r<2$}{$\Delta_\CO =2-\frac{1}{2}r$}\right \}$}&  \multirow{ 2}{*}{$\overline{Q}^{4}\mathcal{O} \in \left\{ \stackanchor{$(0\,;\,r)~,~r<2$}{$\Delta =4-\frac{1}{2}r>3$}\right \}$} & \multirow{ 2}{*}{$F$-Term~$(*)$} \\
&&\\
\hline
 \multirow{ 2}{*}{$L \overline{B}_{1}\left\{ \stackanchor{$(R+2\,;\,r+2)~,~r>0$}{$\Delta_\CO =3+R+\frac{1}{2}r$}\right \}$}&  \multirow{ 2}{*}{$Q^{4}\overline{Q}^{2}\mathcal{O} \in \left\{ \stackanchor{$(R\,;\,r)~,~r>0$}{$\Delta =6+R+\frac{1}{2}r>6+R$}\right \}$} & \multirow{ 2}{*}{$(\dagger)$} \\
&&\\
\hline
 \multirow{ 2}{*}{$B_{1} \overline{L}\left\{ \stackanchor{$(R+2\,;\,r-2)~,~r<0$}{$\Delta_\CO =3+R-\frac{1}{2}r$}\right \}$}&  \multirow{ 2}{*}{$Q^{2}\overline{Q}^{4}\mathcal{O} \in \left\{ \stackanchor{$(R\,;\,r)$}{$\Delta =6+R-\frac{1}{2}r>6+R$}\right \}$} & \multirow{ 2}{*}{$(\dagger)$} \\
&&\\
\hline
 \multirow{ 2}{*}{$L\overline{L} \left\{ \stackanchor{$(R\,;\,r)$}{$\Delta_\CO >2+R+\frac{1}{2}|r|$}\right \}$}&  \multirow{ 2}{*}{$Q^{4}\overline{Q}^{4}\mathcal{O} \in \left\{ \stackanchor{$(R\,;\,r)$}{$\Delta >6+R+\frac{1}{2}|r|$}\right \}$} & \multirow{ 2}{*}{$D$-term} \\
&&\\
\hline
\end{tabular}
  \caption{Deformations of four-dimensional~$\CN=2$ SCFTs. The~$\frak{su}(2)_R$ Dynkin label~$R \in \Z_{\geq 0}$ and the~$\frak{u}(1)_R$ charge~$r \in \R$ denote the~$R$-symmetry representation of the deformation.}
  \label{tab:4DN2D}
\end{table}

\subsubsection{$d=4$,~$\CN=3$}

The~$\CN=3$ superconformal algebra is~$\frak{su}(2,2|3)$,\footnote{~A standard argument (based on the single-particle representations of the~$\CN$-extended super-Poincar\'e algebras and the CPT theorem, see for instance~\cite{Wess:1992cp}) shows that weakly coupled~$\CN=3$ SCFTs must actually be~$\CN=4$ theories. However, no known argument rules out the existence of strongly-coupled~$\CN=3$ SCFTs that are not~$\CN=4$ theories. Aspects of such theories were recently discussed in~\cite{Aharony:2015oyb,Garcia-Etxebarria:2015wns}.}  with~$R$-symmetry~$\frak{su}(3)_R \times \frak{u}(1)_R$. The~$R$-charges of an operator are denoted by~$(R_1, R_2\,;\,r)$. Here~$R_1, R_2 \in \Z_{\geq 0}$ are~$\frak{su}(3)_R$ Dynkin labels, e.g.~$(1,0)$ denotes the fundamental~$\bf 3$ and~$(0,1)$ the anti-fundamental~$\b {\bf 3}$. The~$\frak{u}(1)_R$ charge is given by~$r \in \R$. The~$Q$-supersymmetries transform as \begin{equation}
Q \in [1;0]_{1/2}^{(1,0\,;\,-1)}~, \qquad \b Q \in [0;1]_{1/2}^{(0,1\,;\,1)}~, \qquad N_Q = 12~.
\end{equation}
Superconformal multiplets obey unitarity bounds and shortening conditions with respect to both~$Q$ and~$\b Q$, summarized in tables~\ref{tab:4DN3C} and~\ref{4DN3AC}, and hence they are labeled by a pair of capital letters. For instance, $B_1 \b B_1 [0;0]^{(1,1\,; \,0)}_2$ is the stress-tensor multiplet.

\medskip 

\renewcommand{\arraystretch}{1.5}
\renewcommand\tabcolsep{8pt}
\begin{table}[H]
  \centering
  \begin{tabular}{ |c|lr| l|l| }
\hline
{\bf Name} &  \multicolumn{2}{c}{\bf Primary} &  \multicolumn{1}{|c|}{\bf Unitarity Bound} & \multicolumn{1}{c|}{\bf $Q$~Null State } \\
\hline
\hline
$L$ & $[\,j;\overline{j}\,]_{\Delta}^{(R_1,R_2\,;\,r)}$&  &$\Delta>2 + j+{2 \over 3} \left(2R_1+R_2\right) - {1 \over 6}r$ & \multicolumn{1}{c|}{$-$} \\
\hline
\hline
$A_{1}$ & $[\,j;\overline{j}\,]_{\Delta}^{(R_1,R_2\,;\,r)}~,$& $j\geq1$ &$\Delta= 2+j+{2 \over 3} \left(2R_1+R_2\right) - {1 \over 6}r$ & $[\,j-1;\overline{j}\,]_{\Delta+1/2}^{(R_1+1,R_2\,;\,r-1)}$ \\
\hline 
$A_{2}$ & $[\,0;\overline{j}\,]_{\Delta}^{(R_1,R_2\,;\,r)}$& $$ &$\Delta=2+{2 \over 3} \left(2R_1+R_2\right) - {1 \over 6}r$ & $[\,0;\overline{j}\,]_{\Delta+1}^{(R_1+2,R_2\,;\, r-2)}$ \\
\hline
\hline
$B_{1}$ & $[\,0;\overline{j}\,]_{\Delta}^{(R_1,R_2\,;\,r)}$& $$ &$\Delta={2 \over 3} \left(2R_1+R_2\right) - {1 \over 6}r$ & $[\,1;\overline{j}\,]_{\Delta+1/2}^{(R_1+1, R_2\,;\, r-1)}$ \\
\hline
\end{tabular}
  \caption{$Q$ shortening conditions in four-dimensional~$\CN=3$ SCFTs.}
  \label{tab:4DN3C}
\end{table} 

\renewcommand{\arraystretch}{1.5}
\renewcommand\tabcolsep{8pt}
\begin{table}[H]
  \centering
  \begin{tabular}{ |c|lr| l|l| }
\hline
{\bf Name} &  \multicolumn{2}{c}{\bf Primary} &  \multicolumn{1}{|c|}{\bf Unitarity Bound} & \multicolumn{1}{c|}{\bf $\b Q$ Null State } \\
\hline
\hline
$\overline{L}$ & $[\,j;\overline{j}\,]_{\Delta}^{(R_1,R_2\,;\,r)}$&  &$\Delta>2+ \overline{j}+{2 \over 3}\left(R_1+2R_2\right)+{1 \over 6} r$ & \multicolumn{1}{c|}{$-$} \\
\hline
\hline
$\overline{A}_{1}$ & $[\,j;\overline{j}\,]_{\Delta}^{(R_1,R_2\,;\,r)}~,$& $\overline{j}\geq1$ &$\Delta=2+ \overline{j}+{2 \over 3}\left(R_1+2R_2\right)+{1 \over 6} r$ & $[\,j;\overline{j}-1\,]_{\Delta+1/2}^{(R_1,R_2+1\,;\, r+1)}$ \\
\hline 
$\overline{A}_{2}$ & $[\,j;0\,]_{\Delta}^{(R_1,R_2\,;\,r)}$& $$ &$\Delta=2+{2 \over 3}\left(R_1+2R_2\right)+{1 \over 6} r$ & $[\,j;0\,]_{\Delta+1}^{(R_1,R_2+2\,;\, r+2)}$ \\
\hline
\hline
$\overline{B}_{1}$ & $[\,j;0\,]_{\Delta}^{(R_1,R_2\,;\,r)}$& $$ &$\Delta={2 \over 3}\left(R_1+2R_2\right)+{1 \over 6} r$ & $[\,j;1\,]_{\Delta+1/2}^{(R_1,R_2+1\,;\, r+1)}$ \\
\hline
\end{tabular}
  \caption{$\b Q$ shortening conditions in four-dimensional~$\CN=3$ SCFTs.}
  \label{4DN3AC}
\end{table}

The Lorentz-invariant supersymmetric deformations of four-dimensional~$\CN=3$ SCFTs are summarized in table~\ref{tab:4DN3D}. All entries in this table are irrelevant operators: there are no relevant or marginal deformations. More precisely, there are no multiplets of~$\CN=3$ superconformal symmetry that contain relevant supersymmetric deformations, and there are exactly two such multiplets that contain a marginal deformation:~the~$B_1\b B_1[0;0]^{(2,0\,;\,4)}_2$ multiplet, and its complex conjugate~$B_1\b B_1[0;0]^{(0,2\,;\,-4)}_2$. However, these multiplets also contain additional supersymmetry currents, which enhance~$\CN=3$ to~$\CN=4$, and for this reason they have been omitted from table~\ref{tab:4DN3D}. As is well known, all~$\CN=4$ SCFTs possess exactly marginal deformations that reside in their stress-tensor multiplets (see sections~\ref{sec:4dn4defs} and~\ref{sec:defst}). Therefore, genuine~$\CN=3$ SCFTs admit neither relevant nor marginal supersymmetric deformations, as was also observed in~\cite{Aharony:2015oyb}. However, there is a rich variety of  irrelevant supersymmetric deformations, many of which reside in short multiplets. Pairs of multiplets that share one of the symbols~$(*), (\star), (\dagger), (\ddagger)$ are related by complex conjugation.

\begin{landscape}

\renewcommand{\arraystretch}{1.33}
\renewcommand\tabcolsep{10pt}
\begin{table}[H]
  \centering
  \begin{tabular}{ |c|c|c| }
\hline
 \multicolumn{1}{|c|}{\bf Primary $\mathcal{O}$} &  \multicolumn{1}{c}{\bf Deformation $\delta \SL$} &\multicolumn{1}{|c|}{\bf Comments}\\
\hline
\hline
\multirow{ 2}{*}{$B_{1}\overline{B}_{1} \left\{ \stackanchor{$(R_{1}+4,0\,;\,2R_{1}+8)$}{$\Delta_\CO =4+R_{1}$}\right \}$}&  \multirow{ 2}{*}{$Q^{4}\overline{Q}^{2}\mathcal{O} \in \left\{ \stackanchor{$(R_{1},0\,;\,2R_{1}+6)$}{$\Delta =7+R_{1}$}\right \}$} & \multirow{ 2}{*}{$F$-Term~$(*)$} \\
&&\\
\hline
\multirow{ 2}{*}{$B_{1}\overline{B}_{1} \left\{ \stackanchor{$(0,R_{2}+4\,;\,-2R_{2}-8)$}{$\Delta_\CO =4+R_{2}$}\right \}$}&  \multirow{ 2}{*}{$Q^{2}\overline{Q}^{4}\mathcal{O} \in \left\{ \stackanchor{$(0,R_{2}\,;\,-2R_{2}-6)$}{$\Delta =7+R_{2}$}\right \}$} & \multirow{ 2}{*}{$F$-Term~$(*)$} \\
&&\\
\hline
\multirow{ 2}{*}{$B_{1}\overline{B}_{1} \left\{ \stackanchor{$\big(R_{1}+2,R_{2}+2\,;\,2(R_{1}-R_{2})\big)$}{$\Delta_\CO =4+R_{1}+R_{2}$}\right \}$}&  \multirow{ 2}{*}{$Q^{4}\overline{Q}^{4}\mathcal{O} \in \left\{ \stackanchor{$\big(R_{1},R_{2}\,;\,2(R_{1}-R_{2})\big)$}{$\Delta =8+R_{1}+R_{2}$}\right \}$} & \multirow{ 2}{*}{$-$} \\
&&\\
\hline
 \multirow{ 2}{*}{$L\overline{B}_{1} \left\{ \stackanchor{$(0,0\,; \, r+6)~,~r>0$}{$\Delta_\CO =1+\frac{1}{6}r$}\right \}$}&  \multirow{ 2}{*}{$Q^{6}\mathcal{O} \in \left\{ \stackanchor{$(0,0\,;\,r)~,~r>0$}{$\Delta =4+\frac{1}{6}r>4$}\right \}$} & \multirow{ 2}{*}{$F$-term~$(\star)$} \\
&&\\
\hline
 \multirow{ 2}{*}{$B_1\overline{L} \left\{ \stackanchor{$(0,0\,;\, r-6)~,~r < 0$}{$\Delta_\CO =1-\frac{1}{6}r$}\right \}$}&  \multirow{ 2}{*}{$\overline{Q}^{6}\mathcal{O} \in \left\{ \stackanchor{$(0,0\,;\, r)~,~r < 0$}{$\Delta =4-\frac{1}{6}r> 4$}\right \}$} & \multirow{ 2}{*}{$F$-Term~$(\star)$} \\
&&\\
\hline
 \multirow{ 2}{*}{$L\overline{B}_{1} \left\{ \stackanchor{$(R_{1}+2,0\,;\,r+4)~,~r>2R_1+6$}{$\Delta_\CO =2+\frac{2}{3}R_{1}+\frac{1}{6}r$}\right \}$}&  \multirow{ 2}{*}{$Q^{6}\overline{Q}^{2}\mathcal{O} \in \left\{ \stackanchor{$(R_{1},0\,;\,r)~,~r>2R_1+6$}{$\Delta =6+\frac{2}{3}R_{1}+\frac{1}{6}r>7+R_1$}\right \}$} & \multirow{ 2}{*}{$(\dagger)$} \\
&&\\
\hline
 \multirow{ 2}{*}{$B_1\overline{L} \left\{ \stackanchor{$(0,R_2+2\,;\, r-4)~,~r < -2R_2-6$}{$\Delta_\CO =2+\frac{2}{3}R_2-\frac{1}{6}r$}\right \}$}&  \multirow{ 2}{*}{$Q^{2}\overline{Q}^{6}\mathcal{O} \in \left\{ \stackanchor{$(0,R_{2}\,;\, r)~,~r < -2R_2-6$}{$\Delta =6+\frac{2}{3}R_2-\frac{1}{6}r> 7+R_2$}\right \}$} & \multirow{ 2}{*}{$(\dagger)$} \\
&&\\
\hline
 \multirow{ 2}{*}{$L\overline{B}_{1} \left\{ \stackanchor{$(R_{1},R_{2}+2\,;\,r+2)~,~r>2(R_1-R_2)$}{$\Delta_\CO =3+\frac{2}{3}(R_{1}+2R_{2})+\frac{1}{6}r$}\right \}$}&  \multirow{ 2}{*}{$Q^{6}\overline{Q}^{4}\mathcal{O} \in \left\{ \stackanchor{$(R_{1},R_{2}\,;\,r)~,~r>2(R_1-R_2)$}{$\Delta =8+\frac{2}{3}(R_{1}+2R_{2})+\frac{1}{6}r>8+R_1+R_2$}\right \}$} & \multirow{ 2}{*}{$(\ddagger)$} \\
&&\\
\hline
 \multirow{ 2}{*}{$B_1\overline{L} \left\{ \stackanchor{$(R_1+2,R_2\,;\, r-2)~,~r < 2(R_1-R_2)$}{$\Delta_\CO =3+\frac{2}{3}(2R_1+R_2)-\frac{1}{6}r$}\right \}$}&  \multirow{ 2}{*}{$Q^{4}\overline{Q}^{6}\mathcal{O} \in \left\{ \stackanchor{$(R_{1},R_{2}\,;\, r)~,~r < 2(R_1-R_2)$}{$\Delta =8+\frac{2}{3}(2R_1+R_2)-\frac{1}{6}r> 8 + R_1+R_2$}\right \}$} & \multirow{ 2}{*}{$(\ddagger)$} \\
&&\\
\hline
 \multirow{ 3}{*}{ $L\overline{L} \left\{ \begin{aligned}
(R_1, & R_2\,;\,r) \\[1pt]
\Delta_\CO >2+\text{max} & \left\{\stackanchor{${2 \over 3}(2R_1+R_2) -{1 \over 6} r$}{${2 \over 3}(R_1+2R_2) +{1 \over 6} r$}\right\}
 \end{aligned}
\right \}$} &  \multirow{ 3}{*}{$Q^{6}\overline{Q}^{6}\mathcal{O} \in \left\{\begin{aligned} (R_1, & R_2\,;\,r) \\[1pt]
\Delta > 8+\text{max} & \left\{\stackanchor{${2 \over 3}(2R_1+R_2) -{1 \over 6} r$}{${2 \over 3}(R_1+2R_2) +{1 \over 6} r$}\right\} \end{aligned}
\right \}$} & \multirow{ 3}{*}{$D$-Term} \\[4ex]
&&\\
\hline
\end{tabular}
  \caption{Deformations of four-dimensional~$\mathcal{N}=3$ SCFTs. The~$\frak{su}(3)_R$ Dynkin labels~$R_1,R_2 \in \Z_{\geq 0}$ and the~$\frak{u}(1)_R$ charge~$r \in \R$ denote the~$R$-symmetry representation of the deformation.}
  \label{tab:4DN3D}
\end{table} 

\end{landscape}

\subsubsection{$d = 4$,~$\CN=4$} \label{sec:4dn4defs}

The~$\CN=4$ superconformal algebra is~$ \frak{psu}(2,2|4)$, with~$R$-symmetry~$\frak{su}(4)_R \simeq \frak{so}(6)_R$. The~$R$-charges are denoted by~$\frak{su}(4)_R$ Dynkin labels~$(R_1, R_2, R_3)$ with~$R_1, R_2, R_3 \in \Z_{\geq 0}\,$. For instance, $(1,0,0)$ and~$(0,0,1)$ are the fundamental~$\bf 4$ and the anti-fundamental~$\b {\bf 4}$ of~$\frak{su}(4)_R$, while~$(0,1,0)$ is the fundamental vector representation~$\bf 6$ of~$\frak{so}(6)_R$. The~$Q$-supercharges are 
\begin{equation}
Q \in [1;0]_{1/2}^{(1,0,0)}~, \qquad \b Q \in [0;1]_{1/2}^{(0,0,1)}~, \qquad N_Q = 16~.
\end{equation}
Superconformal multiplets obey unitarity bounds and shortening conditions with respect to both~$Q$ and~$\b Q$, summarized in tables~\ref{tab:4DN4C} and~\ref{4DN4AC}, and are labeled by a pair of capital letters. For instance, the stress-tensor multiplet is given by~$B_1 \b B_1[0;0]^{(0,2,0)}_2$.
\medskip

\renewcommand{\arraystretch}{1.5}
\renewcommand\tabcolsep{8pt}
\begin{table}[H]
  \centering
  \begin{tabular}{ |c|lr| l|l| }
\hline
{\bf Name} &  \multicolumn{2}{c}{\bf Primary} &  \multicolumn{1}{|c|}{\bf Unitarity Bound} & \multicolumn{1}{c|}{\bf $Q$~Null State } \\
\hline
\hline
$L$ & $[\,j;\overline{j}\,]_{\Delta}^{(R_1,R_2,R_3)}$&  &$\Delta>2 + j+\half\left(3R_1 +2R_2 +R_3\right)$ & \multicolumn{1}{c|}{$-$} \\
\hline
\hline
$A_{1}$ & $[\,j;\overline{j}\,]_{\Delta}^{(R_1,R_2,R_3)}~,$& $j\geq1$ &$\Delta= 2+j+\half\left(3R_1 +2R_2 +R_3\right)$ & $[\,j-1;\overline{j}\,]_{\Delta+1/2}^{(R_1+1,R_2,R_3)}$ \\
\hline 
$A_{2}$ & $[\,0;\overline{j}\,]_{\Delta}^{(R_1,R_2,R_3)}$& $$ &$\Delta=2+\half\left(3R_1 +2R_2 +R_3\right)$ & $[\,0;\overline{j}\,]_{\Delta+1}^{(R_1+2,R_2,R_3)}$ \\
\hline
\hline
$B_{1}$ & $[\,0;\overline{j}\,]_{\Delta}^{(R_1,R_2,R_3)}$& $$ &$\Delta=\half\left(3R_1 +2R_2 +R_3\right)$ & $[\,1;\overline{j}\,]_{\Delta+1/2}^{(R_1+1, R_2,R_3)}$ \\
\hline
\end{tabular}
  \caption{$Q$ shortening conditions in four-dimensional~$\CN=4$ SCFTs.}
  \label{tab:4DN4C}
\end{table}

\renewcommand{\arraystretch}{1.5}
\renewcommand\tabcolsep{8pt}
\begin{table}[H]
  \centering
  \begin{tabular}{ |c|lr| l|l| }
\hline
{\bf Name} &  \multicolumn{2}{c}{\bf Primary} &  \multicolumn{1}{|c|}{\bf Unitarity Bound} & \multicolumn{1}{c|}{\bf $\b Q$~Null State } \\
\hline
\hline
$\overline{L}$ & $[\,j;\overline{j}\,]_{\Delta}^{(R_1,R_2,R_3)}$&  &$\Delta>2+ \overline{j}+\half\left(R_1+2R_2 +3 R_3\right)$ & \multicolumn{1}{c|}{$-$} \\
\hline
\hline
$\overline{A}_{1}$ & $[\,j;\overline{j}\,]_{\Delta}^{(R_1,R_2,R_3)}~,$& $\overline{j}\geq1$ &$\Delta=2+ \overline{j}+\half\left(R_1+2R_2 +3 R_3\right)$ & $[\,j;\overline{j}-1\,]_{\Delta+1/2}^{(R_1,R_2,R_3+1)}$ \\
\hline 
$\overline{A}_{2}$ & $[\,j;0\,]_{\Delta}^{(R_1,R_2,R_3)}$& $$ &$\Delta=2+\half\left(R_1+2R_2 +3 R_3\right)$ & $[\,j;0\,]_{\Delta+1}^{(R_1,R_2,R_3+2)}$ \\
\hline
\hline
$\overline{B}_{1}$ & $[\,j;0\,]_{\Delta}^{(R_1,R_2,R_3)}$& $$ &$\Delta=\half\left(R_1+2R_2+3R_3\right)$ & $[\,j;1\,]_{\Delta+1/2}^{(R_1,R_2,R_3+1)}$ \\
\hline
\end{tabular}
  \caption{$\b Q$ shortening conditions in four-dimensional~$\CN=4$ SCFTs.}
  \label{4DN4AC}
\end{table}

The Lorentz-invariant supersymmetric deformations of four-dimensional~$\CN=4$ SCFTs are summarized in table~\ref{tab:4DN4D}. There are no relevant deformations, but every such theory necessarily possesses an exactly marginal deformation residing in its stress-tensor multiplet (see section~\ref{sec:defst} for more detail). There are also~$\half$-BPS ($F$-term) and~${1 \over 4}$-BPS irrelevant deformations in~$B_1 \b B_1$ multiplets, as well as irrelevant deformations in~$L \b B_1$ and~$B_1 \b L$ multiplets. (As indicated by the symbol~$(*)$ in table~\ref{tab:4DN4D}, the latter are related by complex conjugation.)

\begin{landscape}

\vspace*{\fill}

\renewcommand{\arraystretch}{1.8}
\renewcommand\tabcolsep{10pt}
\begin{table}[H]
  \centering
  \begin{tabular}{ |c|c|c| }
\hline
 \multicolumn{1}{|c|}{\bf Primary $\mathcal{O}$} &  \multicolumn{1}{c}{\bf Deformation $\delta \SL$} &\multicolumn{1}{|c|}{\bf Comments}\\
\hline
\hline
\multirow{ 2}{*}{$B_{1}\overline{B}_{1} \left\{ \stackanchor{$(0, 2, 0)$}{$\Delta_\CO =2$}\right \}$}&  \multirow{ 2}{*}{$Q^{4}\mathcal{O} \oplus \overline{Q}^{4} \CO \in \left\{ \stackanchor{$(0, 0, 0)$}{$\Delta =4$}\right \}$} & \multirow{ 2}{*}{Stress Tensor} \\
&&\\
\hline
\multirow{ 2}{*}{$B_{1}\overline{B}_{1} \left\{ \stackanchor{$(0, R_2+4, 0)$}{$\Delta_\CO =R_2+4$}\right \}$}&  \multirow{ 2}{*}{$Q^{4}\overline{Q}^{4}\mathcal{O} \in \left\{ \stackanchor{$(0, R_2, 0)$}{$\Delta =8 + R_2$}\right \}$} & \multirow{ 2}{*}{$F$-Term} \\
&&\\
\hline
\multirow{ 2}{*}{$B_{1}\overline{B}_{1} \left\{ \stackanchor{$(R_1+2, R_2, R_1+2)$}{$\Delta_\CO =4+2 R_1 + R_2$}\right \}$}&  \multirow{ 2}{*}{$Q^{6}\overline{Q}^{6}\mathcal{O} \in \left\{ \stackanchor{$(R_1, R_2, R_1)$}{$\Delta =10 + 2R_1 +R_2$}\right \}$} & \multirow{ 2}{*}{$-$} \\
&&\\
\hline
 \multirow{ 2}{*}{$L\overline{B}_{1} \left\{ \stackanchor{$(R_{1},R_{2}, R_3 + 2)~,~R_1 < R_3$}{$\Delta_\CO =3+\frac{1}{2}(R_1 + 2 R_2 + 3 R_3)$}\right \}$}&  \multirow{ 2}{*}{$Q^{8}\overline{Q}^{6}\mathcal{O} \in \left\{ \stackanchor{$(R_1,R_2,R_3)~,~R_1 < R_3$}{$\Delta =10+\frac{1}{2}(R_1+2 R_2 + 3 R_3) > 10 + 2 R_1 + R_2$}\right \}$} & \multirow{ 2}{*}{$(*)$} \\
&&\\
\hline
 \multirow{ 2}{*}{$B_1\overline{L} \left\{ \stackanchor{$(R_{1}+2,R_{2}, R_3)~,~R_1 > R_3$}{$\Delta_\CO =3+\frac{1}{2}(3R_1 + 2 R_2 + R_3)$}\right \}$}&  \multirow{ 2}{*}{$Q^{6}\overline{Q}^{8}\mathcal{O} \in \left\{ \stackanchor{$(R_1,R_2,R_3)~,~R_1 > R_3$}{$\Delta =10+\frac{1}{2}(3 R_1+2 R_2 + R_3) > 10 +R_2+2 R_3$}\right \}$} & \multirow{ 2}{*}{$(*)$} \\
&&\\
\hline
 \multirow{ 3}{*}{ $L\overline{L} \left\{ \begin{aligned}
(R_1, & R_2 , R_3) \\[1pt]
\Delta_\CO >2+\half\,\text{max} & \left\{\stackanchor{$R_1+2R_2 + 3R_3$}{$3R_1+2R_2 + R_3$}\right\}
 \end{aligned}
\right \}$} &  \multirow{ 3}{*}{$Q^{8}\overline{Q}^{8}\mathcal{O} \in \left\{\begin{aligned} (R_1, & R_2,R_3) \\[1pt]
\Delta > 10+\half\,\text{max} & \left\{\stackanchor{$R_1+2R_2 + 3R_3$}{$3R_1+2R_2 + R_3$}\right\} \end{aligned}
\right \}$} & \multirow{ 3}{*}{$D$-Term} \\[4ex]
&&\\
\hline
\end{tabular}
  \caption{Deformations of four-dimensional~$\mathcal{N}=4$ SCFTs. The~$R$-charges of the deformation are denoted by the~$\frak{su}(4)_R$ Dynkin labels~$R_1, R_2, R_3 \in \Z_{\geq 0}$}
  \label{tab:4DN4D}
\end{table} 

\vspace*{\fill}

\end{landscape}

\subsection{Five Dimensions}  

In this subsection we list all Lorentz-invariant supersymmetric deformations of five-dimensional SCFTs. The unique superconformal algebra in five dimensions is~$\frak{f}(4)$ and corresponds to~$\CN=1$ supersymmetry. The Lorentz algebra is~$\frak{so}(5) = \frak{sp}(4)$ and the~$R$-symmetry is~$\frak{su}(2)_R$. Lorentz representations are denoted by~$\frak{sp}(4)$ Dynkin labels~$j_1, j_2 \in \Z_{\geq 0}$, e.g.~$[1,0]$ and~$[0,1]$ are the spinor~$\bf 4$ and the vector~$\bf 5$ representations of~$\frak{so}(5)$. The~$R$-charges are denoted by~$(R)$, where~$R \in \Z_{\geq 0}$ is an~$\frak{su}(2)_R$ Dynkin label. The quantum numbers of an operator with scaling dimension~$\Delta$ are indicated as follows,
\begin{equation}\label{5dlabeling}
[\, j_1, j_2\,]^{(R)}_\Delta~, \qquad j_1, j_2, R \in \Z_{\geq 0}~.
\end{equation} 
The~$Q$-supersymmetries transform as \begin{equation}
Q \in [1,0]_{1/2}^{(1)}~, \qquad N_Q = 8~.
\end{equation}
The superconformal unitarity bounds and shortening conditions are summarized in table~\ref{tab:5Dmult}.  (See~\cite{Minwalla:1997ka,Bhattacharya:2008zy,multiplets} and references therein for a more detailed discussion.) For instance, $C_1[0,0]^{(1)}_{3/2}$ is a free hypermultiplet, $C_1[0,0]^{(2)}_3$ is a flavor current multiplet, and~$B_2[0,0]^{(0)}_3$ is the stress-tensor multiplet. 

\renewcommand{\arraystretch}{1.7}
\renewcommand\tabcolsep{10pt}
\begin{table}[H]
  \centering
  \begin{tabular}{ |c|lr| l|l| }
\hline
{\bf Name} &  \multicolumn{2}{c}{\bf Primary} &  \multicolumn{1}{|c|}{\bf Unitarity Bound} & \multicolumn{1}{c|}{\bf Null State } \\
\hline
\hline
$L$ & $[\, j_{1},j_{2} \, ]_{\Delta}^{(R)}$&  &$\Delta>j_{1}+j_{2}+\frac{3}{2}\,R+4$ & \multicolumn{1}{c|}{$-$} \\
\hline
\hline
$A_{1}$ &$[\, j_{1}, j_{2} \,]_{\Delta}^{(R)}~,$& $j_{1}\geq1$ &$\Delta=j_{1}+j_{2}+\frac{3}{2}\,R+4$ & $[\,j_{1}-1,j_{2}\,]_{\Delta+1/2}^{(R+1)}$ \\
\hline 
$A_{2}$ & $[\,0,j_{2}\,]_{\Delta}^{(R)}~,$& $j_{2}\geq1$ &$\Delta=j_{2}+\frac{3}{2}\,R+4$ & $[\,0,j_{2}-1\,]_{\Delta+1}^{(R+2)}$ \\
\hline
$A_{4}$ & $[\,0,0\,]_{\Delta}^{(R)}$& $$ &$\Delta=\frac{3}{2}\,R+4$ & $[\,0,0\,]_{\Delta+2}^{(R+4)}$ \\
\hline
\hline
$B_{1}$ & $[\,0,j_{2}\,]_{\Delta}^{(R)}~$& $j_2\geq 1$&$\Delta=j_{2}+\frac{3}{2}\,R+3$ & $[\,1,j_{2}\,]_{\Delta+1/2}^{(R+1)}$ \\
\hline
$B_{2}$ & $[\,0,0\,]_{\Delta}^{(R)}$& &$\Delta=\frac{3}{2}\,R+3$ & $[\,0,0\,]_{\Delta+1}^{(R+2)}$ \\
\hline
\hline
$C_{1}$ & $[\,0,0\,]_{\Delta}^{(R)}$& &$\Delta=\frac{3}{2}\,R$ & $[\,1,0\,]_{\Delta+1/2}^{(R+1)}$ \\
\hline
\end{tabular}
  \caption{Shortening conditions in five-dimensional~$\CN=1$ SCFTs.}
  \label{tab:5Dmult}
\end{table} 

The Lorentz-invariant supersymmetric deformations of five-dimensional~$\CN=1$ SCFTs are summarized in table~\ref{tab:5DD}. The only relevant deformations are flavor masses, which reside in~$C_1[0,0]_3^{(2)}$ flavor current multiplets. (See section~\ref{sec:flavormass} for a more detailed discussion.) There are no marginal deformations. 

\renewcommand{\arraystretch}{1.7}
\renewcommand\tabcolsep{10pt}
\begin{table}[H]
  \centering
  \begin{tabular}{ |c|c|c| }
\hline
 \multicolumn{1}{|c|}{\bf Primary $\mathcal{O}$} &  \multicolumn{1}{c}{\bf Deformation $\delta \SL$} &\multicolumn{1}{|c|}{\bf Comments}\\
\hline
\hline
 \multirow{ 2}{*}{$C_{1} \left\{ \stackanchor{$(2)$}{$\Delta_\CO =3$}\right \}$}&  \multirow{ 2}{*}{$Q^{2}\mathcal{O} \in \left\{ \stackanchor{$(0)$}{$\Delta =4$}\right \}$} & \multirow{ 2}{*}{Flavor Current} \\
&&\\
\hline
 \multirow{ 2}{*}{$C_{1} \left\{ \stackanchor{$(R+4)$}{$\Delta_\CO =6+\frac{3}{2}R$}\right \}$}&  \multirow{ 2}{*}{$Q^{4}\mathcal{O} \in \left\{ \stackanchor{$(R)$}{$\Delta =8+\frac{3}{2}R$}\right \}$} & \multirow{ 2}{*}{$F$-Term} \\
&&\\
\hline
 \multirow{ 2}{*}{$L \left\{ \stackanchor{$(R)$}{$\Delta_\CO >4+\frac{3}{2}R$}\right \}$}&  \multirow{ 2}{*}{$Q^{8}\mathcal{O} \in \left\{ \stackanchor{$(R)$}{$\Delta >8+\frac{3}{2}R$}\right \}$} & \multirow{ 2}{*}{$D$-Term} \\
&&\\
\hline
\end{tabular}
  \caption{Deformations of five-dimensional~$\CN=1$ SCFTs. The~$R$-charge of the deformation is denoted by the~$\frak{su}(2)_R$ Dynkin label~$R \in \Z_{\geq 0}$.}
  \label{tab:5DD}
\end{table}

\subsection{Six Dimensions}

In this subsection we list all Lorentz-invariant deformations of six-dimensional~SCFTs with~$(\CN, 0)$ supersymmetry for~$\CN=1,2$. (Unitarity SCFTs with~$\CN \geq 3$ do not exist, because they do not admit a stress tensor~\cite{multiplets}.) The corresponding superconformal algebras and their unitary representations are briefly summarized below. (See for instance~\cite{Minwalla:1997ka,Ferrara:2000xg,Dobrev:2002dt,Bhattacharya:2008zy,multiplets} and references therein for additional details.) Representations of the~$\frak{so}(6) = \frak{su}(4)$ Lorentz algebra are denoted using~$\frak{su}(4)$ Dynkin labels,
\begin{equation}\label{Lor6d}
[j_1, j_2, j_3]~, \qquad j_1, j_2, j_3 \in \Z_{\geq 0}~.
\end{equation} 
For instance, $[1,0,0]$ and~$[0,0,1]$ are the left- and right-handed chiral spinor representations~${\bf 4}, {\bf 4'}$ of~$\frak{so}(6)$,\footnote{~As representations of~$\frak{su}(4)$, the~$\bf 4'$ is typically denoted by~$\bf \b 4$, which is related to the~$\bf 4$ by complex conjugation. However, in six-dimensional Minkowski space, chiral spinors are not related by complex conjugation.} while~$[0,1,0]$ is the vector representation~$\bf 6$ of~$\frak{so}(6)$. Operators of scaling dimension~$\Delta$ are denoted by~$[j_1,j_2,j_3]_\Delta$.

\subsubsection{$d=6$,~$\CN=(1,0)$}

\label{sec:d6n1defs}

The~$\CN=(1,0)$ superconformal algebra is~$\frak{osp}(8|2)$, hence the~$R$-symmetry is~$\frak{sp}(2)_R \simeq \frak{su}(2)_R$. Its representations are denoted by~$(R)$, where~$R \in \Z_{\geq 0}$ is an~$\frak{su}(2)_R$ Dynkin label. The~$Q$-supersymmetries transform as
 \begin{equation}
Q \in [1,0,0]_{1/2}^{(1)}~~, \qquad N_Q = 8~.
\end{equation}
The superconformal unitarity bounds and shortening conditions are summarized in table~\ref{tab:6DN1}. For instance, $B_1[0,0,0]^{(1)}_{2}$ is a free hypermultiplet and~$C_2[0,0,0]^{(0)}_2$ is a free tensor multiplet, while~$B_1[0,0,0]^{(2)}_{4}$ is a flavor current multiplet and~$B_3[0,0,0]^{(0)}_4$ is the stress-tensor multiplet.  

\renewcommand{\arraystretch}{1.7}
\renewcommand{\tabcolsep}{8pt}
\begin{table}[H]
  \centering
  \begin{tabular}{ |c|lr| l|l| }
\hline
{\bf Name} &  \multicolumn{2}{c}{\bf Primary} &  \multicolumn{1}{|c|}{\bf Unitarity Bound} & \multicolumn{1}{c|}{\bf Null State } \\
\hline
\hline
$L$ & $[j_1,j_2,j_3]_{\Delta}^{(R)}$&  &$\Delta>\frac{1}{2}\left(j_1+2j_2+3j_3\right)+2R+6$ & \multicolumn{1}{c|}{$-$} \\
\hline
\hline
$A_{1}$ &$[j_1,j_2,j_3]_{\Delta}^{(R)}~,$& $j_3\geq1$ &$\Delta=\frac{1}{2}\left(j_1+2j_2+3j_3\right)+2R+6$ & $[j_1,j_2,j_3-1]_{\Delta+1/2}^{(R+1)}$ \\
\hline 
$A_{2}$ &$[j_1,j_2,0]_{\Delta}^{(R)}~,$& $j_2\geq1$ &$\Delta=\frac{1}{2}\left(j_1+2j_2\right)+2R+6$ & $[j_1,j_2-1,0]_{\Delta+1}^{(R+2)}$ \\
\hline
$A_{3}$ &$[j_1,0,0]_{\Delta}^{(R)}~,$& $j_1\geq1$ &$\Delta=\frac{1}{2}\, j_1+2R+6$ & $[j_1-1,0,0]_{\Delta+3/2}^{(R+3)}$ \\
\hline
$A_{4}$ &$[0,0,0]_{\Delta}^{(R)}$& $$ &$\Delta=2R+6$ & $[0,0,0]_{\Delta+2}^{(R+4)}$ \\
\hline
\hline
$B_{1}$ &$[j_1,j_2,0]_{\Delta}^{(R)}~,$& $j_2\geq1$ &$\Delta=\frac{1}{2}\left(j_1+2j_2\right)+2R+4$ & $[j_1,j_2-1,1]_{\Delta+1/2}^{(R+1)}$ \\
\hline
$B_{2}$ &$[j_1,0,0]_{\Delta}^{(R)}~,$& $j_1\geq1$ &$\Delta=\frac{1}{2}\, j_1+2R+4$ & $[j_1-1,0,1]_{\Delta+1}^{(R+2)}$ \\
\hline
$B_{3}$ &$[0,0,0]_{\Delta}^{(R)}$& $$ &$\Delta=2R+4$ & $[0,0,1]_{\Delta+3/2}^{(R+3)}$ \\
\hline
\hline
$C_{1}$ &$[j_1,0,0]_{\Delta}^{(R)}~,$& $j_1 \geq 1$ &$\Delta=\frac{1}{2} \, j_1+2R+2$ & $[j_1-1,1,0]_{\Delta+1/2}^{(R+1)}$ \\
\hline
$C_{2}$ &$[0,0,0]_{\Delta}^{(R)}$& $$ &$\Delta=2R+2$ & $[0,1,0]_{\Delta+1}^{(R+2)}$ \\
\hline
\hline
$D_{1}$ &$[0,0,0]_{\Delta}^{(R)}$& $$ &$\Delta=2R$ & $[1,0,0]_{\Delta+1/2}^{(R+1)}$ \\
\hline
\end{tabular}
  \caption{Shortening conditions in six-dimensional~$\CN=(1,0)$ theories.}
  \label{tab:6DN1}
\end{table} 

The Lorentz-invariant supersymmetric deformations of six-dimensional~$\CN=(1,0)$ SCFTs are summarized in table~\ref{tab:6D1D}. Note that there are neither relevant nor marginal deformations. Therefore, the only possible supersymmetric RG flows out of these fixed points are triggered by moving onto a moduli space of vacua~\cite{Cordova:2015fha}. The fact that there are no marginal deformations was also discussed in~\cite{Louis:2015mka}.

\renewcommand{\arraystretch}{1.7}
\renewcommand\tabcolsep{10pt}
\begin{table}[H]
  \centering
  \begin{tabular}{ |c|c|c| }
\hline
 \multicolumn{1}{|c|}{\bf Primary $\mathcal{O}$} &  \multicolumn{1}{c}{\bf Deformation $\delta \SL$} &\multicolumn{1}{|c|}{\bf Comments}\\
\hline
\hline
 \multirow{ 2}{*}{$D_{1} \left\{ \stackanchor{$(R+4)$}{$\Delta_\CO =8+2R$}\right \}$}&  \multirow{ 2}{*}{$Q^{4}\mathcal{O} \in \left\{ \stackanchor{$(R)$}{$\Delta =10+2R$}\right \}$} & \multirow{ 2}{*}{$F$-Term} \\
&&\\
\hline
 \multirow{ 2}{*}{$L \left\{ \stackanchor{$(R)$}{$\Delta_\CO >6+2R$}\right \}$}&  \multirow{ 2}{*}{$Q^{8}\mathcal{O} \in \left\{ \stackanchor{$(R)$}{$\Delta >10+2R$}\right \}$} & \multirow{ 2}{*}{$D$-Term} \\
&&\\
\hline
\end{tabular}
  \caption{Deformations of six-dimensional~$\CN=(1,0)$ SCFTs. The~$R$-charge of the deformation is denoted by the~$\frak{su}(2)_R$ Dynkin label~$R \in \Z_{\geq 0}$.}
  \label{tab:6D1D}
\end{table} 

\subsubsection{$d=6$,~$\CN=(2,0)$}

The~$\CN=(2,0)$ superconformal algebra is~$\frak{osp}(8|4)$, so that the~$R$-symmetry is~$\frak{sp}(4)_R$. Its representations are denoted by~$(R_1,R_2)$, where~$R_1,R_2 \in \Z_{\geq 0}$ are~$\frak{sp}(4)_R$ Dynkin labels, e.g.~$(1,0)$ and~$(0,1)$ denote the~$\bf 4$ and~$\bf 5$, respectively. The~$Q$-supersymmetries transform as 
 \begin{equation}
Q \in [1,0,0]_{1/2}^{(1,0)}~~, \qquad N_Q = 16~.
\end{equation}
The superconformal unitarity bounds and shortening conditions are summarized in table~\ref{tab:6DN2}. For instance, $D_1[0,0,0]_2^{(0,1)}$ is a free tensor multiplet, while~$D_1[0,0,0]_4^{(0,2)}$ is the stress-tensor multiplet. 


\renewcommand{\arraystretch}{1.7}
\renewcommand\tabcolsep{3.5pt}
\begin{table}[H]
  \centering
  \begin{tabular}{ |c|lr| l|l| }
\hline
{\bf Name} &  \multicolumn{2}{c}{\bf Primary} &  \multicolumn{1}{|c|}{\bf Unitarity Bound} & \multicolumn{1}{c|}{\bf Null State } \\
\hline
\hline
$L$ & $[j_1,j_2,j_3]_{\Delta}^{(R_1,R_2)}$&  &$\Delta>\frac{1}{2}\left(j_1+2j_2+3j_3\right)+2(R_1+R_2)+6$ & \multicolumn{1}{c|}{$-$} \\
\hline
\hline
$A_{1}$ &$[j_1,j_2,j_3]_{\Delta}^{(R_1,R_2)}~,$& $j_3\geq1$ &$\Delta=\frac{1}{2}\left(j_1+2j_2+3j_3\right)+2(R_1+R_2)+6$ & $[j_1,j_2,j_3-1]_{\Delta+1/2}^{(R_1+1,R_2)}$ \\
\hline 
$A_{2}$ &$[j_1,j_2,0]_{\Delta}^{(R_1,R_2)}~,$& $j_2\geq1$ &$\Delta=\frac{1}{2}\left(j_1+2j_2\right)+2(R_1+R_2)+6$ & $[j_1,j_2-1,0]_{\Delta+1}^{(R_1+2,R_2)}$ \\
\hline
$A_{3}$ &$[j_1,0,0]_{\Delta}^{(R_1,R_2)}~,$& $j_1\geq1$ &$\Delta=\frac{1}{2}\, j_1+2(R_1+R_2)+6$ & $[j_1-1,0,0]_{\Delta+3/2}^{(R_1+3,R_2)}$ \\
\hline
$A_{4}$ &$[0,0,0]_{\Delta}^{(R_1,R_2)}$& $$ &$\Delta=2(R_1+R_2)+6$ & $[0,0,0]_{\Delta+2}^{(R_1+4,R_2)}$ \\
\hline
\hline
$B_{1}$ &$[j_1,j_2,0]_{\Delta}^{(R_1,R_2)}~,$& $j_2\geq1$ &$\Delta=\frac{1}{2}\left(j_1+2j_2\right)+2(R_1+R_2)+4$ & $[j_1,j_2-1,1]_{\Delta+1/2}^{(R_1+1,R_2)}$ \\
\hline
$B_{2}$ &$[j_1,0,0]_{\Delta}^{(R_1,R_2)}~,$& $j_1\geq1$ &$\Delta=\frac{1}{2}\, j_1+2(R_1+R_2)+4$ & $[j_1-1,0,1]_{\Delta+1}^{(R_1+2,R_2)}$ \\
\hline
$B_{3}$ &$[0,0,0]_{\Delta}^{(R_1,R_2)}$& $$ &$\Delta=2(R_1+R_2)+4$ & $[0,0,1]_{\Delta+3/2}^{(R_1+3,R_2)}$ \\
\hline
\hline
$C_{1}$ &$[j_1,0,0]_{\Delta}^{(R_1,R_2)}~,$& $j_1 \geq 1$ &$\Delta=\frac{1}{2} \, j_1+2(R_1+R_2)+2$ & $[j_1-1,1,0]_{\Delta+1/2}^{(R_1+1,R_2)}$ \\
\hline
$C_{2}$ &$[0,0,0]_{\Delta}^{(R_1,R_2)}$& $$ &$\Delta=2(R_1+R_2)+2$ & $[0,1,0]_{\Delta+1}^{(R_1+2,R_2)}$ \\
\hline
\hline
$D_{1}$ &$[0,0,0]_{\Delta}^{(R_1,R_2)}$& $$ &$\Delta=2(R_1+R_2)$ & $[1,0,0]_{\Delta+1/2}^{(R_1+1,R_2)}$ \\
\hline
\end{tabular}
  \caption{Shortening conditions in six-dimensional~$\CN=(2,0)$ theories.}
  \label{tab:6DN2}
\end{table} 


The Lorentz-invariant supersymmetric deformations of six-dimensional~$\CN=(2,0)$ SCFTs are summarized in table~\ref{tab:6D2D}. As was already the case for~$\CN=(1,0)$ theories (see section~\ref{sec:d6n1defs}), there are no supersymmetric relevant or marginal deformations. 

\renewcommand{\arraystretch}{1.7}
\renewcommand\tabcolsep{10pt}
\begin{table}[H]
  \centering
  \begin{tabular}{ |c|c|c| }
\hline
 \multicolumn{1}{|c|}{\bf Primary $\mathcal{O}$} &  \multicolumn{1}{c}{\bf Deformation $\delta \SL$} &\multicolumn{1}{|c|}{\bf Comments}\\
\hline
\hline
 \multirow{ 2}{*}{$D_{1} \left\{ \stackanchor{$(0,R_{2}+4)$}{$\Delta_\CO =8+2R_{2}$}\right \}$}&  \multirow{ 2}{*}{$Q^{8}\mathcal{O} \in \left\{ \stackanchor{$(0,R_{2})$}{$\Delta =12+2R_{2}$}\right \}$} & \multirow{ 2}{*}{$F$-Term} \\
&&\\
\hline
 \multirow{ 2}{*}{$D_{1} \left\{ \stackanchor{$(R_{1}+4,R_{2})$}{$\Delta_\CO =8+2(R_1+R_2)$}\right \}$}&  \multirow{ 2}{*}{$Q^{12}\mathcal{O} \in \left\{ \stackanchor{$(R_{1},R_{2})$}{$\Delta =14+2(R_{1}+R_{2})$}\right \}$} & \multirow{ 2}{*}{$-$} \\
&&\\
\hline
 \multirow{ 2}{*}{$L \left\{ \stackanchor{$(R_{1},R_{2})$}{$\Delta_\CO >6+2(R_{1}+R_{2})$}\right \}$}&  \multirow{ 2}{*}{$Q^{16}\mathcal{O} \in \left\{ \stackanchor{$(R_{1},R_{2})$}{$\Delta >14+2(R_{1}+R_{2})$}\right \}$} & \multirow{ 2}{*}{$D$-Term} \\
&&\\
\hline
\end{tabular}
  \caption{Deformations of six-dimensional~$\CN=(2,0)$ SCFTs. The~$R$-symmetry representation of the deformation is denoted by the~$\frak{sp}(4)_R$ Dynkin labels~$R_1,R_2 \in \Z_{\geq 0}$.}
  \label{tab:6D2D}
\end{table}

\section{Deformations Related To Conserved Currents}

\label{sec:currdefs}

In this section we discuss Lorentz-invariant supersymmetric deformations which reside in supermultiplets that also contain conserved currents, focusing on flavor currents and the stress tensor. (As in section~\ref{sec:tables}, we will not discuss multiplets containing additional supersymmetry currents or higher spin currents. A detailed analysis of all superconformal multiplets that contain conserved currents can be found in~\cite{multiplets}.)  Such deformations can lead to the appearance of additional bosonic charges in the supersymmetry algebra, which arise from the currents that reside in the same multiplet as the deformation. We also comment on the fact that some of these deformations can fail to be supersymmetric at higher order, even though they preserve supersymmetry at leading order.

\subsection{Flavor Current Multiplets}

\label{sec:flavormass}

Using the tables in section~\ref{sec:tables}, we can enumerate theories admitting Lorentz-invariant deformations that reside in the same multiplet as a conserved flavor current~$j_\mu^a$, where~$a$ is an adjoint flavor index. (The corresponding flavor charges commute with all supersymmetries.) All of these deformations are relevant, with scaling dimension~$\Delta = d-1$, and all of them carry an adjoint flavor index~$a$, just as the current~$j_\mu^a$. In many weakly-coupled examples, such deformations are fermion mass terms of the schematic form~$m_a \left(\b \psi \psi\right)^a$. (The corresponding bosonic mass terms only arise at~$\CO(m^2)$.) For this reason, we will refer to such deformations as flavor masses and collectively denote them by~$\CM_\text{flav.}^a$. Flavor mass deformations can occur in the following theories:
\begin{itemize}
\item Three-dimensional~$2 \leq \mathcal{N} \leq 4$ SCFTs all admit flavor mass deformations~$\CM_\text{flav.}^a$ that are real. Depending on~$\CN$, they reside in different multiplets and transform differently under the appropriate~$R$-symmetries:
\begin{itemize}
\item[] $\CN=2$: The flavor mass is neutral under the~$\frak{u}(1)_R$ symmetry, i.e.~$\CM^a_\text{flav.} \in [0]_2^{(0)}$. It resides at level two in an~$A_2\b A_2[0]^{(0)}_1$ flavor current multiplet (see tables~\ref{tab:3DN2C}, \ref{tab:3DN2AC},~\ref{tab:3DN2D}). 

\item[] $\CN=3$: The flavor mass transforms as an~$\frak{su}(2)_R$ triplet, i.e.~$\left(\CM^{a}_\text{flav.}\right)^{(ij)} \in [0]_2^{(2)}$. It resides at level two in a~$B_1[0]_1^{(2)}$ flavor current multiplet (see tables~\ref{tab:3DN3} and~\ref{tab:3DN3D}).

\item[] $\CN=4$: Here there are two different flavor mass deformations~$\left(\CM^{a}_\text{flav.}\right)^{(ij)} \in [0]_2^{(2;0)}$ and~$\left(\CM'^{\,a}_\text{flav.}\right)^{(i'j')} \in [0]_2^{(0;2)}$, which are triplets under~$\frak{su}(2)_R$ and~$\frak{su}(2)'_R$, respectively. They reside at level two, in two different flavor current multiplets, $B_1[0]^{(0;2)}_1$ and~$B_1[0]^{(2;0)}$, which are exchanged by mirror symmetry (see tables~\ref{tab:3DN4} and~\ref{tab:3DN4D}). 

\end{itemize}

\item Four-dimensional~$\CN=2$ SCFTs admit complex flavor mass deformations~$\CM^a_\text{flav.}$, which are paired with their complex conjugates~$\b \CM^{\, a}_\text{flav.}$. They are neutral under the~$\frak{su}(2)_{R}$ symmetry, but carry~$\frak{u}(1)_{r}$ charges: $\CM^a_\text{flav.} \in [0;0]^{(0; 2)}_3$ and $\b \CM^{\, a}_\text{flav.} \in [0;0]^{(0;-2)}_3$. Both deformations reside at level two in the same~$B_1\b B_1[0;0]^{(2;0)}_2$ flavor current multiplet (see tables~\ref{tab:4DN2C}, \ref{4DN2AC} and~\ref{tab:4DN2D}).
 
\item Five-dimensional~$\CN=1$ SCFTs admit real flavor mass deformations~$\CM^a_\text{flav.}$ that are neutral under the~$\frak{su}(2)_{R}$ symmetry, i.e.~$\CM^a_\text{flav.} \in [0,0]^{(0)}_4$. They reside at level two in a~$C_1[0,0]^{(2)}_3$ flavor current multiplet. 
\end{itemize}

\noindent All of these deformations arise in myriad well-studied examples. In addition to standard mass terms for scalars and fermions in matter multiplets, which are possible in all cases, we also mention the following two interesting possibilities in three and five dimensions:
\begin{itemize}
\item In three dimensions, FI-terms for abelian gauge fields can be interpreted as flavor masses for the topological current~$\star F$, where~$F$ the abelian field strength. (By contrast, deformations of SCFTs in~$4 \leq d \leq 6$ dimensions do not include FI-terms, see section~\ref{sec:noFIterms}.) Abstractly, such FI flavor masses cannot be distinguished from conventional flavor masses for matter fields, as reflected by the fact that the two can be exchanged by duality (see for instance~\cite{Intriligator:1996ex,Aharony:1997bx}). 

\item In five dimensions, many SCFTs possess flavor symmetries. Upon activating the corresponding flavor masses, such theories often flow to weakly-coupled, generally nonabelian, gauge theories in the IR. Each gauge group gives rise to a topological current~$\star \Tr (F \wedge F)$, which descends from a particular flavor current of the SCFT in the UV. The corresponding Yang-Mills kinetic term descends from the flavor mass deformation related to that UV flavor current (see for instance~\cite{Seiberg:1996bd,Morrison:1996xf,Douglas:1996xp,Intriligator:1997pq}).

\end{itemize}

In all cases discussed above, the Lorentz-invariant flavor mass deformation resides at level two in a flavor current multiplet whose superconformal primary~$\CJ^{a, I}$ is also a Lorentz scalar. (Here~$I$ collectively denotes all~$R$-symmetry labels.) Schematically, we can therefore write the deformation as follows,
\begin{equation}
\delta \SL_\text{flav.} =m_{a,I} \CM^{a, I}_\text{flav.}   + \CO(m^2)=m_{a,I} (Q^{2}\CJ^{a})^{I} + \CO(m^2)~. \label{currentlag}
\end{equation}
Following~\cite{Seiberg:1993vc}, the mass parameters~$m_{a, I}$ can be interpreted as scalars residing in non-dynamical vector multiplets that contain background flavor gauge fields. This makes a variety of non-renormalization theorems manifest. For instance, in four-dimensional~$\CN=2$ theories, the Higgs-branch metric cannot depend on vector multiplet scalars, and hence it also does not depend on flavor mass parameters~\cite{Argyres:1996eh}. 

This perspective also leads to an intuitive picture for the structure of flavor mass deformations in theories with~$N_Q = 8$ supercharges. In six-dimensional~$\CN=(1,0)$ theories, vector multiplets only contain gauge fields and fermions, but no scalars. Correspondingly, there are no scalars that reside in flavor current multiplets, and hence such theories do not admit flavor mass deformations, even though they may possess flavor symmetries. (This is consistent with section~\ref{sec:d6n1defs}, where it was shown that these theories do not possess any relevant deformations.) In~$3 \leq d \leq 5$ dimensions, theories with the same amount of supersymmetry do admit flavor masses.   The corresponding background vector multiplet scalars~$m_{a, I}$ can be viewed as Wilson lines for six-dimensional background gauge fields which wrap one-cycles of a~$(6-d)$-dimensional torus that is used to compactify the six-dimensional theory down to~$d=5,4,3$. This correctly accounts for the fact that each flavor symmetry leads to~$1,2,3$ mass parameters for~$\CN=1,2,4$ theories in the respective dimension. The enhanced~$\frak{u}(1)_R$ or~$\frak{su}(2)'_R$ symmetries of four-dimensional~$\CN=2$ or three-dimensional~$\CN=4$ theories (in addition to the~$\frak{su}(2)_R$ symmetry that is already present in five and six dimensions) can also be understood as arising from dimensional reduction, and this explains why flavor mass deformations in these theories are charged under~$\frak{u}(1)_R$ or~$\frak{su}(2)'_R$. 

\subsection{Stress Tensor Multiplets}

\label{sec:defst}

We now consider Lorentz-invariant supersymmetric deformations that reside in the same multiplet as the stress tensor~$T_{\mu\nu}$. For fixed values of~$d$ and~$\CN$, the stress tensor multiplet of a SCFT is unique. (The only exception occurs in three-dimensional~$\CN=8$ theories, as mentioned around~\eqref{eq:d3n8scs}. See~\cite{multiplets} for a detailed discussion of stress tensor multiplets in all dimensions.) Moreover, every theory must possess such a multiplet, and hence the corresponding deformations are universal, i.e.~they always exist. Universal deformations occur in the following theories: 
\begin{itemize}
\item Four-dimensional~$\mathcal{N}=4$ SCFTs possess a universal marginal deformation~$\CO$, which is complex and paired with its complex conjugate~$\b \CO$.  Both deformations are neutral under the~$\frak{su}(4)_{R}$ symmetry, i.e.~$\CO, \b \CO \in [0;0]_4^{(0,0,0)}$. They reside at the top of the multiplet, at level~$\ell = 4$, together with the stress tensor~$T_{\mu\nu}$. 

\item Three-dimensional SCFTs with~$\mathcal{N} \geq 4$ supersymmetry have universal relevant deformations~$\CM_\text{univ.}$ of dimension~$\Delta = 2$. In analogy with the flavor masses~$\CM^a_\text{flav.}$ discussed in section~\ref{sec:flavormass}, we will refer to them as universal mass deformations. They reside in the middle of the stress-tensor multiplet, at level~$\ell =2$, while~$T_{\mu\nu}$ resides at level~$\ell = 4$. The~$\CM_\text{univ.}$ are generally charged under the~$R$-symmetry (see below), but since they reside in the same multiplet as~$T_{\mu\nu}$, they are neutral under any flavor symmetries. 
\end{itemize}
\noindent We will now comment on these two kinds of universal deformations in turn. 

Since the marginal deformation~$\CO$ in four-dimensional~$\CN=4$ theories resides in the same multiplet as~$T_{\mu\nu}$, many of its properties follow directly from superconformal Ward identities (see for instance~\cite{Petkou:1999fv,Basu:2004nt,Papadodimas:2009eu} and references therein). First, the deformation is necessarily exactly marginal, because its dimension is tethered to that of the stress tensor.\footnote{~Alternatively, we can use the fact that all marginal deformations of four-dimensional~$\CN\geq 2$ theories are exactly marginal, as discussed in section~\ref{sec:d4n2defs}.} It therefore gives rise to a conformal manifold labeled by one complex parameter~$\tau$. The local geometry of this manifold is fixed by Ward identities, which imply that its Zamolodchikov metric has constant negative curvature,
\begin{equation}
ds^{2} = C\,\frac{d\tau d\overline{\tau}}{(\Im \tau)^{2}}~, \qquad C>0~.
\end{equation}
Here~$C$ is proportional to the stress tensor two-point function, which is determined by the Weyl anomaly of the theory. (The four-dimensional Weyl anomalies~$a$ and~$c$ coincide in all~$\CN=4$ SCFTs~\cite{Anselmi:1997am}.)  In four-dimensional~$\CN=4$ SCFTs that have a Lagrangian description as gauge theories, the marginal parameter~$\tau$ is identified with the complexified gauge coupling. However, it is not known whether all~$\CN=4$ theories admit such a description.

The universal mass deformation~$\CM_\text{univ.}$ that exists in all three-dimensional~$\CN\geq 4$ theories is somewhat less familiar. If we use~$V$ to denote the fundamental~$\CN$-dimensional vector representation of the~$\frak{so}(\CN)_R$ symmetry, then~$\CM_\text{univ.}$ transforms as a~$(\CN-4)$-form,
\begin{equation}\label{umassrsym}
\CM_\text{univ.} \in \wedge^{\CN-4} V~.
\end{equation}
For~$\CN=8$, the four-forms~$\Lambda^4 V$ can be decomposed into their self-dual and anti-self-dual parts, $\big(\Lambda^4 V\big)_\pm$, with Dynkin labels~$(0,0,0,2)$ and~$(0,0,2,0)$. They reside, respectively, in two different stress-tensor multiplets~$B_1[0]^{(0,0,2,0)}_{1}$ and~$B_1[0]^{(0,0,0,2)}_{1}$ that are exchanged by the triality subgroup~$T$, as discussed in section~\ref{sec:d3n8defs}. Some early and recent discussions of universal mass deformations in~$\CN=8$ theories and their gravity duals are in~\cite{Intriligator:1999ai, Bena:2004jw, Itzhaki:2005tu, Lin:2005nh,Gomis:2008vc,Agarwal:2008pu}. 

As a simple example, we write down the universal mass deformation for a free~$\CN=4$ hypermultiplet~$(H^i, \psi_\alpha^{i'})$, which itself constitutes a~$B_1[0]^{(1;0)}_{1/2}$ multiplet. The scalars~$H^i$ and the fermions~$\psi^{i'}_\alpha$ are doublets under the first and second factors of the~$\frak{su}(2)_R \times \frak{su}(2)'_R$ symmetry. The action of the supercharges on the scalars is given by~$Q_\alpha^{i i'} H^j \sim \ep^{ij} \psi^{i'}_\alpha$~, where the~$\sim$ indicates that we are not keeping track of convention-dependent proportionality factors. The universal mass deformation resides at level two of the~$A_2[0]^{(0;0)}_1$ stress tensor multiplet, whose bottom component is~$\ep^{ij} \b H_i H_j$. According to~\eqref{umassrsym}, the universal mass is~$R$-symmetry (as well as Lorentz) invariant. Explicitly, 
\begin{equation}\label{n4hypum}
\CM_\text{univ.} \sim m \ep^{\alpha\beta} \ep_{i' j'} \b \psi^{\, i'}_\alpha \psi^{\, j'}_\beta \sim m \, \ep^{\alpha\beta} \ep_{ij} \ep_{i' j'} Q_\alpha^{i i'} Q_\beta^{j j'} \left(\ep^{kl} \b H_k H_l\right)~,
\end{equation}
where~$m$ is a real mass parameter. Note that the two supercharges are contracted to a Lorentz and~$R$-symmetry singlet. As was the case for flavor masses (see~\eqref{currentlag}), the~$\CO(m)$ deformation is a pure fermion mass term, while the corresponding scalar mass term~$m^2 \ep^{ij} \b H_i H_j$ only arises at~$\CO(m^2)$. However, unlike the flavor masses residing in~$B_1[0]_1^{(2;0)}$ and~$B_1[0]_1^{(0;2)}$ flavor current multiplets, which are exchanged by mirror symmetry, the universal mass is inert under the mirror automorphism. In this example, the universal mass~\eqref{n4hypum} leads to a fully gapped theory. As we will review below, this is a general, model-independent property of all universal mass deformations.  
 
\subsection{Deformed Stress-Tensor Multiplets and Supersymmetry Algebras}

\label{sec:defst}

In an SCFT, the stress tensor~$T_{\mu\nu}$ resides in a short superconformal multiplet (whose structure essentially only depends on~$d$ and~$\CN$), together with the~$N_Q$ supersymmetry currents~$S^i_{\mu \alpha}$ and the~$R$-symmetry currents. The multiplet is typically completed by other operators that need not be conserved currents. At the conformal point, both the stress tensor and the supersymmetry currents are traceless. Schematically,
\begin{equation}
T^\mu _\mu = 0~, \qquad \b \sigma^{\mu\alphadot\alpha} S^i_{\mu\alpha} = 0~,\label{trace}
\end{equation}
where we have used four-dimensional notation to indicate the spin-$\half$ projection of the supersymmetry currents in all dimensions. The vanishing traces in~\eqref{trace} allow the definition of dilatation and special conformal generators, as well as the superconformal~$S$-supersymmetries. See~\cite{multiplets} for a unified discussion of superconformal stress-tensor multiplets for all~$d$ and~$\CN$. 

When a CFT is deformed by a scalar operator~$\CO$ of dimension~$\Delta$, i.e.~$\delta \SL = \lambda \CO$, then the trace of the stress tensor is deformed to~$T^\mu_\mu \sim \lambda \left(\Delta - d\right) \CO$, at leading order in~$\lambda$. (More generally, the coefficient of~$\CO$ in~$T^\mu_\mu$ is proportional to the~$\beta$-function of the coupling~$\lambda$.) If the deformation~$\CO$ is marginal, the deformation preserves conformal symmetry at leading order and~$T^\mu_\mu$ remains zero at that order. (A non-zero trace, i.e.~a beta function, may be generated at higher order.) In all other cases, conformal symmetry is broken and~$T_{\mu\nu}$ acquires a trace. For the case of an SCFT deformed by a Lorentz-scalar operator~$\CO$ that preserves the~$Q$-supersymmetries but breaks conformal symmetry (and hence the~$S$-supersymmetries), the currents~$T_{\mu\nu}, S^i_{\mu\alpha}$ remain conserved, but both of the traces that were set to zero in~\eqref{trace} are activated by the deformation. Together with their superpartners, they deform the conformal stress-tensor multiplet into a multiplet of Poincar\'e supersymmetry that contains the currents~$T_{\mu\nu}, S^i_{\mu\alpha}$. Typically, it contains more operators than the conformal stress-tensor multiplet, which are supplied by the multiplet of the deformation~$\CO$. (However, operators from other multiplets can also participate, see below.) 

As an example, consider four-dimensional~$\CN=1$ theories.  At the superconformal point, the stress-tensor resides in an~$A_1 \b A_1[0;0]^{(0)}_2$ multiplet (see tables~\ref{tab:4DN1C} and~\ref{4DN1AC}), whose primary~$\CJ_\mu = \b \sigma_\mu^{\alpha\alphadot} \CJ_{\alpha\alphadot}$ is the~$\frak{u}(1)_R$ current. The superconformal shortening conditions take the form
\begin{equation}
\overline Q^{\dot\alpha}\CJ _{\alpha \dot \alpha}=0~, \qquad Q^\alpha \CJ_{\alpha \dot\alpha}=0~.\label{4dn1conf}\
\end{equation}
After using these to eliminate all null states (including conservation laws), the multiplet has~$8 +8$ bosonic and fermionic operators: the conserved~$R$-current~$\CJ_\mu$, as well as the conserved and traceless supersymmetry currents~$S_{\mu\alpha}, \b S_{\mu\alphadot}\,$ and stress-tensor~$T_{\mu\nu}$. 

Unlike the superconformal case, the representation theory of the Poincar\'e supersymmetry algebra admits several different non-conformal stress-tensor multiplets. (See~\cite{Komargodski:2010rb,Dumitrescu:2011iu} and references therein for a detailed discussion.) The most common such multiplet is Ferrara-Zumino (FZ) multiplet. In addition to~$\CJ_{\alpha\alphadot}$, it also contains a chiral submultiplet~$X$ (and its conjugate anti-chiral multiplet~$\b X$). The superconformal shortening condition~\eqref{4dn1conf} is deformed to 
\begin{equation}
\overline Q^{\dot\alpha}\CJ _{\alpha \dot \alpha}=Q_\alpha X~, \qquad Q^\alpha \CJ_{\alpha \dot\alpha}=\overline Q_{\dot \alpha} \b X~, \qquad \b Q_\alphadot X = Q_\alpha \b X = 0~.\label{FZmultiplet}
\end{equation}
The~$4+4$ bosonic and fermionic operators in~$X$ combine with the~$8+8$ operators in the superconformal multiplet to make a~$12+12$ multiplet. The bottom component~$\CJ_\mu = \sigma_\mu^{\alpha\alphadot}\CJ_{\alpha\alphadot}$ is no longer a conserved current; its non-zero divergence is one of the additional operators residing in~$X$. Therefore, the FZ-multiplet only contains a conserved~$R$-current if~$X$ vanishes and the theory is superconformal. If a non-conformal~$\CN=1$ theory has a~$\frak{u}(1)_R$ symmetry, the associated~$R$-current resides in a different~$12+12$ stress-tensor multiplet, which is related to the FZ-multiplet by an improvement transformation.

The vast majority of~$\CN=1$ theories admit an FZ-multiplet. The only known exceptions are models with FI-terms, or theories that can be obtained from such models by RG flow, e.g.~sigma models whose target spaces contain compact two-cycles~\cite{Komargodski:2009pc,Komargodski:2010rb,Dumitrescu:2011iu}. Since supersymmetry deformations of~$\CN=1$ superconformal theories never give rise to FI-terms (see the more detailed discussion in section~\ref{sec:noFIterms}), we expect that all deformed~$\CN=1$ SCFTs possess an FZ stress-tensor multiplet, whose~$X$ submultiplet is determined by the deformation.\footnote{~In a weakly-coupled theory of chiral superfields with superpotential~$W$ and K\"ahler potential~$K$, the chiral operator~$X$ in the FZ multiplet is a linear combination of~$W$ and~$\b Q^2 K$. Using conformal perturbation theory, it should be possible to find an analogous expression for~$X$ in any~$\CN=1$ SCFT that has been deformed by general~$F$- and~$D$-terms (see table~\ref{tab:4DN1D}), but we will not pursue it here.} 

We will now consider the effect of flavor mass deformations (see section~\ref{sec:flavormass}) on the stress-tensor multiplet. For simplicity, we will focus on theories with~$N_Q = 8$ supercharges in~$d = 5,4,3$ dimensions, where the discussion is fairly uniform.\footnote{~See~\cite{Dumitrescu:2011iu} for a discussion of stress-tensor multiplets in three-dimensional~$\CN=2$ theories, including the effects of flavor masses.} As explained below~\eqref{currentlag}, flavor currents in these theories are conveniently understood in terms of dimensional reduction, staring with a~$D_1[0,0,0]_4^{(2)}$ flavor current multiplet with primary~$\CJ_a^{ij}$ in six dimensions (see table~\ref{tab:6DN1}). Here~$a$ is an adjoint flavor index, and~$i,j$ are~$\frak{su}(2)_R$ doublet indices. This leads to flavor current multiplets with~$8+8$ operators that contain~$0,1,2,3$ scalar mass deformations in~$d = 6,5,4,3$ dimensions. The currents in~$d = 5,4$ reside in~$C_1[0,0]^{(2)}_3$ and~$B_1 \b B_1[0,0]_2^{(2)}$ multiplets (see tables~\ref{tab:5Dmult} and~\ref{tab:4DN2C}, \ref{4DN2AC}), while they can reside in either~$B_1[0]^{(2;0)}_1$ or~$B_1[0]^{(0;2)}_2$ multiplets in~$d = 3$ (see table~\ref{tab:3DN4}). Explicitly, the primaries and shortening conditions of these flavor current multiplets are
\begin{equation}\label{flavcurmults}
\begin{tabular}{ccc}
$Q_{\alpha}^{(i}\CJ_{a}^{jk)}=0~,$ &~~~$Q_{\alpha}^{(i}\CJ_{a}^{jk)}=\overline{Q}_{\dot{\alpha}}^{(i}\CJ_{a}^{jk)}=0~,$&~~~$Q_{\alpha}^{i'(i}\CJ_{a}^{jk)}=0$~~or~~$Q_{\alpha}^{i(i'}\CJ_{a}^{j'k')}=0~.$ \\
$d=5$ & $d=4$ & $d=3$
\end{tabular}
\end{equation}
Here~$\alpha, \alphadot$ are spacetime spinor indices in the respective dimension, while~$i, j, k$ are~$\frak{su}(2)_R$ doublet indices. In three dimensions, $i', j', k'$ are~$\frak{su}(2)'_R$ doublet indices.

The superconformal stress-tensor multiplets in these theories are based on a Lorentz-scalar and~$R$-symmetry neutral primary~$\CT$ of dimension~$\Delta = d-2$. In~$d = 6$ it is a~$B_3[0,0,0]^{(0)}_4$ multiplet, with a null state at~$\ell = 3$, and in~$d = 5,4,3$ the stress tensor multiplets~$B_2[0,0]^{(0)}_3$, $A_2 \b A_2[0;0]^{(0;0)}_2$, $A_2[0]_1^{(0;0)}$ all have null states at~$\ell=2$. Explicitly,\footnote{~In five dimensions, spinors are contracted using the~$\frak{sp}(4)$-invariant symplectic matrix~$\Omega_{\alpha\beta}$.}
\begin{equation}
\begin{tabular}{ccc}
$\Omega^{\alpha \beta}Q_{\alpha}^{(i}Q_{\beta}^{j)}\CT=0~,$ &~~~$\varepsilon^{\alpha \beta}Q_{\alpha}^{(i}Q_{\beta}^{j)}\CT=\varepsilon^{\dot{\alpha} \dot{\beta}}\overline{Q}_{\dot{\alpha}}^{(i}\overline{Q}_{\dot{\beta}}^{j)}\CT=0~,$&$~~~\varepsilon^{\alpha \beta}Q_{\alpha}^{(i'i}Q_{\beta}^{j'j)}\CT=0~.$ \\
$d=5$ & $d=4$ & $d=3$
\end{tabular}
\label{conformalTs}
\end{equation}
In three dimensions, the symmetrization of the~$\frak{su}(2)_R$ doublet indices~$i, j$ and the~$\frak{su}(2)'_R$ doublet indices~$i',j'$ denotes a projection onto the representation with Dynkin labels~$(2\,;2)$. The resulting multiplets contain~$8 (d-1) + 8(d-1)$ operators. Upon reducing the theory from~$d$ to~$d-1$ dimensions, the~$d$-dimensional stress-tensor multiplet becomes reducible and splits into a~$(d-1)$-dimensional stress-tensor multiplet and an~$8+8$ flavor current multiplet associated with the Kaluza-Klein (KK) symmetry, i.e.~the momentum in the reduced direction.

This perspective immediately suggests the form of the non-conformal stress-tensor multiplet in the presence of flavor mass deformations~\eqref{currentlag}, since they can be viewed as Wilson lines that wrap the reduced directions (see the discussion below~\eqref{currentlag}), and hence contribute to the momentum in those directions. We simply interpret a suitable linear combination of the flavor currents~$ \CJ^{ij}_a$ as the KK current, which no longer decouples in the presence of flavor mass deformations. This logic leads to a non-conformal stress-tensor multiplet in~$d = 5,4,3$ that was first described in~\cite{Sohnius:1978pk} for~$\CN=2$ theories in four dimensions, 
\begin{subequations}\label{deformedcurrentalgebra}
\begin{align}
& d = 5~: \quad \Omega^{\alpha \beta}Q_{\alpha}^{(i}Q_{\beta}^{j)}\CT=m_{a}\CJ_{a}^{ij}~,\\
& d=4~: \quad  \varepsilon^{\alpha \beta}Q_{\alpha}^{(i}Q_{\beta}^{j)}\CT=m_{a}\CJ_{a}^{ij}~, \qquad \varepsilon^{\dot{\alpha} \dot{\beta}}\overline{Q}_{\dot{\alpha}}^{(i}\overline{Q}_{\dot{\beta}}^{j)}\CT=\overline{m}_{a}\CJ_{a}^{ij}~,\\
& d=3~:\quad \varepsilon^{\alpha \beta}Q_{\alpha}^{(i'i}Q_{\alpha}^{j'j)}\CT= m_{a}^{i'j'}\CJ_{a}^{ij}+(m')_{a'}^{ij}\CJ_{a'}^{i'j'}~.
\end{align}
\end{subequations}
Here the~$m_a$ are real and~$R$-neutral in~$d=5$, while they are complex (with complex conjugates~$\b m_a$) and charged under~$U(1)_R$ in~$d = 4$. In three dimensions, both~$m_{a}^{i'j'}$ and~$(m')_{a'}^{ij}$ are real and transform as triplets of~$\frak{su}(2)'_R$ and~$\frak{su}(2)_R$, respectively. The conserved flavor currents on the right-hand sides of~\eqref{deformedcurrentalgebra} integrate to Lorentz-scalar central charges in the supersymmetry algebra, which is therefore deformed. For instance, in five dimensions they lead to a term $\Omega_{\alpha \beta}\varepsilon^{ij} Z \subset \{Q_{\alpha}^{i}, Q_{\beta}^{j}\}$, where the real central charge~$Z = m_a F_a$ is determined by the masses~$m_a$ and the flavor charges~$F_a$ corresponding to the currents~$\CJ_a^{ij}$. This is consistent with the interpretation of~$m_{a}\CJ_{a}^{ij}$ as the KK current, since the central charge~$Z$ can be identified with the momentum in the reduced direction. 

An even more drastic modification of the supersymmetry algebra takes place in the presence of universal mass deformations in three-dimensional theories with~$\CN \geq 4$ supersymmetry. As discussed in section~\ref{sec:defst}, the universal mass deformation resides in the stress-tensor multiplet. This leads to an unusual deformation of the supersymmetry algebra by non-central terms proportional to the unbroken~$R$-symmetry generators, see~\cite{Itzhaki:2005tu, Lin:2005nh,Agarwal:2008pu}. For instance, in~$\CN=4$ theories the universal mass~$m$ preserves the entire~$\frak{su}(2)_R \times \frak{su}(2)'_R$ symmetry, with generators~$R^{ij}, (R')^{i'j'}$, so that the deformed algebra takes the form
\begin{equation}
\big\{ Q_\alpha ^{i' i}, Q_\beta ^{j' j}\big\}=\varepsilon ^{ij}\varepsilon ^{i'j'} P_{(\alpha \beta)}+2 m \, \varepsilon _{\alpha \beta} \, \varepsilon ^{i'j'} R^{ij}-2m\, \varepsilon _{\alpha \beta}\varepsilon ^{ij}\, (R')^{i'j'}\label{mTdeform}~.
\end{equation}
In interacting theories, the appearance of the non-central~$R$-symmetry generators on the right-hand side explicitly contradicts the supersymmetric extension~\cite{Haag:1974qh} of the Coleman-Mandula theorem~\cite{Coleman:1967ad}, which follows from certain analyticity assumptions on the~$S$-matrix. (In free theories, such as~\eqref{n4hypum}, the~$S$-matrix is trivial and there is no contradiction.) As we will review below, the deformed theory is necessarily gapped. However, it may contain massive anyons, which can lead to an~$S$-matrix with non-standard analytic properties (see for instance~\cite{Jain:2014nza} for a recent discussion). In Lagrangian theories based on hypermultiplets interacting with gauge fields, the universal mass deformation triggers same-sign real mass terms for all hypermultiplet fermions, as in~\eqref{n4hypum}, which leads to a gapped theory with induced Chern-Simons terms in the deep IR.

It follows directly from the algebra~\eqref{mTdeform} that~$\CN=4$ theories with a universal mass deformation are gapped, i.e.~the low-energy effective theory must be empty or topological and does not contain massless particles. To see this, we follow~\cite{Lin:2005nh} and consider a massless particle with lightcone momentum~$P_+ = E$ and~$P_- = P_3 = 0$, on which~\eqref{mTdeform} reduces to
\begin{subequations}
\begin{align}\label{masslessi}
& \big\{ Q_+ ^{i' i}, Q_+ ^{j' j}\big\}=\varepsilon ^{ij}\varepsilon ^{i'j'}E~,\\\label{masslessii}
& \big\{ Q_- ^{i' i}, Q_- ^{j' j}\big\}=0~,\\
& \big\{ Q_+ ^{i' i}, Q_- ^{j' j}\big\}=2m\varepsilon ^{i'j'}R^{ij}-2m\varepsilon^{ij}(R')^{i'j'}~. \label{masslessiii}
\end{align}
\end{subequations}
In Lorentzian signature, the supercharges~$Q_{-}^{ii'}$ are Hermitian, so~\eqref{masslessii} implies that they are represented trivially, $Q_{-}^{ii'} = 0$. It then follows from~\eqref{masslessiii} that the~$R$-charges~$R^{ij}, (R')^{i'j'}$ must also act trivially. Since~$Q_{+}^{ii'}$ transforms as a bifundamental under these generators, this is only consistent if~$Q_{+}^{ii'} = 0$, so that the entire representation is trivial, and in particular~$E = 0$. Therefore, the only massless single-particle states are vacua, and hence the theory is gapped. The argument straightforwardly generalizes to all~$\CN \geq 5$ theories with universal masses.

\subsection{Comments on Preserving Supersymmetry at Higher Order}

Throughout this paper, we have focused on deformations~$\delta \SL  = \lambda \CO^{(1)}$ that preserve supersymmetry at leading order in the deformation parameter~$\lambda$, i.e.~the operator~$\CO^{(1)}$ is annihilated by the~$Q$-supercharges of the undeformed theory at~$\lambda = 0$, up to a total derivative. However, once the deformation has been activated, it typically does not preserve supersymmetry at~$\CO(\lambda^2)$. Resorting to weakly-coupled, Lagrangian intuition, this is due to the fact that we used the equations of motion of the undeformed theory to show that~$\CO^{(1)}$ is annihilated by the undeformed supercharges. Therefore the supercharges themselves are corrected at~$\CO(\lambda)$ and may no longer annihilate~$\CO^{(1)}$, leading to a remainder term at~$\CO(\lambda^2)$. If this term can be cancelled by the supersymmetry variation of a local operator~$\CO^{(2)}$, then the following Lagrangian is supersymmetric up to and including~$\CO(\lambda^2)$,
\begin{equation}\label{deformeddl}
\delta \SL = \lambda \CO^{(1)} + \lambda^2 \CO^{(2)}~.
\end{equation}
If such an operator~$\CO^{(2)}$ does not exist, then~$\delta \SL$ breaks supersymmetry at~$\CO(\lambda^2)$, even though the leading-order deformation~$\CO^{(1)}$ was supersymmetric. (An explicit example where this happens will be discussed below.) 

Even if~$\CO^{(2)}$ exists, the deformation~$\delta \SL$ may fail to be supersymmetric at~$\CO(\lambda^3)$. We must then repeat the procedure and look for a local operator~$\CO^{(3)}$ that can be added to~$\delta \SL$ to preserve supersymmetry at this order. This procedure may terminate after a finite number of steps, or it can continue indefinitely.  A simple example in the former category arises by deforming the free SCFT consisting of a single chiral superfield~$\Phi = \left(\phi, \psi_\alpha\right)$ by a superpotential~$\lambda W(\Phi)$.\footnote{~Here we are using an on-shell formalism without auxiliary fields, since it more closely resembles the situation encountered when deforming an abstract SCFT.} The~$\CO(\lambda)$ leading deformation takes the form (see table~\ref{tab:4DN1D}),
\begin{equation}\label{superpotdef}
\CO^{(1)} =  Q^2 \, W + \left(\text{h.c.}\right)\sim  W''(\phi) \, \psi^2 + \left(\text{h.c.}\right)~,
\end{equation}
where the~$\sim$ means that we are omitting convention-dependent numerical factors. We see from~\eqref{superpotdef} that the~$\CO(\lambda)$ deformation~$\CO^{(1)}$ only contains the Yukawa couplings, but not the scalar potential. The latter must be added to restore supersymmetry at~$\CO(\lambda^2)$, i.e.~we must choose~$\CO^{(2)} \sim |W'(\phi)|^2$, after which the procedure terminates.  A deformation that requires corrections at all orders in~$\lambda$ is the Born-Infeld-like higher-derivative~$F$-term that arises on the Coulomb branch of four-dimensional~$\CN=4$ Yang-Mills theories (see section~\ref{sec:higherderiv}). 

We will not attempt to systematically determine which deformations preserve supersymmetry beyond leading order.\footnote{\label{ft:contactterms}~One possible approach, which was mentioned in~\cite{Cordova:2015vwa}, is to examine the OPE of two or more leading-order deformations~$\CO^{(1)}$. The supersymmetry Ward identities may require the presence of certain operator-valued contact terms that can be identified with the higher-order corrections~$\CO^{(n \geq 2)}$ to the deformation.} Rather, we will use the flavor mass deformations discussed in sections~\ref{sec:flavormass} and~\ref{sec:defst} above to give a simple example of an allowed leading-order deformation that breaks supersymmetry at second order. Following the discussion around~\eqref{currentlag}, we consider flavor mass deformations of~$\CN=1,2,4$ theories in~$d =5,4,3$. The mass parameters~$m_{a, I}$ carry an adjoint index~$a$, as well as a suitable~$R$-symmetry index~$I$. In~$d = 5$, the masses are real and the index~$I$ is absent; in~$d = 4$ the masses are complex and~$I = \pm$ indicates the sign of their~$\frak{u}(1)_R$ charge, so that~$m_{a, -} = \b{ \left(m_{a, +}\right)}$\,; in~$d = 3$ they are real and~$I$ is an~$\frak{su}(2)'_R$ triplet index. (We can also consider twisted masses, for which~$I$ is an~$\frak{su}(2)_R$ triplet index.) As explained below~\eqref{currentlag}, it is useful to think of the~$m_{a, I}$ as flat Wilson lines wrapping the one-cycles of a~$(6-d)$-dimensional torus, which always preserve supersymmetry. (The torus directions are labeled by~$I$.) The vanishing of the field strength leads to the requirement\footnote{~Equivalently, \eqref{msqcons} can be derived by embedding the masses in non-dynamical background vector multiplets and demanding that the corresponding background gauginos are annihilated by all supercharges.}
\begin{equation}\label{msqcons}
f^{abc} m_{a, I} m_{b, J} = 0~,
\end{equation} 
where~$f^{abc}$ are the totally antisymmetric structure constants of the flavor Lie algebra~$\frak{g}$. Therefore, supersymmetry requires all of the~$m_{a, I}$ to reside in a Cartan subalgebra of~$\frak{g}$. This requirement is trivial in~$d = 5$, but not so in~$d = 4,3$. For instance, if~$f^{abc} m_{a, -} m_{+, b} \neq 0$ in four dimensions, then $\mathcal{N}=2$ supersymmetry is explicitly broken to $\mathcal{N}=1$~\cite{Argyres:1996eh}. Note that this is a quadratic constraint, which occurs at second order in the deformation parameters~$m_{a, I}$, all of which preserve supersymmetry at leading order. 

\section{Applications and Examples}

\label{sec:examples}

\subsection{Constraints on Moduli-Space Effective Actions}

\label{sec:moduli}

Many supersymmetric theories in~$d>2$ dimensions have a moduli space of vacua~$\CM$ that is parametrized by the expectation values of massless, scalar moduli fields~$\Phi^I$, which are nearly free in the deep IR. The low-energy effective Lagrangian~$\SL_\CM$ on the moduli space includes a sigma model for the~$\Phi^I$ with target space~$\CM$, whose kinetic terms determine the metric~$g_{IJ}(\Phi)$ on the moduli space.\footnote{~In two dimensions, there are no moduli spaces of vacua due to the strong IR fluctuations of massless scalars. However, most of our discussion concerns supersymmetric sigma models, which also exist in two dimensions.} The moduli-space Lagrangian~$\SL_\CM$ is constrained by supersymmetry, and the constraints depend on the supermultiplets in which the scalars~$\Phi^I$ reside. (Such constraints are often referred to as non-renormalization theorems.) In this subsection we will explain how our results on irrelevant supersymmetric deformations can be used to understand the supersymmetry constraints on~$\SL_{\CM}$, including the moduli-space metric~$g_{IJ}(\Phi)$ (section~\ref{sec:modmet}), as well as higher-derivative terms (section~\ref{sec:higherderiv}). We will follow the standard scaling rules for moduli-space effective actions, which assign scaling weight~$0$ to the moduli~$\Phi^I$ and weight~$\half$ to the~$Q$-supercharges, so that a derivative~$\d_\mu$ has weight~$1$. This fixes the scaling weights of all superpartners of the~$\Phi^I$. 

Our starting point is an expression for~$\SL_\CM$ as a sum of terms with moduli-dependent coefficient functions~$f_i\big(\Phi\big)$ multiplying operators~$\CO_i$ (constructed from~$\Phi^I$ and its superpartners) that can be organized according to their scaling weight,\footnote{~We can also contemplate couplings of the moduli fields to other degrees of freedom that may be present at low energy, but we will not do it here.}
\begin{equation}
\SL _\CM = \sum _i f_i(\Phi) \CO _i \supset g_{IJ}(\Phi)\, \d_\mu \Phi^I \d^\mu \Phi^J~. \label{modeflag}
\end{equation}
Here we have explicitly indicated the kinetic terms of the~$\Phi^I$, which determine the moduli space metric~$g_{IJ}(\Phi)$. They carry scaling weight~$2$ and are often the terms of lowest scaling weight in~\eqref{modeflag}.\footnote{~Some supersymmetric theories allow scalar potentials (e.g.~superpotentials in four-dimensional~$\CN=1$ theories), which carry weight~$0$. Note that such terms can lift all or part of the moduli space.} We would like to understand how the functions~$f_i(\Phi)$ and the possible~$\CO_i$ are constrained by supersymmetry. Our strategy will be to focus on a neighborhood of a point~$\langle \Phi^I\rangle$ on~$\CM$ and to consider the fluctuations~$\delta \Phi^I$ around that point. In the deep IR, the~$\delta \Phi^I$ are free fields with canonical kinetic terms, and we will reorganize the moduli-space Lagrangian in terms of higher-derivative, irrelevant corrections to this free theory. We therefore expand 
\begin{equation}
\Phi ^I = \langle \Phi ^I \rangle +\delta \Phi^{I}, \qquad f_i(\Phi)=f_i|_{\langle \Phi  \rangle}+\partial _I f_i |_{\langle \Phi \rangle} \, \delta \Phi ^I+\half \, \partial _I\partial _J f_i|_{\langle \Phi  \rangle} \, \delta \Phi ^I\delta \Phi ^I+\cdots~,\label{fexpand}
\end{equation}
and substitute into~\eqref{modeflag}. This leads to a series of irrelevant operators in the fluctuations~$\delta \Phi^I$ and their superpartners, such as
\begin{equation}
\partial_{I_{1}}\partial_{I_{2}}\cdots \partial_{I_{n}}f_{i}|_{\langle \Phi \rangle} \, \delta \Phi^{I_{1}}\delta \Phi^{I_{2}}\cdots \delta \Phi^{I_{n}} \, \mathcal{O}_{i}~.
\end{equation}
If such a term cannot be interpreted as an irrelevant supersymmetric deformation of the free IR theory consisting of the~$\delta \Phi^I$ and their superpartners, it must be absent, which leads to a differential constraint on the coefficient functions~$f_i(\Phi)$. Some higher-order terms that should be absent according to this rule can in fact be generated by supersymmetrically completing a lower-order deformation, as in the discussion around~\eqref{deformeddl}. In this case the higher-order deformation is not independent: its coefficient is completely determined by the lower-order deformation that induces it. Some examples of this phenomenon are described in~\cite{Cordova:2015vwa, Cordova:2015fha}.

The preceding discussion offers an alternative perspective on known moduli-space non-renormalization theorems, and can be used to derive new ones. We will discuss several examples below, relying on our understanding of irrelevant supersymmetric deformations in free SCFTs. In some cases, the free IR theory of the fluctuations~$\delta \Phi^I$ is not an SCFT. For instance, this happens on the Coulomb branches of maximally supersymmetric Yang-Mills theories in~$d \geq 5$, because free vector fields in these dimensions are not conformally invariant (see for instance~\cite{ElShowk:2011gz}). In such cases, our classification of irrelevant deformations does not apply and must be worked out separately.\footnote{~For maximally supersymmetric Yang-Mills theories in all dimensions, this was done in~\cite{Movshev:2009ba, Bossard:2010pk,Chang:2014kma}.}

\subsubsection{Kinetic Terms and the Moduli-Space Metric}

\label{sec:modmet}

We will now apply the general procedure outlined above to constrain the weight-$2$ kinetic terms of the moduli-space sigma model in~\eqref{modeflag}, and hence the moduli-space metric~$g_{IJ}(\Phi)$. Expanding in Riemann normal coordinates~$\delta \Phi^I$ around a point~$\langle \Phi^I\rangle$, we can express
\begin{equation}  \label{riemannnormal}
g_{IJ}(\Phi)\, \d_\mu \Phi^I \d^\mu \Phi^J = \delta_{IJ} \, \partial_\mu ( \delta \Phi^{I})\partial^\mu (\delta \Phi^{J})+\frac{1}{3} \, R_{IKJL} \, \delta \Phi^{K}\delta \Phi^{L} \, \partial_\mu (\delta \Phi^{I}) \partial^\mu (\delta \Phi^{J})+\cdots~,
\end{equation}
where~$R_{IKJL}$ is the Riemann curvature tensor, evaluated at the point~$\langle \Phi^I\rangle$, and the ellipsis denotes terms with five or more powers of~$\delta \Phi^I$. The term proportional to~$R_{IKJL}$ contains four powers of~$\delta \Phi^I$ and two derivatives, i.e.~it has weight~$2$. It must therefore be accounted for by a weight-$2$ deformation tabulated in section~\ref{sec:tables} that involves a product of four fields. (In a free theory, the number of fields is preserved by the action of the supercharges.) In the examples discussed below, the number of fields is simply related to the~$R$-symmetry quantum numbers of the superconformal primary that gives rise to the deformation.

By examining the tables in section~\ref{sec:tables}, we see that a large class of theories does not admit any supersymmetric deformations that satisfy these requirements:
\begin{itemize}

\item In three-dimensional~$\CN=5$ theories (see table~\ref{tab:3DN5D}) all deformations involving four or more fields take the form~$Q^n \CO$, with~$n \geq 6$, and therefore have weight~$\geq 3$. They can therefore not account for the term proportional to the Riemann tensor in~\eqref{riemannnormal}, and hence the moduli-space metric must be flat. 

\item Repeating the argument for~$\CN=6,8$ theories in three dimensions (see tables~\ref{tab:3DN6D} and~\ref{tab:3DN8}), and~$\CN=3$ theories in four dimensions (see table~\ref{tab:4DN3D}) immediately shows that these theories must also have flat moduli-space metrics. 

\item Four-dimensional~$\CN=4$ theories have a flat metric, because there are no weight-$2$ terms with four fields, but they do admit weight-$2$ terms with two fields (see table~\ref{tab:4DN4D}). These are the exactly marginal deformations that change the gauge coupling multiplying the kinetic terms. The latter have been canonically normalized in~\eqref{riemannnormal}.

\item  The metric on the tensor branch of six-dimensional~$\CN = (1,0)$ theories must also be flat. Although there is a candidate~$F$-term deformation of weight-$2$ that involves four fields (see table~\ref{tab:6D1D}), it requires fields that carry~$\frak{su}(2)_R$ charge, e.g.~hypermultiplets (see below). By contrast, the tensor-multiplet scalars are~$R$-symmetry neutral, because they reside in a~$C_2[0,0,0]_2^{(0)}$~multiplet (see table~\ref{tab:6DN1}). 

\item In six-dimensional~$\CN=(2,0)$ theories, the entire moduli space must be flat, because there are no candidate weight-$2$ deformations (see table~\ref{tab:6D2D}). 
 
\end{itemize}

\noindent Intuitively, tensor branches in six dimensions must have flat metrics because self-dual two-form gauge fields do not admit continuous (and hence also not moduli-dependent) couplings. 

Theories with less supersymmetry allow richer possibilities for the moduli-space metric. For instance, it is a classic result~\cite{Zumino:1979et} that~$g_{IJ}(\Phi)$ must be K\"ahler in theories with~$N_Q = 4$ supercharges. In four-dimensional~$\CN=2$ theories, the Coulomb branch is parametrized by complex scalars~$\Phi^I$ residing in~$A_2 \b B_1[0;0]_1^{(0;2)}$ vector multiplets (see tables~\ref{tab:4DN2C} and~\ref{4DN2AC}), and their complex conjugates. Supersymmetry requires the Coulomb-branch metric to obey the constraints of rigid special K\"ahler geometry (see for instance~\cite{Seiberg:1994rs,Seiberg:1994aj} and references therein).  This can be understood in terms of the fact that the only weight-$2$ deformations that can be constructed out of the~$\frak{su}(2)_R$ neutral vector multiplets are chiral and anti-chiral~$F$-terms residing in~$L\b B_1$ and~$B_1 \b L$ multiplets (see table~\ref{tab:4DN2D}). The bottom components of these multiplets furnish the holomorphic prepotential and its anti-holomorphic complex conjugate that give rise to special geometry.

Finally, we will discuss the geometry of Higgs branches parametrized by hypermultiplets, which exist in all theories with~$N_Q = 8$ supercharges.\footnote{~In~$d = 6,5,4,3$ they reside in~$D_1[0,0,0]_2^{(1)}$, $C_1[0,0]_{3/2}^{(1)}$, $B_1 \b B_1[0;0]_1^{(1;0)}, B_1[0]_{1/2}^{(1;0)}$ multiplets (see tables~\ref{tab:6DN1}, \ref{tab:5Dmult}, \ref{tab:4DN2C}, \ref{4DN2AC}, \ref{tab:3DN4}). In three-dimensional~$\CN=4$ theories, there are also~$B_1[0]_{1/2}^{(0;1)}$ twisted hypermultiplets, which are related to conventional hypermultiplets by mirror symmetry.}  It is well-known that supersymmetry requires the Higgs-branch metric to be hyperk\"ahler~\cite{AlvarezGaume:1981hm,Bagger:1983tt}. A hyperk\"ahler manifold of quaternionic dimension~$n$ has real dimension~$4n$, and the Riemannian holonomy of its metric must lie in~$Sp(2n) \subset SO(4n)$. Here we will briefly outline how the hyperk\"ahler constraint arises from the perspective of the free hypermultiplets that constitute the deep IR of the Higgs-branch effective theory. 

For simplicity, we will focus on Higgs branches of rank one, which are described by a single hypermultiplet~$H^i$ and its complex conjugate~$\b H_i  = \b{\left(H^i\right)}$. Here~$i =1,2$ is an~$\frak{su}(2)_R$ doublet index, which is raised and lowered with the~$\frak{su}(2)_R$ invariant~$\ep$-symbol. The real target space index~$I$ which appears in~\eqref{riemannnormal} is now replaced by a pair~$i, \b i$ of complex indices. Here~$\b i$ is also an~$\frak{su}(2)_R$ doublet index, which refers to the components of~$\b H^i$. The barred and unbarred indices transform in the same way under~$\frak{su}(2)_R$, so that~$\ep_{i \b j}$ is also an invariant symbol. For instance, the kinetic terms in~\eqref{riemannnormal} are now proportional to~$\ep_{\b i j} \d_\mu \b H^i \d^\mu H^j$. Note that the indices~$i,\b i$ are not standard holomorphic indices on the Higgs branch, because the corresponding K\"ahler form is proportional to~$\ep_{i \b j}$ and hence~$\frak{su}(2)_R$ invariant. By contrast, the usual~$\frak{su}(2)_R$ triplet of hyperk\"ahler forms is proportional to the Pauli matrices~$\sigma^a_{i \b j}$. 

By examining tables~\ref{tab:6D1D}, \ref{tab:5DD}, \ref{tab:4DN2D}, \ref{tab:3DN4D}, which list the supersymmetric deformations of theories with~$N_Q=8$ supercharges in~$d = 6,5,4,3$ dimensions, we see that these theories admit a unique irrelevant deformation of weight~$2$ constructed out of four hypermultiplet fields. In each case, the deformation is obtained by acting with four~$Q$-supercharges on four hypermultiplet scalars in a totally symmetric representation of the~$\frak{su}(2)_R$ symmetry with Dynkin label $(4)$, while the deformation itself is an~$\frak{su}(2)_R$ singlet.\footnote{~In~$d = 6,5,4,3$ dimensions, these deformations are the unique top components of~$D_1[0,0,0]_8^{(4)}$, $C_1[0,0]_6^{(4)}$, $B_1 \b B_1[0;0]_4^{(4;0)}$, $B_1[0]_2^{(4;0)}$ multiplets (see tables~\ref{tab:6D1D}, \ref{tab:5DD}, \ref{tab:4DN2D}, \ref{tab:3DN4D}).} 

Comparing with the normal coordinate expansion in~\eqref{riemannnormal}, we see that the Riemann tensor must be constructed using combinations of~$\frak{su}(2)_{R}$ invariant~$\ep$-symbols, in accord with its usual algebraic symmetry properties. This leads to (here~$\sim$ means that we are not keeping track of numerical coefficients)
\begin{equation}
R_{ikjl}\sim \ep_{ik}\ep_{jl}~,\qquad R_{ikj\bar{l}}\sim \varepsilon_{ik}\varepsilon_{j\bar{l}}~,\qquad R_{ik\bar{j}\bar{l}}\sim \varepsilon_{ik}\varepsilon_{\bar{j}\bar{l}}~, \label{Rforms}
\end{equation}
together with similar expressions for other components, which are related to those in~\eqref{Rforms} by complex conjugation or algebraic symmetries. The highly constrained form of the Riemann tensor in~\eqref{Rforms} implies that the local holonomy of the metric lies in~$\frak{su}(2) \simeq \frak{sp}(2) \subset \frak{so}(4)$. This is reflected by the fact that the Riemann tensor -- viewed as a map from two-forms to two-forms -- annihilates the~$\frak{su}(2)_R$ triplet of hyperk\"ahler forms proportional to~$\sigma^a_{i \b j}$ (see for instance~\cite{swann1991hyperkahler}). Therefore, the metric is locally hyperk\"ahler.  Note that our local analysis does not allow us to conclude that the Higgs-branch metric should be globally hyperk\"ahler, as is in fact required by supersymmetry~\cite{AlvarezGaume:1981hm,Bagger:1983tt}.

\subsubsection{Higher Derivative Terms}

\label{sec:higherderiv}

The approach to moduli-space effective actions described above can also be used to constrain higher-derivative terms. We will focus on a representative example: the four-derivative Born-Infeld-like deformation that exists on the Coulomb branch of four-dimensional~$\CN=4$ theories.\footnote{~Similar deformations exist in all theories with~$N_Q = 16$ (see~\cite{Paban:1998ea,Paban:1998mp,Paban:1998qy,Sethi:1999qv,Maxfield:2012aw,Lin:2015ixa,Cordova:2015vwa} and references therein).} Perhaps surprisingly, these terms are strongly constrained by supersymmetry, as was first observed in the context of BFSS matrix quantum mechanics~\cite{Paban:1998ea}. 

For simplicity, we will consider a Coulomb branch of rank one, which is described by a free abelian~$\CN=4$ vector multiplet, which constitutes a~$B_1\b B_1[0;0]_1^{(0,1,0)}$ representation of the superconformal algebra (see tables~\ref{tab:4DN4C} and~\ref{4DN4AC}). In particular, the moduli space is parametrized by six real scalars~$\Phi^I$, which transform in the vector representation~$\bf 6$ of the~$\frak{so}(6)_R \simeq \frak{su}(4)_R$ symmetry.  Schematically, the term of interest takes the following form,
\begin{equation}\label{biterm}
\delta \SL_\CM = f(\Phi) \left( F^4 +\left(\d \Phi\right)^4 +  \cdots\right)~, 
\end{equation}
where~$F$ denotes the abelian field strength and the ellipsis indicates terms with fermions. For our purposes, it will be sufficient to know that we are looking for a term of weight~$4$ (there are four derivatives and four abelian field-strengths) that involves four fields. By comparing with table~\ref{tab:4DN4D}, we see that there is a unique deformation with these properties. It is the~$F$-term deformation obtained by acting with~$Q^4 \b Q^4$ on a~$B_1 \b B_1[0;0]_4^{(0,4,0)}$ multiplet constructed out of four symmetrized vector-multiplet scalars. The deformation itself is~$R$-symmetry neutral, and hence the function~$f(\Phi)$ can only depend on the radial variable~$\varphi = \sqrt{\delta_{IJ} \Phi^I \Phi^J}$.   
 
In order to determine the function~$f(\varphi)$, we expand~\eqref{biterm} in fluctuations~$\delta \Phi^I$ around a fixed expectation value~$\langle \Phi^I\rangle$. This leads to an infinite number of terms of the schematic form 
\begin{equation}
\partial_{I_{1}}\partial_{I_{2}}\cdots \partial_{I_{n}}f|_{\langle \Phi \rangle} \, \delta \Phi^{I_{1}}\delta \Phi^{I_{2}}\cdots \delta \Phi^{I_{n}} \, \left(F^4 + \left(\d \Phi\right)^4 + \cdots\right)~. \label{series}
 \end{equation}
All of these terms have weight~$4$ and involve~$n+4$ fields. It follows from table~\ref{tab:4DN4D} that the only such deformation is an~$F$-term that arises by acting with~$Q^4 \b Q^4$ on a~$B_1 \b B_1[0;0]_{n+4}^{(0,n+4,0)}$ multiplet constructed out of~$n+4$ symmetrized vector-multiplet scalars. The deformation itself has~$R$-symmetry quantum numbers~$(0,n,0)$, i.e.~it is a totally symmetric, traceless tensor of~$\frak{so}(6)_R$. This immediately implies that the coefficients~$\partial_{I_{1}}\partial_{I_{2}}\cdots \partial_{I_{n}}f|_{\langle \Phi \rangle}$ must also be traceless. Taking~$n = 2$, we obtain 
\begin{equation}
\delta^{IJ}\partial_{I}\partial_{J}f(\Phi)=0~.
\end{equation}
Therefore~$f(\Phi)$ is a harmonic function, as was first pointed out in a quantum mechanical context~\cite{Paban:1998ea}. Since~$f(\Phi)$ only depends on the radial variable~$\varphi = \sqrt{\delta_{IJ} \Phi^I \Phi^J}$, it is fixed in terms of two constants
\begin{equation}
f(\varphi) = A + {B \over \varphi^4}~.
\end{equation}
Note that the constant~$A$ is dimensionful and must vanish on the moduli space of a conformal theory. However, our discussion did not assume that the theory whose Coulomb branch we are discussing is conformal.\footnote{~We did, however, need the fact that the free~$\CN=4$ vector multiplet constitutes an SCFT, in order to use our classification of irrelevant supersymmetric deformations in table~\ref{tab:4DN4D}.}

\subsection{Fayet-Iliopoulos Terms}
  
\label{sec:noFIterms}
 
In this section, we use the classification of supersymmetric deformations in section~\ref{sec:tables} to comment on the status of Fayet-Iliopoulos (FI) terms in different dimensions. As we will see, FI-terms cannot arise as deformations of SCFTs unless~$d = 3$. These results complement the restrictions on field-theoretic FI-terms discussed in~\cite{Komargodski:2009pc,Komargodski:2010rb,Dumitrescu:2011iu}. We will only consider theories with~$N_Q = 4,8$ supercharges. In these theories, vector multiplets have an off-shell formulation and contain a Lorentz-scalar auxiliary field~$D$. This~$D$-component is~$R$-symmetry neutral in theories with~$N_Q = 4$ and transforms under the~$R$-symmetry in theories with~$N_Q = 8$. (For brevity, we suppress the~$R$-symmetry indices.) In all cases, the FI-term is a deformation by the auxiliary~$D$-component of an abelian vector multiplet, which is gauge invariant,
\begin{equation}\label{FIterm}
\delta \SL= \xi D~.
\end{equation}
The~$R$-symmetry representation of the FI-parameter~$\xi$ is conjugate to that of the~$D$-term, e.g.~it is an~$R$-symmetry singlet in four-dimensional~$\CN=1$ theories and an~$\frak{su}(2)_R$ triplet in four-dimensional~$\CN=2$ theories. 

Since the~$D$-term resides in a vector multiplet, and at the same level as the abelian field strength~$F$, its scaling dimension is~$\Delta = 2$ in every spacetime dimension. By examining tables~\ref{tab:4DN1D}, \ref{tab:5DD}, \ref{tab:6D1D}, we see that SCFTs in~$d = 4,5,6$ do not admit relevant deformations of scaling dimension~$\Delta = 2$. Thus, despite their common appearance in supersymmetric theories with abelian gauge fields, FI-terms cannot arise as deformations of UV-complete SCFTs. Conversely, an abelian gauge theory with FI-terms cannot have a UV fixed point. In~$d = 4$, these statements can also be understood from the following alternative point of view: interacting CFTs with abelian gauge fields necessarily require electrically and magnetically charged degrees of freedom~\cite{Argyres:1995xn}. However, it was shown in~\cite{Dumitrescu:2011iu} that the magnetic current identically vanishes in theories with FI-terms. This leaves only the free abelian vector multiplet, for which the deformation~\eqref{FIterm} vanishes on-shell and can be removed by a field redefinition, up to an innocuous shift of the vacuum energy. 

In three dimensions, the situation is different: as can be seen from tables~\ref{tab:3DN1D}, \ref{tab:3DN2D}, \ref{tab:3DN3D}, \ref{tab:3DN4D}, \ref{tab:3DN5D}, \ref{tab:3DN6D}, \ref{tab:3DN8D}, all SCFTs in~$d = 3$ admit relevant deformations of scaling dimension~$\Delta = 2$, and hence the FI-term is not ruled out. On the contrary, FI-terms exist and can be interpreted as flavor mass deformations associated with the topological current~$\star F$, as was already discussed in section~\ref{sec:flavormass}. 

\subsection{Lorentz Non-Invariant Deformations}

\label{sec:nolorentz}

Throughout this paper, we have focused on deformations that preserve the full super-Poincar\'{e} algebra. We can use the same techniques to enumerate deformations that preserve supersymmetry but break Lorentz invariance. These are much less restricted, e.g.~all operators at the highest level~$\ell=\ell_\text{max}$ of any multiplet furnish such deformations. Here we briefly mention a well-studied example, which arises in the context of non-commutative gauge theories (see~\cite{Douglas:2001ba} and references therein). Using the Seiberg-Witten map \cite{Seiberg:1999vs}, gauge theories on a non-commutative geometry with coordinates~$\big[x^\mu, x^\nu\big]=i\theta^{\mu \nu}$ can be described as ordinary gauge theories deformed by a series of irrelevant operators weighted by powers of~$\theta ^{\mu \nu}$. These operators break Lorentz-invariance, but they may preserve supersymmetry. For instance, the leading non-commutative deformation of a four-dimensional~$\CN=4$  gauge theory is~(see~\cite{Intriligator:1999ai}),
\begin{equation}\label{noncomm}
\delta \SL = \theta ^{\alpha\beta}[Q^2\bar Q^4 \CO]_{\alpha \beta}+\theta ^{\dot\alpha\dot\beta}[Q^2\bar Q^4 \CO]_{\dot\alpha\dot\beta}~,
\end{equation}
where~$\theta^{\alpha\beta}$ and $\theta ^{\dot \alpha\dot\beta}$ are the self-dual and anti-self-dual components of~$\theta^{\mu\nu}$, and~$\CO$ is a~$\half$-BPS operator in a~$B_1 \b B_1[0;0]_3^{(0,3,0)}$ multiplet (see tables~\ref{tab:4DN4C}, \ref{4DN4AC}) that can be constructed out of three nonabelian vector multiplet scalars, $\CO \sim \Tr \big( \Phi^{(I} \Phi^J \Phi^{K)}\big)$. The deformations in~\eqref{noncomm} then take the following schematic form,
\begin{subequations}
\begin{align}
& [Q^2\bar Q^4 \CO]_{\alpha \beta} \sim ({\Tr} ~F^3)_{\alpha\beta}+~\cdots~\in~ [2;0]_6^{(0,0,0)}~, \\
& [Q^2\bar Q^4 \CO]_{\dot\alpha\dot\beta} \sim  ({\Tr}~ F^3)_{\dot\alpha\dot\beta}+~\cdots ~\in ~[0;2]_6^{(0,0,0)}~. 
\end{align}
\end{subequations}
They reside at the highest level~$\ell_\text{max} = 6$ of the multiplet and therefore preserve all~$N_Q = 16$ Poincar\'e supersymmetries. 

\bigskip

\section*{Acknowledgements}\noindent We are grateful to P.~Argyres, V.~Dobrev, J. Maldacena, S.~Raju, N. Seiberg, and C.~Vafa for helpful discussions.  We would like to thank the authors of~\cite{Argyres:2015ffa,Louis:2015mka} for communicating their results to us prior to publication. The work of CC was supported by a Junior Fellowship at the Harvard Society of Fellows, a Schmidt fellowship at the Institute for Advanced Study, and DOE grant DE-SC0009988. TD is supported by the Fundamental Laws Initiative at Harvard University, as well as DOE grant DE-SC0007870 and NSF grant PHY-1067976.  KI is supported by DOE grant DE-SC0009919 and the Dan Broida Chair.


\bibliographystyle{utphys}
\bibliography{deform}
\end{document}